\documentclass[a4paper,11pt]{article}
\usepackage[margin=0pt]{geometry} 

\usepackage{pgfplots}
\pgfplotsset{compat=1.15}
\usepackage{mathrsfs}
\usetikzlibrary{arrows}

\usepackage{pstricks,pst-node}

\usetikzlibrary{shapes.misc}
\tikzset{cross/.style={cross out, draw=black, fill=none, minimum size=2*(#1-\pgflinewidth), inner sep=0pt, outer sep=0pt}, cross/.default={2pt}}

\usepackage[T1]{fontenc}

\usepackage[]{amsmath, amssymb, amsthm, tabularx}

\usepackage[]{a4, xcolor, here}

\usepackage[]{pstricks, pst-text, pst-node, pst-tree, gastex}

\usepackage{tikz}

\usepackage[tikz]{bclogo}

\usepackage{comment}

\usepackage{boites,graphicx}

\usepackage{soul}

\definecolor{pastelyellow}{rgb}{0.99, 0.99, 0.59}
\definecolor{aqua}{rgb}{0.0, 1.0, 1.0} 
\definecolor{aquamarine}{rgb}{0.5, 1.0, 0.83} 
\definecolor{bananayellow}{rgb}{1.0, 0.88, 0.21}
\definecolor{burgundy}{rgb}{0.5, 0.0, 0.13}
\definecolor{ao(english)}{rgb}{0.0, 0.5, 0.0}

\setul{}{0.2ex}
\setulcolor{bananayellow}

\usepackage[normalem]{ulem}

\usepackage{mathrsfs}

\usepackage{stmaryrd}

\usepackage{enumerate}

\usepackage{paralist}

\usepackage{multienum}

\usepackage{multicol}

\usepackage{fancyhdr}

\usepackage{datetime}

\usepackage{everypage}
\usepackage[contents={},opacity=1,scale=1.6,
color=gray!90]{background}
\usepackage{ifthen}

\usepackage[]{hyperref} 

\hypersetup{
	colorlinks = true,
	linkcolor = red,
	anchorcolor = black,
	citecolor = blue, 
	filecolor = cyan,
	menucolor = red,
	runcolor = cyan,
	urlcolor = blue,
	linkbordercolor = {white},
	linktocpage = true
}

\newtheorem{theorem}{Theorem}[section]
\newtheorem{proposition}[theorem]{Proposition}
\newtheorem{lemma}[theorem]{Lemma}
\newtheorem{corollary}[theorem]{Corollary}

\theoremstyle{definition}
\newtheorem{definition}[theorem]{Definition}

\newtheorem{example}[theorem]{Example}
\newtheorem{remark}[theorem]{Remark}

\makeatletter
\def\thmhead@plain#1#2#3{%
	\thmname{#1}\thmnumber{\@ifnotempty{#1}{ }\@upn{#2}}%
	\thmnote{ {\the\thm@notefont#3}}}
\let\thmhead\thmhead@plain
\makeatother

\newcommand{\cC}{\mathcal{C}}

\newcommand{\cF}{\mathcal{F}}
\newcommand{\cG}{\mathcal{G}}

\newcommand{\cR}{\mathcal{R}}

\newcommand{\cU}{\mathcal{U}}
\newcommand{\cV}{\mathcal{V}}

\newcommand{\bbZ}{{\mathbb Z}} 
\newcommand{\bbF}{{\mathbb F}} 

\renewcommand{\geq}{\geqslant}
\renewcommand{\leq}{\leqslant}

\begin{document}

	\renewcommand{\headrulewidth}{0pt}
	
	\rhead{ }
	\chead{\scriptsize A Combinatorial Approach to Flag Codes }
	\lhead{ }

	\title{A Combinatorial Approach to Flag Codes}

	\author{\renewcommand\thefootnote{\arabic{footnote}}
		Clementa Alonso-Gonz\'alez\footnotemark[1] \  and  Miguel \'Angel Navarro-P\'erez\footnotemark[1]}

	\footnotetext[1]{Dpt.\ de Matem\`atiques, Universitat d'Alacant, 
		Sant Vicent del Raspeig, Ap.\ Correus 99, E -- 03080 Alacant.\\
		E-mail adresses: \href{mailto:clementa.alonso@ua.es}{clementa.alonso@ua.es}, \href{mailto:miguelangel.np@ua.es}{miguelangel.np@ua.es}		
		}

	{\small \date{\usdate{\today}}} 
	
	\maketitle
	
	\begin{abstract}
		In network coding, a \emph{flag code} is a collection of  \emph{flags}, that is, sequences of nested subspaces of a vector space over a finite field.  Due to its definition as the sum of the corresponding subspace distances, the \emph{flag distance} parameter encloses a hidden combinatorial structure. To bring it to light, in this paper, we interpret flag distances by means of \emph{distance paths} drawn in a convenient \emph{distance support}. The shape of such a support allows us to create an \textit{ad hoc} associated \emph{Ferrers diagram frame} where we develop a combinatorial approach to flag codes by relating the possible realizations of their minimum distance to different partitions of appropriate integers. This novel viewpoint permits to establish noteworthy connections between the flag code parameters and the ones of its projected codes in terms of well known concepts coming from the classical partitions theory.
	\end{abstract}
	
	

	\section{Introduction}\label{sec:Introduction}
	
	Random network coding, firstly introduced in \cite{AhlsCai00CHAP9}, has proven to be the most efficient way to send data across a non-coherent network with multiple sources and sinks. Despite its efficiency, it is very susceptible to error propagation and erasures. In order to overwhelm this weakness, in \cite{KoetKschi08CHAP9} the authors propose an algebraic approach by simply considering subspaces of $\mathbb{F}_q^n$ as codewords and \emph{subspace codes} as collections of subspaces.  Since this pioneering paper, much research has been made in constructing large subspace codes, and also in determining their properties. In case all the subspaces in the code have the same dimension, we speak about \emph{constant dimension codes}. To have an overview of the most important works in this subject, consult \cite{TrautRosen18CHAP9} and the references therein.
	
	In \cite{LiebNebeVaz18CHAP9} the authors developed techniques for using \emph{flags} of constant type, that is, sequences of nested subspaces of prescribed dimensions, in network coding. In this context, collections of flags are denominated \emph{flag codes} and they appear as a generalization of constant dimension codes. The recent works \cite{ConsistentesCHAP9, cotasCHAP9, CasoPlanarCHAP9,CasoNoPlanarCHAP9, Kurz20CHAP9} among others, show a growing interest in this topic. 
	
	If we consider all the subspaces of a given dimension of all the flags in a flag code, we obtain a constant dimension code called \emph{projected code}. In the study of flag codes, a central problem is the one of unraveling to what extent is it possible to get the parameters of a flag code from the ones of its projected codes and conversely.  In \cite{OrbitODFCCHAP9, CasoPlanarCHAP9, CasoNoPlanarCHAP9, MA-XCHAP9} this problem has been addressed for the family of flag codes attaining the maximum distance (\emph{optimum distance flag codes}) whereas in \cite{ConsistentesCHAP9}, the authors define \emph{consistent} flag codes as precisely those ones whose projected codes completely determine their parameters. In the paper at hand, we deal with this question for general full flag codes from an innovative combinatorial perspective. 
	
	When investigating the parameters of a flag code, one of the main difficulties lies in the definition of the distance between flags: it is obtained as the sum of their subspaces distances, which causes that many different combinations of them can give the same flag distance value. To succinctly represent these possible fluctuations, in \cite{cotasCHAP9}, the authors introduce the notion of \emph{distance vector} (associated to a given distance value). Here, we draw distance vectors in the \emph{distance support} to obtain the so-called \emph{distance paths}. This simple geometrical idea allows us to focus on the \emph{codistance} of the flag code (the complement of the distance) and hence, naturally associate to a flag code different combinatorial objects coming from the classical theory of partitions that result very convenient for our purposes.
	
	The remain of the paper is organized as follows. In Section \ref{sec: preliminaries} we remember some basics on partitions and Ferrers diagrams. We also recall some background on subspace codes and flag codes. In Section \ref{sec: on the flag distance} we address a deep study on the flag distance parameter by defining the \emph{distance support} of the full flag variety which allows us to graphically represent the \emph{distance path} of a couple of flags. We analyze the properties of such paths and we define the new concept of \emph{codistance} of a flag code. In Section \ref{sec: combinatorial approach} we translate the information that can be read in the distance support into information encoded in a combinatorial scenario. To this end, we enrich the distance support to create a \emph{Ferrers diagram frame} where each distance path will be read as a set of \emph{Ferrers subdiagrams}, that is, as a set of integer partitions. At the same time, we associate to each of such partitions its \emph{underlying distribution} that gives a particular \emph{splitting} of the corresponding codistance. In this way, we establish a one-to-one correspondence between the set of distance paths associated to a distance value and the set of splittings associated to the corresponding codistance value. Finally, in Section \ref{sec: applications}, we take advantage of the dictionary established in the previous section and, with the help of specific objects coming from the partitions world, as \emph{Durfee rectangles}, we exhibit different results that precisely relate the parameters of a flag code to the ones of its projected codes. We finish the paper with some representative examples that illustrate our results, one of them giving rise to a combinatorial characterization of full flag codes of the maximum distance.

	\section{Preliminaries}\label{sec: preliminaries}
	In this section we briefly recall the main definitions and results on partitions, Ferrers diagrams, subspace codes and flag codes that will be needed along the paper.
	\subsection{Partitions and Ferrers diagrams}
	Let us first fix some notation on integer partitions and their representation by Ferrers diagrams. Our basic reference
	related to this subject is \cite{AndCHAP9}.
	\begin{definition}
		Given a positive integer $s$, a \emph{partition} of $s$ is a sequence of non-increasing positive integers $\lambda= (\lambda_1, \ldots, \lambda_m)$ such that $\lambda_1+\dots+\lambda_m=s$. Each value $\lambda_i$ is called a \emph{part} of  $\lambda$ and we say that $m$ is the \emph{length} of $\lambda$.
	\end{definition}
	The number of partitions, usually denoted by $p(s)$, was determined asymptotically by Hardy and Ramanujan \cite{HardRamCHAP9}. A remarkable expansion by Rademacher that permits calculate $p(s)$ more accurately can be found in \cite[Chapter 5]{ApostolCHAP9}. 
	\begin{example}\label{ex: partitions of 5}
		Here we give the possible seven partitions of  $s=5$:
		\begin{equation} \label{eq1}
			\begin{split}
				5 & = 5\\
				& = 4+1\\
				& = 3+2\\
				& = 3+1+1\\
				& = 2+2+1\\
				& = 2+1+1+1\\
				& = 1+1+1+1+1.
			\end{split}
		\end{equation}
	\end{example}
	The Ferrers diagram of an integer partition provides a very useful tool for geometrically visualizing
	partitions and to extract relevant properties about them in some cases.
	\begin{definition}
		Given a partition $\lambda= (\lambda_1, \ldots, \lambda_m)$, its \emph{associated Ferrers dia\-gram} $\mathfrak{F}_{\lambda}$ is constructed by stacking  right-justified  $m$ rows of dots, where the number of dots in each row corresponds to the size $\lambda_i$ of the corresponding part. The dot at the top right position is called the \emph{corner} of the Ferrers diagram.
		
	\end{definition}
	
	\begin{example}
		The next picture shows the Ferrers diagrams associated to the partitions of $s=5$ given in Example \ref{ex: partitions of 5}.
		
		\vspace{0.5cm}
		\begin{figure}[H]
			\centering
			\begin{tikzpicture}[line cap=round,line join=round,>=triangle 45,x=0.2cm,y=0.2cm]
				\begin{scriptsize}
					\draw [fill=black] (2,4) circle (1.5pt);
					\draw [fill=black] (4,4) circle (1.5pt);
					\draw [fill=black] (6,4) circle (1.5pt);
					\draw [fill=black] (8,4) circle (1.5pt);
					\draw [fill=black] (10,4) circle (1.5pt);
					\draw [fill=black] (14,4) circle (1.5pt);
					\draw [fill=black] (16,4) circle (1.5pt);
					\draw [fill=black] (18,4) circle (1.5pt);
					\draw [fill=black] (20,4) circle (1.5pt);
					\draw [fill=black] (20,2) circle (1.5pt);
					\draw [fill=black] (24,4) circle (1.5pt);
					\draw [fill=black] (26,4) circle (1.5pt);
					\draw [fill=black] (28,4) circle (1.5pt);
					\draw [fill=black] (26,2) circle (1.5pt);
					\draw [fill=black] (28,2) circle (1.5pt);
					\draw [fill=black] (32,4) circle (1.5pt);
					\draw [fill=black] (34,4) circle (1.5pt);
					\draw [fill=black] (36,4) circle (1.5pt);
					\draw [fill=black] (36,2) circle (1.5pt);
					\draw [fill=black] (36,0) circle (1.5pt);
					\draw [fill=black] (40,4) circle (1.5pt);
					\draw [fill=black] (42,4) circle (1.5pt);
					\draw [fill=black] (40,2) circle (1.5pt);
					\draw [fill=black] (42,2) circle (1.5pt);
					\draw [fill=black] (42,0) circle (1.5pt);
					\draw [fill=black] (46,4) circle (1.5pt);
					\draw [fill=black] (48,4) circle (1.5pt);
					\draw [fill=black] (48,2) circle (1.5pt);
					\draw [fill=black] (48,0) circle (1.5pt);
					\draw [fill=black] (48,-2) circle (1.5pt);
					\draw [fill=black] (52,4) circle (1.5pt);
					\draw [fill=black] (52,2) circle (1.5pt);
					\draw [fill=black] (52,0) circle (1.5pt);
					\draw [fill=black] (52,-2) circle (1.5pt);
					\draw [fill=black] (52,-4) circle (1.5pt);
				\end{scriptsize}
			\end{tikzpicture}
			\caption{Ferrers diagrams with $5$ dots.}
		\end{figure}
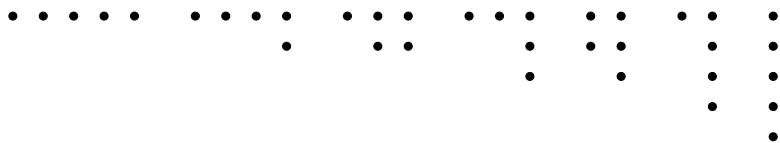
	\end{example}

	There are other important elements naturally associated to the Ferrers diagram of a partition, as their Durfee rectangles and squares, that will be helpful for our purposes as we will see in Section \ref{sec: applications}. Let us briefly recall the precise definition.

	\begin{definition}
		Given the Ferrers diagram $\mathfrak{F}_{\lambda}$ of a partition $\lambda$, we call the \emph{Durfee $k$-rectangle} associated to $\mathfrak{F}_{\lambda}$, and denote it by $D_k(\lambda)$, the largest-sized rectangle within $\mathfrak{F}_{\lambda}$ with top right vertex at the corner of $\mathfrak{F}_\lambda$ and such that its number of columns exceeds its number of rows by $k$. In particular, if $k=0$, the Durfee $k$-rectangle will be just called the \emph{Durfee square} associated to $\mathfrak{F}_{\lambda}$ and simply denoted by $D(\lambda)$.
	\end{definition}
	
	\begin{figure}[H]
		\centering
		\begin{tikzpicture}[line cap=round,line join=round,>=triangle 45,x=0.3cm,y=0.3cm]
			\fill[line width=2pt,color=blue,fill=blue,fill opacity=0.1] (19,9) -- (25,9) -- (25,3) -- (19,3) -- cycle;
			\begin{scriptsize}
				\draw [fill=black] (16,8) circle (1.5pt);
				\draw [fill=black] (18,8) circle (1.5pt);
				\draw [fill=black] (20,8) circle (1.5pt);
				\draw [fill=black] (22,8) circle (1.5pt);
				\draw [fill=black] (24,8) circle (1.5pt);
				\draw [fill=black] (20,6) circle (1.5pt);
				\draw [fill=black] (18,6) circle (1.5pt);
				\draw [fill=black] (22,6) circle (1.5pt);
				\draw [fill=black] (24,6) circle (1.5pt);
				\draw [fill=black] (24,4) circle (1.5pt);
				\draw [fill=black] (20,4) circle (1.5pt);
				\draw [fill=black] (22,4) circle (1.5pt);
				\draw [fill=black] (24,2) circle (1.5pt);
				\draw [fill=black] (20,8) circle (1.5pt);
				\draw [fill=black] (24,8) circle (1.5pt);
				\draw [fill=black] (24,4) circle (1.5pt);
				\draw [fill=black] (20,4) circle (1.5pt);
			\end{scriptsize}
		\end{tikzpicture}
		\hspace{1cm}
		\begin{tikzpicture}[line cap=round,line join=round,>=triangle 45,x=0.3cm,y=0.3cm]
			\fill[line width=2pt,color=blue,fill=blue,fill opacity=0.1] (17,9) -- (17,5) -- (25,5) -- (25,9) -- cycle;
			\begin{scriptsize}
				\draw [fill=black] (16,8) circle (1.5pt);
				\draw [fill=black] (18,8) circle (1.5pt);
				\draw [fill=black] (20,8) circle (1.5pt);
				\draw [fill=black] (22,8) circle (1.5pt);
				\draw [fill=black] (24,8) circle (1.5pt);
				\draw [fill=black] (20,6) circle (1.5pt);
				\draw [fill=black] (18,6) circle (1.5pt);
				\draw [fill=black] (22,6) circle (1.5pt);
				\draw [fill=black] (24,6) circle (1.5pt);
				\draw [fill=black] (24,4) circle (1.5pt);
				\draw [fill=black] (20,4) circle (1.5pt);
				\draw [fill=black] (22,4) circle (1.5pt);
				\draw [fill=black] (24,2) circle (1.5pt);
				\draw [fill=black] (20,8) circle (1.5pt);
				\draw [fill=black] (24,8) circle (1.5pt);
				\draw [fill=black] (24,4) circle (1.5pt);
				\draw [fill=black] (20,4) circle (1.5pt);
			\end{scriptsize}
		\end{tikzpicture}
		\caption{Durfee square and $2$-Durfee rectangle $D_2(\lambda)$ for $\lambda=(5,4,3,1)$.}
	\end{figure}
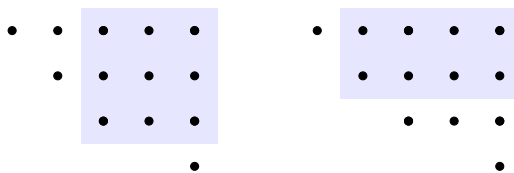
	
	In \cite{AndrewsCHAP9, GorHou68CHAP9}, the reader can find more information on these objects. 
	

	\subsection{Subspace codes and flag codes}
	
	Throughout the paper $q$ will denote a fixed prime power and $k$, $n$ two integers
	with $1 \leq k < n$. Consider $\bbF_q$ the finite field with $q$ elements and denote by $\cG_q(k, n)$ the \emph{Grassmannian}, that is,  the set of $k$-dimensional subspaces of $\bbF_q^n$. The set of vector subspaces of $\bbF_q^n$ can be equipped with different metrics but, in the current paper, we will use the so-called \emph {injection distance.}
	\begin{definition}
		The \emph{injection distance} between two subspaces $\cU, \cV \subseteq \bbF_q^n$ is defined as 
		\begin{equation}\label{eq: subspace distance}
			d_I(\cU, \cV)= \max\{\dim(\cU), \dim(\cV)\}-\dim(\cU\cap \cV). 
		\end{equation}	
		In particular, if $\cU, \cV\in\cG_q(k, n)$, then we have
		\begin{equation}\label{def: subspace distance in the Grassmannian}
			d_I(\cU, \cV)= k-\dim(\cU\cap \cV).
		\end{equation}
	\end{definition}
	Using this distance, we can define error-correcting codes in the Grassmannian as follows.
	\begin{definition}
		A \emph{constant dimension code}  $\cC$ of length $n$ and dimension $k$ is a nonempty subset of $\cG_q(k, n)$. The \emph{minimum distance} of $\cC$ is defined as 
		$$
		d_I(\cC)=\min\{ d_I(\cU, \cV) \ | \ \cU, \cV \in \cC, \ \cU \neq \cV \}
		$$	
		whenever $|\cC| \geq 2$. In case $|\cC|=1$, we put $	d_I(\cC)=0$.
	\end{definition}
	
	\begin{remark}
		Another frequent metric used when working with codes whose codewords are subspaces of $\bbF_q^n$ is the \emph{subspace distance}. It is given by
		\begin{equation}\label{eq: subspace distance ds}
			d_S(\cU, \cV)=\dim(\cU+\cV)-\dim(\cU\cap\cV). 
		\end{equation}
		Observe that, if $\cU, \cV\in\cG_q(k, n)$, then
		$$
		d_S(\cU, \cV)=2(k-\dim(\cU\cap\cV))=2d_I(\cU, \cV).
		$$
		Hence, in the context of constant dimension codes, the injection distance and the subspace distance are equivalent metrics. Consult \cite{TrautRosen18CHAP9} and the references therein for more information on this class of codes.
	\end{remark}
	
	The concept of constant dimension code can be extended when considering flags of constant type on $\bbF_q^n$, that is, sequences of nested subspaces of $\bbF_q^n$ where the list of corresponding dimensions is fixed.  The use of flags in network coding as a generalization of constant dimension codes was first proposed in \cite{LiebNebeVaz18CHAP9}. Let us recall some basic background on flag codes.
	
	\begin{definition}
		A {\em flag} $\mathcal{F}=(\mathcal{F}_1,\ldots,  \mathcal{F}_r)$ on $\mathbb{F}_{q}^n$ is a sequence of nested $\bbF_q$-vector subspaces
		$$
		\{0\}\subsetneq \mathcal{F}_1 \subsetneq \cdots \subsetneq \mathcal{F}_r \subsetneq \mathbb{F}_q^n
		$$
		of $\mathbb{F}_{q}^n$.
		The type of $\mathcal{F}$ is the vector $(\dim(\cF_1), \dots, \dim(\cF_r))$. In particular, if the type vector is $(1, 2, \ldots, n-1),$ we say that ${\cF}$ is a {\em full flag}. The subspace $\mathcal{F}_i$ is called the {\em $i$-th subspace} of $\cF$. 
	\end{definition}

	The set of all the flags on $\mathbb{F}_{q}^n$ of a fixed  type vector $(t_1, \dots, t_r)$ is said to be  the \emph{flag variety} $\mathcal{F}_q((t_1,\dots,t_r),n) \subseteq \cG_q(t_1, n) \times \cdots \times \cG_q(t_r, n)$ and, for every $i=1, \ldots, r$, we define the \emph{$i$-projection} as the map 
	\begin{equation}\label{eq: i-projection} 
		\begin{array}{rccc}
			{p_i}: & \cF_q((t_1, \ldots, t_r), n)  & \longrightarrow & \mathcal{G}_q(t_i,n)  \\
			& & &\\
			&  \cF=(\cF_1,\dots,\cF_r)   & \longmapsto   & p_i(\cF)= \cF_i.
		\end{array}
	\end{equation}
	
	The flag variety $\mathcal{F}_q((t_1,\dots,t_r),n)$ can be endowed with a metric by a natural extension of the injection  distance defined in (\ref{eq: subspace distance}). More precisely, given two flags $\cF=(\mathcal{F}_1,\dots,  \mathcal{F}_r)$ and $\cF'=(\mathcal{F}'_1,\ldots,  \mathcal{F}'_r)$ in $\mathcal{F}_q( (t_1, \ldots, t_r),n)$, the \emph{(injection) flag distance} between them is the value
	\begin{equation}\label{eq: flag distance}
		d_f(\cF,\cF')= \sum_{i=1}^r d_I(\mathcal{F}_i, \mathcal{F}'_i).  
	\end{equation}
	\begin{remark}
		Observe that the subspace distance $d_S$ defined in (\ref{eq: subspace distance ds}) can also be extended to the flag variety. Given $\cF$ and $\cF'$ as above, the sum of subspace distances
		$$
		\sum_{i=1}^{r} d_S(\cF_i, \cF'_i) = 2 d_f(\cF, \cF') 
		$$
		is an equivalent distance to $d_f$.  Due to the approach followed in this paper, it is more convenient for us to choose the injection flag distance.	
	\end{remark}	
	
	\begin{definition}
		A \emph{flag code} of type $(t_1, \dots, t_r)$ on $\bbF_{q}^n$ is a non-empty subset $\cC\subseteq \cF_q((t_1, \dots, t_r), n)$. Its {\em minimum distance} is given by
		$$
		d_f(\cC)=\min\{d_f(\cF,\cF')\ |\ \cF,\cF'\in\cC, \ \cF\neq \cF'\}.
		$$
		when $|\cC| \geq 2$. If $|\cC|=1$, we put $d_f(\cC)=0$. The \emph{$i$-projected code} of $\cC$ is the set $$\mathcal{C}_i=\{p_i(\cF)\,| \,\cF \in \cC\} \subseteq \cG_q(t_i, n).$$
	\end{definition}

	\begin{example}\label{ex: comparing flag distance with subspace distances}
		Let $\{ e_1, e_2, e_3, e_4, e_5, e_6\}$ be the canonical $\bbF_q$-basis of $\bbF_q^6$ and consider the flag code $\cC$ of type $(1,3,5)$ on $\bbF_q^6$ given by the set of flags:
		$$
		\begin{array}{ccc}
			\cF^1 & = & ( \langle e_1 \rangle, \langle e_1, e_2, e_3 \rangle, \langle e_1, e_2, e_3, e_4, e_5 \rangle),\\
			\cF^2 & = & ( \langle e_4 \rangle, \langle e_4, e_5, e_6 \rangle, \langle e_1, e_2, e_4, e_5, e_6 \rangle),\\
			\cF^3 & = & ( \langle e_5 \rangle, \langle e_4, e_5, e_6 \rangle, \langle e_2, e_3, e_4, e_5, e_6 \rangle).
		\end{array}
		$$
		Its projected codes are
		$$
		\begin{array}{ccl}
			\cC_1 & = & \left\lbrace \langle e_1 \rangle, \langle e_4 \rangle, \langle e_5 \rangle \right\rbrace,\\
			\cC_2 & = & \left\lbrace \langle e_1, e_2, e_3 \rangle, \langle e_4, e_5, e_6 \rangle\right\rbrace,\\
			\cC_3 & = & \left\lbrace \langle e_1, e_2, e_3, e_4, e_5 \rangle, \langle e_1, e_2, e_4, e_5, e_6 \rangle, \langle e_2, e_3, e_4, e_5, e_6 \rangle\right\rbrace,
		\end{array}
		$$
		with minimum distances $d_I(\cC_1)=d_I(\cC_3)=1$ and $d_I(\cC_2)=3$. Moreover, it holds
		$$
		d_f(\cC)=d_f(\cF^2, \cF^3)=1+0+1=2.
		$$
	\end{example}

	\begin{remark}\label{rem: relation distance projected}
		Note that the $i$-projected code $\cC_i$ of $\cC$ is a constant dimension code in the Grassmannian $\cG_q(t_i, n)$. At this point it is important to underline that, albeit the projected codes are constant dimension codes closely related to a flag code, they do not determine it at all; different flag codes can share the same set of projected codes. On the other hand, the cardinality of $|\cC_i|$ always satisfies $\vert \cC_i\vert \leq \vert \cC \vert$, whereas concerning the distance, we can have $d_f(\cC)> d_I(\cC_i)$, $d_f(\cC)=d_I(\cC_i)$ or even $d_f(\cC) < d_I(\cC_i)$. It suffices to see that, if $\cC$ is the flag code given in Example \ref{ex: comparing flag distance with subspace distances}, we have $d_f(\cC)=2 > 1= d_I(\cC_i)$ for $i=1, 3$, but $d_f(\cC)=2<3=d_I(\cC_2)$. This range of possibilities in the distance parameter behaviour comes from the flag distance definition itself; a fixed distance value can be obtained by adding different configurations of the subspaces distances. For instance, if we take flags $\cF^2,\cF^3$ as in Example \ref{ex: comparing flag distance with subspace distances} and consider 
		$$\cF^4= ( \langle e_3 \rangle, \langle e_3, e_5, e_6 \rangle, \langle e_2, e_3, e_4, e_5, e_6 \rangle),
		$$
		then we have 
		$$
		d_f(\cF^2, \cF^3)=  1 + 0 +1 = 2 = 1 + 1 + 0 = d_f(\cF^3, \cF^4).
		$$ 
		In \cite{cotasCHAP9} the authors deal algebraically with this question  by capturing such a variability with the so-called \emph{distance vectors}. The \emph{distance paths} defined in Section \ref{sec: on the flag distance} are a geometrical version of such distance vectors.
	\end{remark}
	
	In light of the previous remark, it naturally arises the problem of obtaining the parameters of a flag code from the ones of its projected codes and conversely. In Section \ref{sec: applications} we tackle this problem with the help of new techniques based on the combinatorial objects that we will describe along Sections \ref{sec: on the flag distance} and \ref{sec: combinatorial approach}.

	\section{Flag distance and distance paths} \label{sec: on the flag distance}
	
	In this section we deepen the study of the flag distance parameter describing its particular quirks from a brand-new combinatorial viewpoint. In the remain of the paper we will always work with full flags.
	
	As said in Section  \ref{sec: preliminaries}, the flag distance defined in (\ref{eq: flag distance}) extends the subspace distance given in (\ref{eq: subspace distance}) in the following way: the flag distance between two given flags on a vector space is exactly the sum of the distances between their subspaces.  This fact implies that, contrary to what happens with subspaces distances, flag distances conceal certain complexity in the sense that a fixed value for the flag distance can be attained from different combinations of the corresponding subspace distances.

	\begin{remark} Observe that every full flag $\cF=(\cF_1, \dots, \cF_{n-1})$ of length $n-1$ can be ``enlarged'' to the sequence of $n+1$ nested subspaces of $\bbF_q^n$ given by
		
		\begin{equation}\label{eq:extended flags}
			\bar{\cF}=(\{0\}, \cF_1, \dots, \cF_{n-1}, \bbF_q^n)  
		\end{equation}
		just by adding $\cF_0=\{0\}$ and $\cF_n=\bbF_q^n$. Now, for every pair of full flags $\cF, \cF'$, it clearly holds
		$$
		d_f(\cF, \cF')=\sum_{i=1}^{n-1} d_I(\cF_i, \cF_i') = \sum_{i=0}^{n} d_I(\cF_i, \cF_i')= d_f(\bar{\cF}, \bar{\cF}').
		$$
		\noindent So that the distance between two full flags $\cF$, $\cF'$ does not change if we extend them respectively to $\bar{\cF}, \bar{\cF}'$. Taking this fact into account, and for technical reasons, our study of the flag distance parameter will be undertaken by using extended full flags as in (\ref{eq:extended flags}). However,  observe that, when we consider an ``extended'' full flag code $\cC$, two new and trivial projected codes arise: $\cC_0=\{0\}$ and $\cC_n=\{\bbF_q^n\}$. These codes do not give any relevant information about $\cC$. Consequently, in our study, we will just take into account the projected codes $\cC_i$ with $1\leq i\leq n-1$, as usual.
	\end{remark}
	
	The injection flag distance between two (extended) full flags $\cF, \cF'$ on $\bbF_q^n$ is an integer that satisfies $0\leq d_f(\cF, \cF') \leq  D^n$, where 
	\begin{equation}\label{eq: Dn}
		D^n= \left\lfloor \frac{n^2}{4}\right\rfloor = 
		\left\lbrace   
		\begin{array}{cccl}
			\frac{n^2}{4}   & \text{if} & n & \text{is even,} \\
			& & & \\[-1em]
			\frac{n^2-1}{4} & \text{if} & n & \text{is odd.}
		\end{array}
		\right.
	\end{equation}
	This expression is a direct consequence of the possible values that the injection distance between $i$-dimensional subspaces of $\bbF_q^n$ can reach. For every $0 \leq i \leq n$, let us write $\cR(i,n)$ to denote the set of attainable injection subspace distances by subspaces in $\cG_q(i, n)$. It is clear that  $\cR(0,n)=\cR(n,n)=\{0\}$ and 
	\begin{equation}\label{def: admisibble dist}
		\cR(i,n) = \{ 0,1, \dots, \min\{ i, (n-i) \}  \}, \text{ for } i\in \{1,2, \ldots, n-1\}. 
	\end{equation}
	Hence, we deduce straightforwardly the next lemma.
	\begin{lemma}\label{lemma: distance ranges}
		The following statements hold:
		\begin{enumerate}
			\item $\cR(i,n)=\cR(n-i, n)$ for every $0\leq i\leq n$.
			\item $\cR(0, n) \subset \cR(1, n) \subset \cR(2, n) \subset \dots \subset \cR(\lfloor \frac{n}{2}\rfloor, n)$.
		\end{enumerate}
	\end{lemma}	
	
	Using this notation, for every value of $0\leq i\leq n$, we consider the set of points $\mathrm{S}(i,n)$ of $\mathbb{Z}^2$  given by
	\begin{equation}\label{def: injection distance support}
		\mathrm{S}(i,n)=\{i\} \times  \cR(i,n)= \{ (i, \delta) \in \bbZ^2 \ | \ \delta \in \cR(i,n) \}.
	\end{equation}
	
	\begin{definition}
		For every dimension $0\leq i\leq n$, the set $\mathrm{S}(i,n)$ defined in (\ref{def: injection distance support}) is called \emph{the distance support} of the Grassmannian $\cG_q(i, n)$.
	\end{definition}
	
	This geometrical representation can be generalized to the full flag variety  as follows.
	\begin{definition}
		The \emph{distance support} of the  full flag variety on $\bbF_q^n$ is the set
		\begin{equation}\label{def: injection distance support2}
			\mathrm{S}(n) = \bigcup_{i=0}^{n} \mathrm{S}(i, n) \subset \mathbb{Z}^2.
		\end{equation}
		
	\end{definition}
	
	Graphically, the distance support $\mathrm{S}(n)$ has the following representation.
	\begin{figure}[H]
		\centering
		
		\begin{tikzpicture}[line cap=round,line join=round,>=triangle 45,x=1cm,y=1cm]
			\begin{axis}[
				x=0.6cm,y=0.3cm,
				axis lines=middle,
				xmin=-1,
				xmax=8,
				ymin=-1,
				ymax=9,
				xtick={0,1,...,7},
				ytick={0,2,...,8},
				yticklabels={0,1,...,4}]
				\begin{scriptsize}
					\draw [color=black] (0,0) node[cross] {};
					\draw [color=black] (1,0) node[cross] {};
					\draw [color=black] (2,0) node[cross] {};
					\draw [color=black] (3,0) node[cross] {};
					\draw [color=black] (4,0) node[cross] {};
					\draw [color=black] (5,0) node[cross] {};
					\draw [color=black] (6,0) node[cross]{} ;
					\draw [color=black] (7,0) node[cross] {};
					\draw [fill=black] (1,2) circle (1.5pt);
					\draw [fill=black] (2,2) circle (1.5pt);
					\draw [fill=black] (2,4) circle (1.5pt);
					\draw [fill=black] (3,2) circle (1.5pt);
					\draw [fill=black] (3,4) circle (1.5pt);
					\draw [fill=black] (3,6) circle (1.5pt);
					\draw [fill=black] (4,2) circle (1.5pt);
					\draw [fill=black] (4,4) circle (1.5pt);
					\draw [fill=black] (4,6) circle (1.5pt);
					\draw [fill=black] (5,2) circle (1.5pt);
					\draw [fill=black] (5,4) circle (1.5pt);
					\draw [fill=black] (6,2) circle (1.5pt);
				\end{scriptsize}
			\end{axis}
		\end{tikzpicture}
		\hspace{0.5cm}
		\begin{tikzpicture}[line cap=round,line join=round,>=triangle 45,x=1cm,y=1cm]
			\begin{axis}[
				x=0.6cm,y=0.3cm,
				axis lines=middle,
				xmin=-1,
				xmax=9,
				ymin=-1,
				ymax=9,
				xtick={0,1,...,8},
				ytick={0,2,...,8},
				yticklabels={0,1,...,4}]
				\begin{scriptsize}
					\draw [color=black] (0,0) node[cross] {};
					\draw [color=black] (1,0) node[cross] {};
					\draw [color=black] (2,0) node[cross] {};
					\draw [color=black] (3,0) node[cross] {};
					\draw [color=black] (4,0) node[cross] {};
					\draw [color=black] (5,0) node[cross] {};
					\draw [color=black] (6,0) node[cross] {};
					\draw [color=black] (7,0) node[cross] {};
					\draw [color=black] (8,0) node[cross] {};
					\draw [fill=black] (1,2) circle (1.5pt);
					\draw [fill=black] (2,2) circle (1.5pt);
					\draw [fill=black] (2,4) circle (1.5pt);
					\draw [fill=black] (3,2) circle (1.5pt);
					\draw [fill=black] (3,4) circle (1.5pt);
					\draw [fill=black] (3,6) circle (1.5pt);
					\draw [fill=black] (4,2) circle (1.5pt);
					\draw [fill=black] (4,4) circle (1.5pt);
					\draw [fill=black] (4,6) circle (1.5pt);
					\draw [fill=black] (4,8) circle (1.5pt);
					\draw [fill=black] (5,2) circle (1.5pt);
					\draw [fill=black] (5,4) circle (1.5pt);
					\draw [fill=black] (5,6) circle (1.5pt);
					\draw [fill=black] (6,2) circle (1.5pt);
					\draw [fill=black] (6,4) circle (1.5pt);
					\draw [fill=black] (7,2) circle (1.5pt);
				\end{scriptsize}
			\end{axis}
		\end{tikzpicture}
		\caption{Distance supports $\mathrm{S}(7)$ and $\mathrm{S}(8).$}\label{fig: dist support n=7, 8}
	\end{figure}
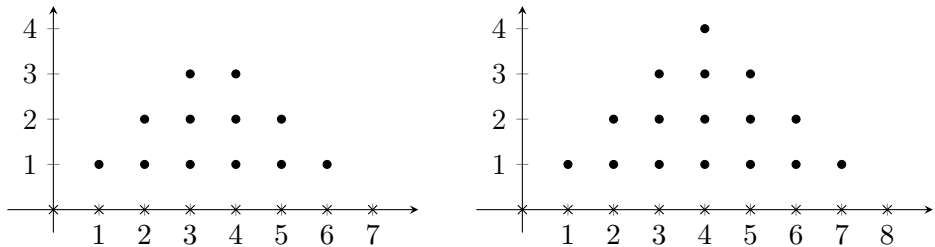
	
	\begin{remark}\label{rem: number of points in S(n)}
		The reader can appreciate that two different node styles have been used in Figure \ref{fig: dist support n=7, 8}. This is due to the fact that not every point contributes equally when we use the support to represent flag distances. On one side, crossed dots denote null distances. On the other side, there are exactly  $D^n$ circle dots representing positive distances. This dichotomy will be very useful when representing distances between pairs of flags in  $\mathrm{S}(n)$. 
		
		It is clear that the $i$-th column of the distance support $\mathrm{S}(n)$ is just the distance support $\mathrm{S}(i, n)$ of $\cG_q(i, n)$. Hence, as a consequence of Lemma \ref{lemma: distance ranges},  the support $\mathrm{S}(n)$ is symmetric with respect to the vertical line $x=\frac{n}{2}$ and the columns heights grow as the dimension gets closer to $\frac{n}{2}$. Observe that there is a remarkable difference in  the distance support shape depending on the parity of $n$: when $n$ is even, the value $\frac{n}{2}$ is a dimension on the type vector. In this case, $\mathrm{S}(n)$ has a \textit{peak} in this central dimension. In contrast, when $n$ is odd, $\mathrm{S}(n)$ presents a \textit{plateau} on its top. The higher attainable distances in this latter case are placed at dimensions $\lfloor \frac{n}{2}\rfloor$ and $\lceil \frac{n}{2}\rceil$, i.e.,  the closest integers (from left and right, respectively) to the value $\frac{n}{2}$.
	\end{remark}	
	
	\begin{remark}\label{rem: S(n-1) from S(n)}
		Concerning also the distance support shape, notice that the distance support $\mathrm{S}(n-1)$ can be obtained from $\mathrm{S}(n)$ just by removing the set of points with coordinates $(i, (n-i))$ whenever $2i\geq n$. These points are the ones in the ``right-roof'' as in the next figure. 
	\end{remark}

	\begin{figure}[H]
		\centering
		\begin{tikzpicture}[line cap=round,line join=round,>=triangle 45,x=1cm,y=1cm]
			\begin{axis}[
				x=0.6cm,y=0.3cm,
				axis lines=middle,
				xmin=-1,
				xmax=9,
				ymin=-1,
				ymax=9,
				xtick={0,1,...,8},
				ytick={0,2,...,8},
				yticklabels={0,1,...,4},]
				\begin{scriptsize}
					\draw [color=black] (0,0) node[cross] {};
					\draw [color=black] (1,0) node[cross] {};
					\draw [color=black] (2,0) node[cross] {};
					\draw [color=black] (3,0) node[cross] {};
					\draw [color=black] (4,0) node[cross] {};
					\draw [color=black] (5,0) node[cross] {};
					\draw [color=black] (6,0) node[cross] {};
					\draw [color=black] (7,0) node[cross] {};
					\draw [fill=white] (8,0) circle (1.5pt);
					\draw [fill=black] (1,2) circle (1.5pt);
					\draw [fill=black] (2,2) circle (1.5pt);
					\draw [fill=black] (2,4) circle (1.5pt);
					\draw [fill=black] (3,2) circle (1.5pt);
					\draw [fill=black] (3,4) circle (1.5pt);
					\draw [fill=black] (3,6) circle (1.5pt);
					\draw [fill=black] (4,2) circle (1.5pt);
					\draw [fill=black] (4,4) circle (1.5pt);
					\draw [fill=black] (4,6) circle (1.5pt);
					\draw [fill=white] (4,8) circle (1.5pt);
					\draw [fill=black] (5,2) circle (1.5pt);
					\draw [fill=black] (5,4) circle (1.5pt);
					\draw [fill=white] (5,6) circle (1.5pt);
					\draw [fill=black] (6,2) circle (1.5pt);
					\draw [fill=white] (6,4) circle (1.5pt);
					\draw [fill=white] (7,2) circle (1.5pt);
				\end{scriptsize}
			\end{axis}
		\end{tikzpicture}
		\hspace{0.25cm}
		\begin{tikzpicture}[line cap=round,line join=round,>=triangle 45,x=1cm,y=1cm]
			\begin{axis}[
				x=0.6cm,y=0.3cm,
				axis lines=middle,
				xmin=-1,
				xmax=9,
				ymin=-1,
				ymax=9,
				xtick={0,1,...,8},
				ytick={0,2,...,8},
				yticklabels={0,1,...,4}]
				\begin{scriptsize}
					\draw [color=black] (0,0) node[cross] {};
					\draw [color=black] (1,0) node[cross] {};
					\draw [color=black] (2,0) node[cross] {};
					\draw [color=black] (3,0) node[cross] {};
					\draw [color=black] (4,0) node[cross] {};
					\draw [color=black] (5,0) node[cross] {};
					\draw [color=black] (6,0) node[cross] {};
					\draw [color=black] (7,0) node[cross] {};
					\draw [fill=black] (1,2) circle (1.5pt);
					\draw [fill=black] (2,2) circle (1.5pt);
					\draw [fill=black] (2,4) circle (1.5pt);
					\draw [fill=black] (3,2) circle (1.5pt);
					\draw [fill=black] (3,4) circle (1.5pt);
					\draw [fill=black] (3,6) circle (1.5pt);
					\draw [fill=black] (4,2) circle (1.5pt);
					\draw [fill=black] (4,4) circle (1.5pt);
					\draw [fill=black] (4,6) circle (1.5pt);
					\draw [fill=black] (5,2) circle (1.5pt);
					\draw [fill=black] (5,4) circle (1.5pt);
					\draw [fill=black] (6,2) circle (1.5pt);
				\end{scriptsize}
			\end{axis}
		\end{tikzpicture}
		\caption{Getting $\mathrm{S}(7)$ (right) from $\mathrm{S}(8)$ (left).}
	\end{figure}
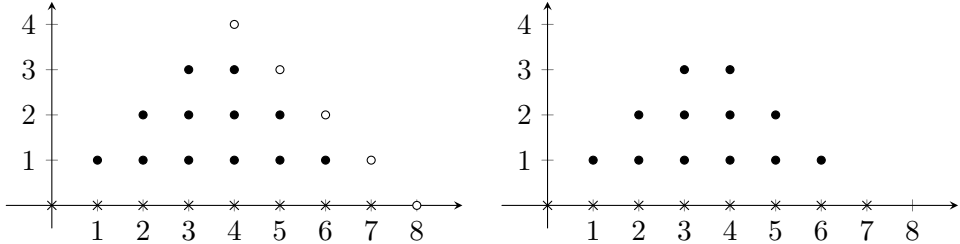

	At this point, if we consider two full flags on $\bbF_q^n$, their flag distance can be geometrically represented by means of a collection of $n+1$ points in the distance support $\mathrm{S}(n)$, each one of them in a different column $\mathrm{S}(i, n)$ as follows.
	
	\begin{definition}\label{def: distance path}
		Given a pair of full flags $\cF$ and $\cF'$ on $\bbF_q^n$, we define their \emph{distance path} $\Gamma(\cF, \cF')$ as the directed polygonal path whose vertices are the points $(i, d_I(\cF_i, \cF'_i))$ for every $0\leq i\leq n$. 
	\end{definition} 
	
	\begin{figure}[H]
		\centering
		\begin{tikzpicture}[line cap=round,line join=round,>=triangle 45,x=1cm,y=1cm]
			\begin{axis}[
				x=0.6cm,y=0.3cm,
				axis lines=middle,
				xmin=-1,
				xmax=8,
				ymin=-1,
				ymax=7,
				xtick={0,1,...,7},
				ytick={0,2,...,6},
				yticklabels={0,1,...,3}]
				\draw [line width=1pt,color=red] (0,0)-- (1,2);
				\draw [line width=1pt,color=red] (1,2)-- (2,4);
				\draw [line width=1pt,color=red] (2,4)-- (3,2);
				\draw [line width=1pt,color=red] (3,2)-- (4,2);
				\draw [line width=1pt,color=red] (4,2)-- (5,2);
				\draw [line width=1pt,color=red] (5,2)-- (6,0);
				\draw [line width=1pt,color=red] (6,0)-- (7,0);
				\draw [line width=1pt,color=blue] (0,0)-- (1,2);
				\draw [line width=1pt,color=blue] (1,2)-- (2,2);
				\draw [line width=1pt,color=blue] (2,2)-- (3,0);
				\draw [line width=1pt,color=blue] (3,0)-- (4,2);
				\draw [line width=1pt,color=blue] (4,2)-- (5,4);
				\draw [line width=1pt,color=blue] (5,4)-- (6,2);
				\draw [line width=1pt,color=blue] (6,2)-- (7,0);
				\begin{scriptsize}
					\draw [color=black] (0,0) node[cross] {};
					\draw [color=black] (1,0) node[cross] {};
					\draw [color=black] (2,0) node[cross] {};
					\draw [color=black] (3,0) node[cross] {};
					\draw [color=black] (4,0) node[cross] {};
					\draw [color=black] (5,0) node[cross] {};
					\draw [color=black] (6,0) node[cross] {};
					\draw [color=black] (7,0) node[cross] {};
					\draw [fill=black] (1,2) circle (1.5pt);
					\draw [fill=black] (2,2) circle (1.5pt);
					\draw [fill=black] (2,4) circle (1.5pt);
					\draw [fill=black] (3,2) circle (1.5pt);
					\draw [fill=black] (3,4) circle (1.5pt);
					\draw [fill=black] (3,6) circle (1.5pt);
					\draw [fill=black] (4,2) circle (1.5pt);
					\draw [fill=black] (4,4) circle (1.5pt);
					\draw [fill=black] (4,6) circle (1.5pt);
					\draw [fill=black] (5,2) circle (1.5pt);
					\draw [fill=black] (5,4) circle (1.5pt);
					\draw [fill=black] (6,2) circle (1.5pt);
				\end{scriptsize}
			\end{axis}
		\end{tikzpicture}
		\caption{Examples of distance paths in $\mathrm{S}(7)$.}\label{fig: distance path examples}
	\end{figure}
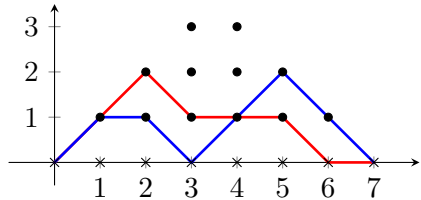
	
	Similarly, given a full flag code, we can consider a collection of distance paths associated to it.
	\begin{definition}\label{def: distance path of a code}
		Let $\cC$ be a full flag code on $\bbF_q^n$. The set of \emph{distance paths of $\cC$} is the set
		$$
		\Gamma(\cC) = \{ \Gamma(\cF, \cF') \ | \ \cF, \cF'\in\cC, \ \cF\neq \cF'\}.
		$$
	\end{definition} 
	
	Notice that every distance path in $\mathrm{S}(n)$ starts at the point $(0,0)$ and arrives to $(n, 0)$. Nevertheless, it is important to point out that not every polygonal path with vertices in the support $\mathrm{S}(n)$ satisfying this condition represents a potential distance path of a pair of flags. In order to characterize the polygonal paths in $\mathrm{S}(n)$ also being distance paths between a couple of flags in $\bbF_q^n$, in the following result we see that, for a given pair of full flags $\cF, \cF'$ on $\bbF_q^n$, the value of $d_I(\cF_i, \cF_i')$ completely determines the range of possibilities for $d_I(\cF_{i+1}, \cF_{i+1}').$
	
	\begin{theorem}\label{prop: allowed pattern}
		Consider $\cF, \cF'$ full flags  on $\bbF_q^n$ and denote  $\delta_i=d_I(\cF_i, \cF'_i)$ where $i \in \{0,1,\ldots, n\}$. Then, for any $0\leq i < n$, it holds
		$$
		\delta_{i+1}\in \{ \delta_i-1, \delta_i, \delta_i+1\}.
		$$
	\end{theorem}
	\begin{proof}
		The proof is based on the flags nested structure. Consider full flags $\cF$ and $\cF'$ on $\bbF_q^n$. For every $1\leq i< n-1$, we have
		$$
		\cF_i\cap \cF'_i \subseteq \cF_{i+1}\cap \cF'_{i+1} \ \text{and} \ \cF_i + \cF'_i \subseteq \cF_{i+1} + \cF'_{i+1}.
		$$
		The second inclusion leads to the next inequality 
		$$
		2i - \dim(\cF_i\cap \cF'_i) \leq 2(i+1) - \dim(\cF_{i+1}\cap \cF'_{i+1})
		$$
		or, equivalently, 
		$$
		\dim(\cF_{i+1}\cap \cF'_{i+1}) \leq \dim(\cF_i\cap\cF'_i) + 2.
		$$
		Using this fact, and taking into account that $d_I(\cF_{j},\cF'_j)=j-\dim(\cF_j\cap\cF'_j)$ for every $1\leq j\leq n-1$, it follows that
		$$
		i+1 - \dim(\cF_i\cap\cF'_i) -2 \leq d_I(\cF_{i+1}, \cF'_{i+1}) \leq i+1 - \dim(\cF_i\cap\cF'_i),
		$$
		which is equivalent to 
		$$
		i- \dim(\cF_i\cap\cF'_i) -1 \leq d_I(\cF_{i+1}, \cF'_{i+1}) \leq i- \dim(\cF_i\cap\cF'_i)+1.
		$$
		Hence,
		$$ 
		d_I(\cF_{i}, \cF'_{i}) -1 \leq d_I(\cF_{i+1}, \cF'_{i+1})\leq d_I(\cF_{i}, \cF'_{i}) +1.
		$$
		In other words, we have that the value $\delta_{i+1}=d_I(\cF_{i+1}, \cF'_{i+1})$ is an element in $\{\delta_i-1, \delta_i, \delta_i+1\}$ as we wanted to prove.
	\end{proof}
	
	\begin{example}
		In view of Theorem \ref{prop: allowed pattern}, the next figure illustrates how a distance path is allowed to continue once we have fixed one of its points. From some points, for instance $(3,1)$, we have three options to continue. On the other hand, there are just two possibilities if we fix either the point $(0,0)$ or the point $(5,1)$. Last, paths passing through $(4,3)$ must contain the point $(5,2)$.
		
		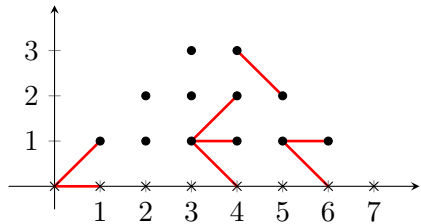
\begin{figure}[H]\label{fig: trident pattern}
			\centering 
			\begin{tikzpicture}[line cap=round,line join=round,>=triangle 45,x=1cm,y=1cm]
				\begin{axis}[
					x=0.6cm,y=0.3cm,
					axis lines=middle,
					xmin=-1,
					xmax=8,
					ymin=-1,
					ymax=8,
					xtick={0,1,...,7},
					ytick={0,2,...,6},
					yticklabels={0,1,...,3}]
					\draw [line width=1pt,color=red] (3,2)-- (4,4);
					\draw [line width=1pt,color=red] (3,2)-- (4,2);
					\draw [line width=1pt,color=red] (3,2)-- (4,0);
					\draw [line width=1pt,color=red] (5,2)-- (6,2);
					\draw [line width=1pt,color=red] (5,2)-- (6,0);
					\draw [line width=1pt,color=red] (0,0)-- (1,0);
					\draw [line width=1pt,color=red] (0,0)-- (1,2);
					\draw [line width=1pt,color=red] (4,6)-- (5,4);
					\begin{scriptsize}
						\draw [color=black] (0,0) node[cross] {};
						\draw [color=black] (1,0) node[cross] {};
						\draw [color=black] (2,0) node[cross] {};
						\draw [color=black] (3,0) node[cross] {};
						\draw [color=black] (4,0) node[cross] {};
						\draw [color=black] (5,0) node[cross] {};
						\draw [color=black] (6,0) node[cross] {};
						\draw [color=black] (7,0) node[cross] {};
						\draw [fill=black] (1,2) circle (1.5pt);
						\draw [fill=black] (2,2) circle (1.5pt);
						\draw [fill=black] (2,4) circle (1.5pt);
						\draw [fill=black] (3,2) circle (1.5pt);
						\draw [fill=black] (3,4) circle (1.5pt);
						\draw [fill=black] (3,6) circle (1.5pt);
						\draw [fill=black] (4,2) circle (1.5pt);
						\draw [fill=black] (4,4) circle (1.5pt);
						\draw [fill=black] (4,6) circle (1.5pt);
						\draw [fill=black] (5,2) circle (1.5pt);
						\draw [fill=black] (5,4) circle (1.5pt);
						\draw [fill=black] (6,2) circle (1.5pt);
					\end{scriptsize}
				\end{axis}
			\end{tikzpicture}
			\caption{Allowed movements from some points in $\mathrm{S}(7)$.}
		\end{figure}
	\end{example}
	
	In general, given the point $(i, \delta_i)$ in $\mathrm{S}(i, n)$, distance paths passing through it can only come from a point $(i-1, \delta_{i-1}) \in \mathrm{S}(i-1, n)$ with 
	\begin{equation}\label{eq: rule 1}
		\delta_{i-1}\in\{\delta_i-1, \delta_i, \delta_i+1\}.   
	\end{equation}
	At the same time, these paths can only continue through points $(i+1, \delta_{i+1}) \in \mathrm{S}(i+1, n)$
	\begin{equation}\label{eq: rule 2}
		\delta_{i+1}\in\{\delta_i-1, \delta_i, \delta_i+1\}.    
	\end{equation}
	All in all, distance paths are, graphically, oriented polygonal paths passing through points
	\begin{equation}\label{eq: distance path}
		(0,0) \rightarrow (1, \delta_1) \rightarrow (2, \delta_2) \rightarrow \dots \rightarrow (n-1, \delta_{n-1}) \rightarrow (n,0)    
	\end{equation}
	such that consecutive vertices $(i, \delta_i)$ and $(i+1, \delta_{i+1})$ are related according to the \textit{trident} rules given by (\ref{eq: rule 1}) and (\ref{eq: rule 2}).

	\begin{remark}
		Observe that distance paths defined as above are the graphic representation of the notion of the \emph{distance vector} associated to a couple flags introduced in \cite{cotasCHAP9}. Moreover, our Theorem \ref{prop: allowed pattern} is a geometric version of Theorem $3.9$ in  \cite{cotasCHAP9}, for the special case of full flags. In particular, as a consequence of that result, we can assure that, given a path $\Gamma$ in $S(n)$ satisfying (\ref{eq: rule 1}), (\ref{eq: rule 2}) and (\ref{eq: distance path}), there is always a couple of full flags $\cF, \cF'$ on $\bbF_q^n$ such that $\Gamma=\Gamma(\cF, \cF')$. 
	\end{remark}
	
	In view of the previous remark, from now on, a path $\Gamma$ in $S(n)$ described by (\ref{eq: rule 1}), (\ref{eq: rule 2}) and (\ref{eq: distance path}) will be said a \textit{distance path} given that it represents the flag distance value  $d_\Gamma=\sum_{i=0}^{n}\delta_i$, attained by a couple of full flags $\cF$ and $\cF'$ on $\bbF_q^n$ such that $d_I(\cF_i, \cF'_i)=\delta_i$. Conversely, this tool allows us to geometrically determine how an arbitrary distance flag value $d$ can suitably split into the $n-1$ terms that are not trivially zero (recall that $\delta_0=\delta_n=0)$. For instance, looking again at paths $\Gamma$ and $\Gamma'$ in $\mathrm{S}(7)$ in Figure \ref{fig: distance path examples}, we can say that
	$$
	6 =0+1+2+1+1+1+0+0 = 0+1+1+0+1+2+1+0
	$$
	are permitted subspace distance combinations for the distance $d_\Gamma= d_{\Gamma'}=6.$ In terms of the language used in \cite{cotasCHAP9}, these two paths correspond, respectively, to the distance vectors $(0,1,2,1,1,1,0,0)$ and $(0,1,1,0,1,2,1,0)$.

	Note that any distance path ${\Gamma}$ in $\mathrm{S}(n)$ consists of $n$ segments of lines with slope equal to  $-1$, $0$ or $1$. Moreover, recall that every distance path starts at the origin and ends at the point $(n,0)$, both with null height. Hence, the number of edges with positive slope in a distance path must coincide with the one of edges with negative slope. As a result, the number of horizontal segments appearing in $\Gamma$ has the same parity than $n$. These latter edges will play an important role in the following section. Let us define them more precisely.
	
	\begin{definition}\label{def: plateu}
		A \emph{plateau of height $\delta$} in a distance path ${\Gamma}$ is a sequence of two consecutive vertices $(i, \delta), (i+1, \delta)$ on it. In other words, if we consider two full flags $\cF, \cF'$ on $\bbF_q^n$ with ${\Gamma}=\Gamma(\cF, \cF')$, a \textit{plateau} appears when $d_I(\cF_i, \cF'_i)=d_I(\cF_{i+1}, \cF'_{i+1})$ for some $0\leq i\leq n-1$. We denote by $p_{\Gamma}$ the number of \textit{plateaus} on a given distance path ${\Gamma}$.
	\end{definition}
	\begin{example}
		The distance path in red represented in Figure \ref{fig: distance path examples} has two \textit{plateaus} of  height $1$. As said before, if $n$ is an even (resp. odd) positive integer, then every distance path $\Gamma$ in $\mathrm{S}(n)$ contains an even (resp. odd) number of \textit{plateaus}. In particular, for odd values of $n$, distance paths must contain at least one \textit{plateau}. 
	\end{example}
	
	\begin{remark}\label{rem: distance 2 number of dots}
		Let us briefly come back to the support $\mathrm{S}(n)$. In the following sections it will be important to compute the number of dots associated to a given distance path. Notice that in $\mathrm{S}(i,n)$, a point $(i, \delta_i)$ leaves exactly $\delta_i$ circle dots and one crossed dot bellow it (including the point $(i, \delta_i)$ itself).  Using this idea, we can compute the associated distance of a given distance path $\Gamma$ by simply counting the number of circle dots on $\Gamma$ or bellow $\Gamma$. Moreover, we can relate the value $d_\Gamma$ to the area of the polygons determined by $\Gamma$ together with the abscissa axis. In Figure  \ref{fig: distance path examples}, the red path on $\mathrm{S}(7)$ determines  a single polygon having the points $(0,0)$ and $(6,0)$ as  vertices whereas the path in blue determines two of them. On the other hand, the path in Figure \ref{fig: path distance and codistance} forms a unique polygon in $\mathrm{S}(7)$ with vertices on the points $(0,0)$ and $(7,0)$.
	\end{remark}
	
	\begin{theorem}\label{prop: distance = area}
		Let $\Gamma$ be a distance path in $S(n)$ such that $\Gamma$ determines a unique polygon $P_\Gamma$ with the abscissa axis. Then the flag distance $d_{\Gamma}$ is exactly the area of $P_\Gamma$.  
	\end{theorem}
	\begin{proof}
It is enough to point out that $P_\Gamma$ is a reticulated polygon with vertices in the integer lattice $\bbZ^2$. We start assuming that the points $(0,0)$ and $(n,0)$ are vertices of $P_\Gamma$. In this case, if we write $I$ and $B$ to denote the set of lattice points in the interior and the boundary of $P_\Gamma$, respectively, then we have that $I$ is a set of circle dots. However, $B$ contains $n-1$ circle dots and $n+1$ crossed ones. Consequently, according to Remark \ref{rem: distance 2 number of dots}, it holds $d_\Gamma= |I|+ n-1$.
		
		By means of Pick's Theorem, the area of $P_\Gamma$ can be computed in terms of $I$ and $B$ as
		$$
		A(P_\Gamma) =  |I| + \frac{|B|}{2} - 1 = |I| + n-1 = d_\Gamma.
		$$
On the other hand, if $\Gamma$ determines a unique polygon $P_\Gamma$ with vertices $(i, 0)$ and $(j, 0)$, for some $0 \leq i < j \leq n$, then the result follows by interpreting $P_\Gamma$ as a polygon in a smaller distance support $\mathrm{S}(j-i)$ with the points $(0,0)$ and $(j-i, 0)$ as its vertices and arguing as above.
	\end{proof}
	The following corollary follows then straightforwardly:
	\begin{corollary}
		Let $\Gamma$ be a distance path in $S(n)$ such that $\Gamma$ determines the polygons $P_\Gamma^1$,\ldots, $P_\Gamma^k$ with the abscissa axis. Then the flag distance $d_{\Gamma}$ is exactly the sum of the areas of $P_\Gamma^1,\ldots, P_\Gamma^k$.
	\end{corollary}
	
	By means of Proposition \ref{prop: allowed pattern} and Theorem \ref{prop: distance = area}, we can remove the coordinate axes when representing flag distances in the support $\mathrm{S}(n)$. In fact, we just need to study paths constructed by chaining the \textit{trident} moves represented in Figure \ref{fig: trident pattern}, and to count how many circle dots remain in or under such paths. Of course, different distance paths can provide the same flag distance, i.e., they leave the same amount of circle dots below them or, equivalently, above them. Next we introduce the notion of \emph{(flag) codistance} as a complementary value associated to a flag distance which will be crucial in the remain of the paper.
	
	\begin{definition}
		Given a flag distance value $d$, i.e., an integer such that $0\leq d\leq D^n$, we define its \emph{(injection flag) codistance} as the value  $\bar{d} = D^n - d$.  Similarly, given a full flag code $\cC$ on $\bbF_q^n$, we define its associated \emph{codistance} as the value $\bar{d}_f(\cC)= D^n - d_f(\cC)$.
	\end{definition}
	
	Notice that both $d$ and $\bar{d}$ provide exactly the same information since every flag distance value determines a unique codistance value and conversely.  Arguing as in Remark \ref{rem: distance 2 number of dots}, the codistance can be read in the distance support as follows.
	
	\begin{corollary}\label{cor: codistance}
		The number of dots over a distance path $\Gamma$ in  $\mathrm{S}(n)$  is equal to the codistance $\bar{d}_\Gamma=D^n-d_\Gamma$ associated to it. 
	\end{corollary}
	
	\begin{example}
		Take $n=7$ and consider the next distance path $\Gamma$ in $\mathrm{S}(7)$ 
		\begin{equation} \label{eq: example distance path}
			\Gamma : \ (0,0) \rightarrow (1,1) \rightarrow (2, 2)  \rightarrow (3, 1) \rightarrow (4, 1) \rightarrow (5, 2) \rightarrow (6,1) \rightarrow (7,0).    
		\end{equation}
		There are $D^7=12$ circle dots in the distance support $\mathrm{S}(7)$. The distance path passes exactly through $6$ of them (in black) and the associated polygon $P_\Gamma$ contains $2$ black circle dots in its interior. The area of such a polygon is exactly $8$ (see the picture below). Hence, the distance path in (\ref{eq: example distance path}) represents a possible distribution to obtain the flag distance $d=8$. Indeed,  $8=1+2+1+1+2+1$. The corresponding codistance is $12-8=4$, which is, as said in Corollary \ref{cor: codistance}, is the number of points over the path (white circle dots).
		
		\vspace{0.5cm}
		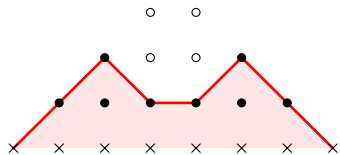
\begin{figure}[H]
			\centering
			\begin{tikzpicture}[line cap=round,line join=round,>=triangle 45,x=0.6cm,y=0.3cm]
				\draw [line width=1pt,color=red] (1,2)-- (0,0);
				\draw [line width=1pt,color=red] (1,2)-- (2,4);
				\draw [line width=1pt,color=red] (2,4)-- (3,2);
				\draw [line width=1pt,color=red] (3,2)-- (4,2);
				\draw [line width=1pt,color=red] (4,2)-- (5,4);
				\draw [line width=1pt,color=red] (5,4)-- (6,2);
				\draw [line width=1pt,color=red] (6,2)-- (7,0);
				\fill[line width=2pt,color=red,fill=red,fill opacity=0.1] (0,0) -- (2,4) -- (3,2) -- (4,2) --(5,4) --(7,0) -- cycle;
				\begin{scriptsize}
					\draw [color=black] (0,0) node[cross] {};
					\draw [color=black] (1,0) node[cross] {};
					\draw [color=black] (2,0) node[cross] {};
					\draw [color=black] (3,0) node[cross] {};
					\draw [color=black] (4,0) node[cross] {};
					\draw [color=black] (5,0) node[cross] {};
					\draw [color=black] (6,0) node[cross] {};
					\draw [color=black] (7,0) node[cross] {};
					\draw [fill=black] (1,2) circle (1.5pt);
					\draw [fill=black] (2,2) circle (1.5pt);
					\draw [fill=black] (2,4) circle (1.5pt);
					\draw [fill=black] (3,2) circle (1.5pt);
					\draw [color=black] (3,4) circle (1.5pt);
					\draw [color=black] (3,6) circle (1.5pt);
					\draw [fill=black] (4,2) circle (1.5pt);
					\draw [color=black] (4,4) circle (1.5pt);
					\draw [color=black] (4,6) circle (1.5pt);
					\draw [fill=black] (5,2) circle (1.5pt);
					\draw [fill=black] (5,4) circle (1.5pt);
					\draw [fill=black] (6,2) circle (1.5pt);
				\end{scriptsize}
			\end{tikzpicture}
			\caption{A distance path $\Gamma$ with associated distance and codistance.}\label{fig: path distance and codistance}
		\end{figure}
	\end{example}
	
	\section{Combinatorial perspective}\label{sec: combinatorial approach}
	
	In this section, we introduce some combinatorial objects closely related to the distance support. To do so, we need to enrich it with an auxiliary collection of (red) points  that will allow us to obtain a Ferrers diagram. With the help of such a diagram we will establish a round trip dictionary that will allows us to obtain information, in Section \ref{sec: applications}, about the distance of a flag code in terms of classical concepts related to partitions of integers.
	
	Let us start describing the enriched version of the distance support $\mathrm{S}(n)$. We complete it by adding suitable auxiliary red points as in the next picture. The resultant two-colored set of points is called \emph{enriched flag distance support} or just \emph{enriched  support} for short. We denote it by $\hat{\mathrm{S}}(n).$

	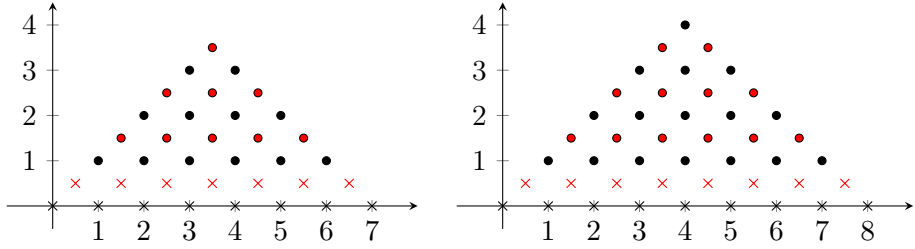
\begin{figure}[H]
		\centering
		\begin{tikzpicture}[line cap=round,line join=round,>=triangle 45,x=1cm,y=1cm]
			\begin{axis}[
				x=0.6cm,y=0.3cm,
				axis lines=middle,
				xmin=-1,
				xmax=8,
				ymin=-1,
				ymax=9,
				xtick={0,1,...,7},
				ytick={0,2,...,8},
				yticklabels={0,1,...,4},]
				\begin{scriptsize}
					\draw [color=black] (0,0) node[cross] {};
					\draw [color=black] (1,0) node[cross] {};
					\draw [color=black] (2,0) node[cross] {};
					\draw [color=black] (3,0) node[cross] {};
					\draw [color=black] (4,0) node[cross] {};
					\draw [color=black] (5,0) node[cross] {};
					\draw [color=black] (6,0) node[cross] {};
					\draw [color=black] (7,0) node[cross] {};
					\draw [fill=black] (1,2) circle (1.5pt);
					\draw [fill=black] (2,2) circle (1.5pt);
					\draw [fill=black] (2,4) circle (1.5pt);
					\draw [fill=black] (3,2) circle (1.5pt);
					\draw [fill=black] (3,4) circle (1.5pt);
					\draw [fill=black] (3,6) circle (1.5pt);
					\draw [fill=black] (4,2) circle (1.5pt);
					\draw [fill=black] (4,4) circle (1.5pt);
					\draw [fill=black] (4,6) circle (1.5pt);
					\draw [fill=black] (5,2) circle (1.5pt);
					\draw [fill=black] (5,4) circle (1.5pt);
					\draw [fill=black] (6,2) circle (1.5pt);
					\draw [color=red] (0.5,1) node[cross, red] {};
					\draw [color=red] (1.5,1) node[cross, red] {};
					\draw [color=red] (2.5,1) node[cross, red] {};
					\draw [color=red] (3.5,1) node[cross, red] {};
					\draw [color=red] (4.5,1) node[cross, red] {};
					\draw [color=red] (5.5,1) node[cross, red] {};
					\draw [color=red] (6.5,1) node[cross, red] {};
					\draw [fill=red] (1.5,3) circle (1.5pt);
					\draw [fill=red] (2.5,3) circle (1.5pt);
					\draw [fill=red] (2.5,5) circle (1.5pt);
					\draw [fill=red] (3.5,3) circle (1.5pt);
					\draw [fill=red] (3.5,5) circle (1.5pt);
					\draw [fill=red] (3.5,7) circle (1.5pt);
					\draw [fill=red] (4.5,3) circle (1.5pt);
					\draw [fill=red] (4.5,5) circle (1.5pt);
					\draw [fill=red] (5.5,3) circle (1.5pt);
				\end{scriptsize}
			\end{axis}
		\end{tikzpicture}
		\hspace{0.25cm}
		\begin{tikzpicture}[line cap=round,line join=round,>=triangle 45,x=1cm,y=1cm]
			\begin{axis}[
				x=0.6cm,y=0.3cm,
				axis lines=middle,
				xmin=-1,
				xmax=9,
				ymin=-1,
				ymax=9,
				xtick={0,1,...,8},
				ytick={0,2,...,8},
				yticklabels={0,1,...,4}]
				\begin{scriptsize}
					\draw [color=black] (0,0) node[cross] {};
					\draw [color=black] (1,0) node[cross] {};
					\draw [color=black] (2,0) node[cross] {};
					\draw [color=black] (3,0) node[cross] {};
					\draw [color=black] (4,0) node[cross] {};
					\draw [color=black] (5,0) node[cross] {};
					\draw [color=black] (6,0) node[cross] {};
					\draw [color=black] (7,0) node[cross] {};
					\draw [color=black] (8,0) node[cross] {};
					\draw [fill=black] (1,2) circle (1.5pt);
					\draw [fill=black] (2,2) circle (1.5pt);
					\draw [fill=black] (2,4) circle (1.5pt);
					\draw [fill=black] (3,2) circle (1.5pt);
					\draw [fill=black] (3,4) circle (1.5pt);
					\draw [fill=black] (3,6) circle (1.5pt);
					\draw [fill=black] (4,2) circle (1.5pt);
					\draw [fill=black] (4,4) circle (1.5pt);
					\draw [fill=black] (4,6) circle (1.5pt);
					\draw [fill=black] (4,8) circle (1.5pt);
					\draw [fill=black] (5,2) circle (1.5pt);
					\draw [fill=black] (5,4) circle (1.5pt);
					\draw [fill=black] (5,6) circle (1.5pt);
					\draw [fill=black] (6,2) circle (1.5pt);
					\draw [fill=black] (6,4) circle (1.5pt);
					\draw [fill=black] (7,2) circle (1.5pt);
					\draw [color=red] (0.5,1) node[cross, red] {};
					\draw [color=red] (1.5,1) node[cross, red] {};
					\draw [color=red] (2.5,1) node[cross, red] {};
					\draw [color=red] (3.5,1) node[cross, red] {};
					\draw [color=red] (4.5,1) node[cross, red] {};
					\draw [color=red] (5.5,1) node[cross, red] {};
					\draw [color=red] (6.5,1) node[cross, red] {};
					\draw [color=red] (7.5,1) node[cross, red] {};
					\draw [fill=red] (1.5,3) circle (1.5pt);
					\draw [fill=red] (2.5,3) circle (1.5pt);
					\draw [fill=red] (2.5,5) circle (1.5pt);
					\draw [fill=red] (3.5,3) circle (1.5pt);
					\draw [fill=red] (3.5,5) circle (1.5pt);
					\draw [fill=red] (3.5,7) circle (1.5pt);
					\draw [fill=red] (4.5,3) circle (1.5pt);
					\draw [fill=red] (4.5,5) circle (1.5pt);
					\draw [fill=red] (4.5,7) circle (1.5pt);
					\draw [fill=red] (5.5,3) circle (1.5pt);
					\draw [fill=red] (5.5,5) circle (1.5pt);
					\draw [fill=red] (6.5,3) circle (1.5pt);
				\end{scriptsize}
			\end{axis}
		\end{tikzpicture}
		\caption{Enriched flag distance diagrams $\hat{\mathrm{S}}(7)$ and $\hat{\mathrm{S}}(8).$}\label{fig: enriched diagrams n=7, 8}
	\end{figure}

	Recall that the silhouette of the distance support $\mathrm{S}(n)$ depends on the parity of $n$. By contrast, the enriched version $\hat{\mathrm{S}}(n)$ has always the same triangular shape. However, the position of black/red points changes depending on the parity  of $n$. For instance, the top vertex, which has coordinates $(n/2, n)$, is black (resp. red) when $n$ is even (resp. odd).
	
	\begin{remark}
		As stated in Remark \ref{rem: S(n-1) from S(n)}, the distance support $\mathrm{S}(n-1)$ can be obtained from $\mathrm{S}(n)$ by deleting the set of points in the right-roof. Similarly, the two-colored diagram $\hat{\mathrm{S}}(7)$ can be constructed from $\hat{\mathrm{S}}(8)$ by performing the same operation. 
	\end{remark}
	
	In order to give a systematic and convenient construction of the two-colored   enriched support $\hat{\mathrm{S}}(n)$, we proceed as follows. First, we fix the set of points in $\mathrm{S}(n)$ and plot them in black.  Next, we consider the distance support $\mathrm{S}(n-1)$ whose points we plot in red, and translate it with the vector  $\left( \frac{1}{2},  \frac{1}{2}\right)$. We obtain the set
	$$
	\mathrm{S}(n-1) + \left( \frac{1}{2},  \frac{1}{2} \right) =\left\lbrace \left( i+\frac{1}{2}, \delta +  \frac{1}{2} \right) \ \Big| \ (i, \delta)\in\mathrm{S}(n-1)\right\rbrace.
	$$

	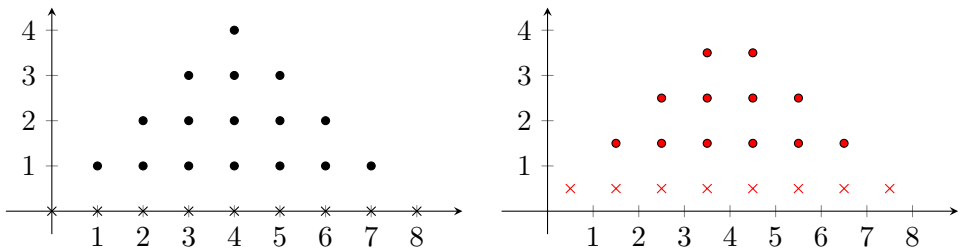
\begin{figure}[H]
		\centering
		\begin{tikzpicture}[line cap=round,line join=round,>=triangle 45,x=1cm,y=1cm]
			\begin{axis}[
				x=0.6cm,y=0.3cm,
				axis lines=middle,
				xmin=-1,
				xmax=9,
				ymin=-1,
				ymax=9,
				xtick={0,1,...,8},
				ytick={0,2,...,8},
				yticklabels={0,1,...,4}]
				\begin{scriptsize}
					\draw [color=black] (0,0) node[cross] {};
					\draw [color=black] (1,0) node[cross] {};
					\draw [color=black] (2,0) node[cross] {};
					\draw [color=black] (3,0) node[cross] {};
					\draw [color=black] (4,0) node[cross] {};
					\draw [color=black] (5,0) node[cross] {};
					\draw [color=black] (6,0) node[cross] {};
					\draw [color=black] (7,0) node[cross] {};
					\draw [color=black] (8,0) node[cross] {};
					\draw [fill=black] (1,2) circle (1.5pt);
					\draw [fill=black] (2,2) circle (1.5pt);
					\draw [fill=black] (2,4) circle (1.5pt);
					\draw [fill=black] (3,2) circle (1.5pt);
					\draw [fill=black] (3,4) circle (1.5pt);
					\draw [fill=black] (3,6) circle (1.5pt);
					\draw [fill=black] (4,2) circle (1.5pt);
					\draw [fill=black] (4,4) circle (1.5pt);
					\draw [fill=black] (4,6) circle (1.5pt);
					\draw [fill=black] (4,8) circle (1.5pt);
					\draw [fill=black] (5,2) circle (1.5pt);
					\draw [fill=black] (5,4) circle (1.5pt);
					\draw [fill=black] (5,6) circle (1.5pt);
					\draw [fill=black] (6,2) circle (1.5pt);
					\draw [fill=black] (6,4) circle (1.5pt);
					\draw [fill=black] (7,2) circle (1.5pt);
				\end{scriptsize}
			\end{axis}
		\end{tikzpicture}
		\hspace{0.25cm}
		\begin{tikzpicture}[line cap=round,line join=round,>=triangle 45,x=1cm,y=1cm]
			\begin{axis}[
				x=0.6cm,y=0.3cm,
				axis lines=middle,
				xmin=-1,
				xmax=9,
				ymin=-1,
				ymax=9,
				xtick={0,1,...,8},
				ytick={0,2,...,8},
				yticklabels={0,1,...,4}]
				\begin{scriptsize}
					\draw [color=red] (0.5,1) node[cross, red] {};
					\draw [color=red] (1.5,1) node[cross, red] {};
					\draw [color=red] (2.5,1) node[cross, red] {};
					\draw [color=red] (3.5,1) node[cross, red] {};
					\draw [color=red] (4.5,1) node[cross, red] {};
					\draw [color=red] (5.5,1) node[cross, red] {};
					\draw [color=red] (6.5,1) node[cross, red] {};
					\draw [color=red] (7.5,1) node[cross, red] {};
					\draw [fill=red] (1.5,3) circle (1.5pt);
					\draw [fill=red] (2.5,3) circle (1.5pt);
					\draw [fill=red] (2.5,5) circle (1.5pt);
					\draw [fill=red] (3.5,3) circle (1.5pt);
					\draw [fill=red] (3.5,5) circle (1.5pt);
					\draw [fill=red] (3.5,7) circle (1.5pt);
					\draw [fill=red] (4.5,3) circle (1.5pt);
					\draw [fill=red] (4.5,5) circle (1.5pt);
					\draw [fill=red] (4.5,7) circle (1.5pt);
					\draw [fill=red] (5.5,3) circle (1.5pt);
					\draw [fill=red] (5.5,5) circle (1.5pt);
					\draw [fill=red] (6.5,3) circle (1.5pt);
				\end{scriptsize}
			\end{axis}
		\end{tikzpicture}
		\caption{The distance support $\mathrm{S}(8)$ (left) and the set  $\mathrm{S}(7)+(1/2, 1/2)$ (right).}
	\end{figure}
	
	The overlap of $\mathrm{S}(8)$ (left) and  $\mathrm{S}(7)+(1/2, 1/2)$ leads to the two-colored enriched support $\hat{\mathrm{S}}(8)$ given in Figure \ref{fig: enriched diagrams n=7, 8} (right). Let us give a precise definition.
	
	\begin{definition}\label{prop: enriched diagram as union}
		For every $n\geq 2$, the \emph{enriched distance support} $\hat{\mathrm{S}}(n)$ is given by the set of points in
		$$
		\mathrm{S}(n) \dot\cup \left(  \mathrm{S}(n-1) + \left(\frac{1}{2}, \frac{1}{2}\right) \right).
		$$
	\end{definition}

	\begin{remark}\label{cor: number circle dots enriched diagram}
		One can easily compute the number of dots (both black and red) included in the enriched  support $\hat{\mathrm{S}}(n)$. By Proposition \ref{prop: enriched diagram as union} along with Remark \ref{rem: number of points in S(n)}, it is clear that $\hat{\mathrm{S}}(n)$ contains $D^n$ circle black points and $D^{n-1}$ red ones. Hence,  using the explicit value of $D^n$ given in formula (\ref{eq: Dn}), we conclude that for every $n\geq 2$, the enriched support $\hat{\mathrm{S}}(n)$ contains $\frac{n(n-1)}{2}$ circle dots. It also clearly contains $2n+1$ crossed dots.
	\end{remark}

	As done with distance supports in the previous part, we can remove the axis and work with the two-colored enriched support $\hat{\mathrm{S}}(n)$ without specifying the coordinates of each point.
	
	\subsection{Associated Ferrers diagrams}
	
	This subsection is devoted to describe the flag distance between full flags on $\bbF_q^n$ through the concept of distance path, by using suitable Ferrers diagrams. To do so, fixed a positive integer $n$, we consider the enriched distance support $\hat{\mathrm{S}}(n)$ just introduced  and rotate it around the point $(n,0)$ it as in the next figure. 
	
	\begin{figure}[H]
		\centering
		\begin{tikzpicture}[rotate=-45, line cap=round,line join=round,>=triangle 45,x=0.6cm,y=0.3cm]
			\begin{scriptsize}
				\draw [color=black] (0,0) node[cross] {};
				\draw [color=black] (1,0) node[cross] {};
				\draw [color=black] (2,0) node[cross] {};
				\draw [color=black] (3,0) node[cross] {};
				\draw [color=black] (4,0) node[cross] {};
				\draw [color=black] (5,0) node[cross] {};
				\draw [color=black] (6,0) node[cross] {};
				\draw [color=black] (7,0) node[cross] {};
				\draw [color=black] (8,0) node[cross] {}; 
				\draw [fill=black] (1,2) circle (1.5pt);
				\draw [fill=black] (2,2) circle (1.5pt);
				\draw [fill=black] (2,4) circle (1.5pt);
				\draw [fill=black] (3,2) circle (1.5pt);
				\draw [fill=black] (3,4) circle (1.5pt);
				\draw [fill=black] (3,6) circle (1.5pt);
				\draw [fill=black] (4,2) circle (1.5pt);
				\draw [fill=black] (4,4) circle (1.5pt);
				\draw [fill=black] (4,6) circle (1.5pt);
				\draw [fill=black] (4,8) circle (1.5pt);
				\draw [fill=black] (5,2) circle (1.5pt);
				\draw [fill=black] (5,4) circle (1.5pt);
				\draw [fill=black] (5,6) circle (1.5pt);
				\draw [fill=black] (6,2) circle (1.5pt);
				\draw [fill=black] (6,4) circle (1.5pt);
				\draw [fill=black] (7,2) circle (1.5pt);
				\draw [color=red] (0.5,1) node[cross, red] {};
				\draw [color=red] (1.5,1) node[cross, red] {};
				\draw [color=red] (2.5,1) node[cross, red] {};
				\draw [color=red] (3.5,1) node[cross, red] {};
				\draw [color=red] (4.5,1) node[cross, red] {};
				\draw [color=red] (5.5,1) node[cross, red] {};
				\draw [color=red] (6.5,1) node[cross, red] {};
				\draw [color=red] (7.5,1) node[cross, red] {};
				\draw [fill=red] (1.5,3) circle (1.5pt);
				\draw [fill=red] (2.5,3) circle (1.5pt);
				\draw [fill=red] (2.5,5) circle (1.5pt);
				\draw [fill=red] (3.5,3) circle (1.5pt);
				\draw [fill=red] (3.5,5) circle (1.5pt);
				\draw [fill=red] (3.5,7) circle (1.5pt);
				\draw [fill=red] (4.5,3) circle (1.5pt);
				\draw [fill=red] (4.5,5) circle (1.5pt);
				\draw [fill=red] (4.5,7) circle (1.5pt);
				\draw [fill=red] (5.5,3) circle (1.5pt);
				\draw [fill=red] (5.5,5) circle (1.5pt);
				\draw [fill=red] (6.5,3) circle (1.5pt);
			\end{scriptsize}
		\end{tikzpicture}
		\hspace{0.5cm}
		\begin{tikzpicture}[rotate=-45, line cap=round,line join=round,>=triangle 45,x=0.6cm,y=0.3cm]
			\begin{scriptsize}
				\draw [color=black] (0,0) node[cross] {};
				\draw [color=black] (1,0) node[cross] {};
				\draw [color=black] (2,0) node[cross] {};
				\draw [color=black] (3,0) node[cross] {};
				\draw [color=black] (4,0) node[cross] {};
				\draw [color=black] (5,0) node[cross] {};
				\draw [color=black] (6,0) node[cross] {};
				\draw [color=black] (7,0) node[cross] {};
				\draw [fill=black] (1,2) circle (1.5pt);
				\draw [fill=black] (2,2) circle (1.5pt);
				\draw [fill=black] (2,4) circle (1.5pt);
				\draw [fill=black] (3,2) circle (1.5pt);
				\draw [fill=black] (3,4) circle (1.5pt);
				\draw [fill=black] (3,6) circle (1.5pt);
				\draw [fill=black] (4,2) circle (1.5pt);
				\draw [fill=black] (4,4) circle (1.5pt);
				\draw [fill=black] (4,6) circle (1.5pt);
				\draw [fill=black] (5,2) circle (1.5pt);
				\draw [fill=black] (5,4) circle (1.5pt);
				\draw [fill=black] (6,2) circle (1.5pt);
				\draw [color=red] (0.5,1) node[cross, red] {};
				\draw [color=red] (1.5,1) node[cross, red] {};
				\draw [color=red] (2.5,1) node[cross, red] {};
				\draw [color=red] (3.5,1) node[cross, red] {};
				\draw [color=red] (4.5,1) node[cross, red] {};
				\draw [color=red] (5.5,1) node[cross, red] {};
				\draw [color=red] (6.5,1) node[cross, red] {};
				\draw [fill=red] (1.5,3) circle (1.5pt);
				\draw [fill=red] (2.5,3) circle (1.5pt);
				\draw [fill=red] (2.5,5) circle (1.5pt);
				\draw [fill=red] (3.5,3) circle (1.5pt);
				\draw [fill=red] (3.5,5) circle (1.5pt);
				\draw [fill=red] (3.5,7) circle (1.5pt);
				\draw [fill=red] (4.5,3) circle (1.5pt);
				\draw [fill=red] (4.5,5) circle (1.5pt);
				\draw [fill=red] (5.5,3) circle (1.5pt);
			\end{scriptsize}
		\end{tikzpicture}
		\caption{Rotated enriched supports $\hat{\mathrm{S}}(8)$ and $\hat{\mathrm{S}}(7)$.}\label{fig: rotated enriched diagrams n=7, 8}
	\end{figure}
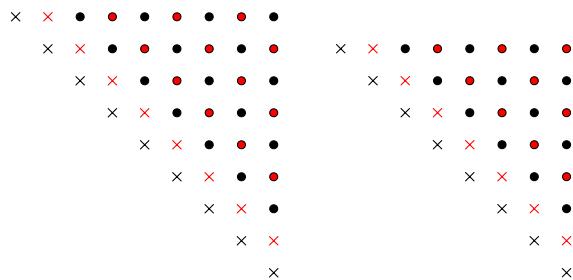

	Note that after rotation of $\hat{\mathrm{S}}(n)$ it arises  a Ferrers diagram.
	
	\begin{definition}
		Given $n\geq 2$, the \emph{primary Ferrers diagram frame} associated to the full flag variety on $\bbF_q^n$ is the diagram $\overline{\mathrm{FF}}(n)$ obtained from the enriched support $\hat{\mathrm{S}}(n)$ after a rotation with center $(n, 0)$ and angle $-\frac{\pi}{4}$. The set of circle dots (both black and red) in $\overline{\mathrm{FF}}(n)$ is called \emph{Ferrers diagram frame} and denoted by $\mathrm{FF}(n)$.
	\end{definition}
	
	\begin{remark}
		Once again, the reason to distinguish these two  Ferrers diagrams is the null contribution of crossed dots to the flag distance. On the other hand, observe that $\mathrm{FF}(n)$  is a Ferrers diagram associated to the partition $(n-1, n-2, \dots, 1)$ of the integer $n(n-1)/2$ that, as said in Remark \ref{cor: number circle dots enriched diagram}, is the exactly number of circle points in $\hat{\mathrm{S}}(n)$.
	\end{remark}
	
	There are some partitions of positive integers $r \leq n(n-1)/2$ that can be represented by Ferrers diagrams  contained in the Ferrers diagram frame (then in the primary one). The following definition precises this idea.
	
\begin{figure}[H]
		\centering
		\begin{tikzpicture}[rotate=-45, line cap=round,line join=round,>=triangle 45,x=0.6cm,y=0.3cm]
			\fill[line width=2pt,color=red,fill=red,fill opacity=0.1] (0.5,2) -- (1,1) -- (1.5,2) -- (2,1) --(4,5) --(4.5,4) --(5,5) --(6,3) --(6.5,4) --(4,9) -- cycle;
			\begin{scriptsize}
				\draw [line width=1pt,color=red] (2,0)-- (3,2);
				\draw [line width=1pt,color=red] (3,2)-- (4, 4);
				\draw [line width=1pt,color=red] (5,4)-- (6,2);
				\draw [line width=1pt,color=red] (7,2)--(8,0);
				\draw [line width=1pt,color=red] (0,0)-- (0.5,1);
				\draw [line width=1pt,color=red] (1,0)-- (0.5,1);
				\draw [line width=1pt,color=red] (1,0)-- (1.5, 1);
				\draw [line width=1pt,color=red] (2,0)-- (1.5, 1);
				\draw [line width=1pt,color=red] (4,4)-- (4.5,3);
				\draw [line width=1pt,color=red] (6,2)-- (6.5, 3);
				\draw [line width=1pt,color=red] (5,4)-- (4.5, 3);
				\draw [line width=1pt,color=red] (7,2)-- (6.5, 3);
				\draw [color=black] (0,0) node[cross] {};
				\draw [color=black] (1,0) node[cross] {};
				\draw [color=black] (2,0) node[cross] {};
				\draw [color=black] (3,0) node[cross] {};
				\draw [color=black] (4,0) node[cross] {};
				\draw [color=black] (5,0) node[cross] {};
				\draw [color=black] (6,0) node[cross] {};
				\draw [color=black] (7,0) node[cross] {};
				\draw [color=black] (8,0) node[cross] {};
				\draw [fill=black] (1,2) circle (1.5pt);
				\draw [fill=black] (2,2) circle (1.5pt);
				\draw [fill=black] (2,4) circle (1.5pt);
				\draw [fill=black] (3,2) circle (1.5pt);
				\draw [fill=black] (3,4) circle (1.5pt);
				\draw [fill=black] (3,6) circle (1.5pt);
				\draw [fill=black] (4,2) circle (1.5pt);
				\draw [fill=black] (4,4) circle (1.5pt);
				\draw [fill=black] (4,6) circle (1.5pt);
				\draw [fill=black] (4,8) circle (1.5pt);
				\draw [fill=black] (5,2) circle (1.5pt);
				\draw [fill=black] (5,4) circle (1.5pt);
				\draw [fill=black] (5,6) circle (1.5pt);
				\draw [fill=black] (6,2) circle (1.5pt);
				\draw [fill=black] (6,4) circle (1.5pt);
				\draw [fill=black] (7,2) circle (1.5pt);
				\draw [color=red] (0.5,1) node[cross, red] {};
				\draw [color=red] (1.5,1) node[cross, red] {};
				\draw [color=red] (2.5,1) node[cross, red] {};
				\draw [color=red] (3.5,1) node[cross, red] {};
				\draw [color=red] (4.5,1) node[cross, red] {};
				\draw [color=red] (5.5,1) node[cross, red] {};
				\draw [color=red] (6.5,1) node[cross, red] {};
				\draw [color=red] (7.5,1) node[cross, red] {};
				\draw [fill=red] (1.5,3) circle (1.5pt);
				\draw [fill=red] (2.5,3) circle (1.5pt);
				\draw [fill=red] (2.5,5) circle (1.5pt);
				\draw [fill=red] (3.5,3) circle (1.5pt);
				\draw [fill=red] (3.5,5) circle (1.5pt);
				\draw [fill=red] (3.5,7) circle (1.5pt);
				\draw [fill=red] (4.5,3) circle (1.5pt);
				\draw [fill=red] (4.5,5) circle (1.5pt);
				\draw [fill=red] (4.5,7) circle (1.5pt);
				\draw [fill=red] (5.5,3) circle (1.5pt);
				\draw [fill=red] (5.5,5) circle (1.5pt);
				\draw [fill=red] (6.5,3) circle (1.5pt);
			\end{scriptsize}
		\end{tikzpicture}
		\hspace{1cm}
		\begin{tikzpicture}[rotate=-45, line cap=round,line join=round,>=triangle 45,x=0.6cm,y=0.3cm]
			\begin{scriptsize}
				\draw [line width=1pt,color=red] (0,0)--(4,8);
				\draw [line width=1pt,color=red] (4,8)--(8,0);
				\draw [color=black] (0,0) node[cross] {};
				\draw [color=black] (1,0) node[cross] {};
				\draw [color=black] (2,0) node[cross] {};
				\draw [color=black] (3,0) node[cross] {};
				\draw [color=black] (4,0) node[cross] {};
				\draw [color=black] (5,0) node[cross] {};
				\draw [color=black] (6,0) node[cross] {};
				\draw [color=black] (7,0) node[cross] {};
				\draw [color=black] (8,0) node[cross] {};
				\draw [fill=black] (1,2) circle (1.5pt);
				\draw [fill=black] (2,2) circle (1.5pt);
				\draw [fill=black] (2,4) circle (1.5pt);
				\draw [fill=black] (3,2) circle (1.5pt);
				\draw [fill=black] (3,4) circle (1.5pt);
				\draw [fill=black] (3,6) circle (1.5pt);
				\draw [fill=black] (4,2) circle (1.5pt);
				\draw [fill=black] (4,4) circle (1.5pt);
				\draw [fill=black] (4,6) circle (1.5pt);
				\draw [fill=black] (4,8) circle (1.5pt);
				\draw [fill=black] (5,2) circle (1.5pt);
				\draw [fill=black] (5,4) circle (1.5pt);
				\draw [fill=black] (5,6) circle (1.5pt);
				\draw [fill=black] (6,2) circle (1.5pt);
				\draw [fill=black] (6,4) circle (1.5pt);
				\draw [fill=black] (7,2) circle (1.5pt);
				\draw [color=red] (0.5,1) node[cross, red] {};
				\draw [color=red] (1.5,1) node[cross, red] {};
				\draw [color=red] (2.5,1) node[cross, red] {};
				\draw [color=red] (3.5,1) node[cross, red] {};
				\draw [color=red] (4.5,1) node[cross, red] {};
				\draw [color=red] (5.5,1) node[cross, red] {};
				\draw [color=red] (6.5,1) node[cross, red] {};
				\draw [color=red] (7.5,1) node[cross, red] {};
				\draw [fill=red] (1.5,3) circle (1.5pt);
				\draw [fill=red] (2.5,3) circle (1.5pt);
				\draw [fill=red] (2.5,5) circle (1.5pt);
				\draw [fill=red] (3.5,3) circle (1.5pt);
				\draw [fill=red] (3.5,5) circle (1.5pt);
				\draw [fill=red] (3.5,7) circle (1.5pt);
				\draw [fill=red] (4.5,3) circle (1.5pt);
				\draw [fill=red] (4.5,5) circle (1.5pt);
				\draw [fill=red] (4.5,7) circle (1.5pt);
				\draw [fill=red] (5.5,3) circle (1.5pt);
				\draw [fill=red] (5.5,5) circle (1.5pt);
				\draw [fill=red] (6.5,3) circle (1.5pt);
			\end{scriptsize}
		\end{tikzpicture}
		\caption{Staircase paths and corresponding Ferrers subdiagrams.}\label{fig: staircase and diagrams}
	\end{figure}
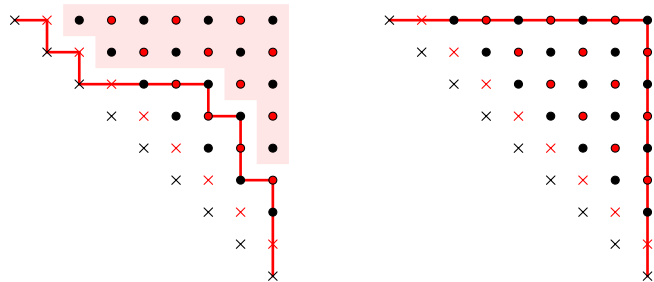

	\begin{definition}
		Every Ferrers diagram contained in $\mathrm{FF}(n)$ is said to be a \emph{Ferrers subdiagram} of $\mathrm{FF}(n)$.  We also say that the partition $\lambda=(\lambda_1, \dots, \lambda_m)$ of the integer $\sum_{i=1}^m \lambda_i$  is an \emph{embedded partition} on $\mathrm{FF}(n)$ if
		\begin{enumerate}
			\item $1\leq m\leq n-1$ and,
			\item for every $1\leq i\leq m$, it holds $\lambda_i\leq n-i$.
		\end{enumerate}
		Due to technical reasons, we also consider the \emph{empty Ferrers subdiagram} $\mathfrak{F}_0$, associated to the \emph{null embedded partition} $\lambda=(0)$ (see the second diagram in Figure \ref{fig: staircase and diagrams}).
	\end{definition}
	Observe that these embedded partitions are exactly those ones whose associated Ferrers diagram fits in $\mathrm{FF}(n)$. With this notation, we conclude directly the next result.
	
	\begin{proposition}
		Let $\mathfrak{F}_\lambda$ be a Ferrers diagram associated to the partition $\lambda=(\lambda_1, \dots, \lambda_m)$. Then the following statements are equivalent. 
		\begin{enumerate}
			\item $\mathfrak{F}_\lambda$ is a Ferrers subdiagram of  $\mathrm{FF}(n)$.
			\item The partition $\lambda=(\lambda_1, \dots, \lambda_m)$ is an embedded partition on $\mathrm{FF}(n)$.
		\end{enumerate}
	\end{proposition}
	
	At this point, and with the purpose of connecting these combinatorial objects with our study on the flag distance, we define a special class of polygonal paths in the primary Ferrers diagram $\overline{\mathrm{FF}}(n)$, closely related to the set of Ferrers subdiagrams in $\mathrm{FF}(n)$.
	
	\begin{definition}
		A  \emph{staircase path}  $\Sigma$ on   $\overline{\mathrm{FF}}(n)$ is just a polygonal directed path whose vertices are dots of $\overline{\mathrm{FF}}(n)$ (crossed or circle ones) such that:
		\begin{itemize}
			\item  it starts (resp. ends) at the highest (resp. lowest) crossed black point and
			\item its directed edges are either vertical segments straight down or horizontal segments from left to right.
		\end{itemize}
	\end{definition}
	
	\begin{remark}\label{rem: staircase silhouettes and Ferrers diagram}
		Observe that, since black and red dots are interspersed in  $\overline{\mathrm{FF}}(n)$, staircase paths travel along the diagram alternating black and red points. Even more, every staircase path contains exactly $n+1$ black dots and $n$ red ones. Moreover, since staircase paths cannot go neither up nor to the left, the collection of points that remains at right of any staircase path satisfy the next property: the number of dots at a given row is always upper bounded by the number of dots at the previous one, that is, any staircase path is the ``silhouette'' of a Ferrers diagram.
	\end{remark}

	\begin{proposition}
		Given $\Sigma$ a staircase path in $\overline{\mathrm{FF}}(n)$, the set of points to the right of $\Sigma$ forms a set $\mathfrak{F}(\Sigma)$ that is a Ferrers subdiagram of $\mathrm{FF}(n)$. Conversely, a Ferrers subdiagram $\mathfrak{F}$ of $\mathrm{FF}(n)$ determines a unique staircase path $\Sigma(\mathfrak{F})$ in $\overline{\mathrm{FF}}(n)$. In this situation we say that $\Sigma$ (resp. $\Sigma(\mathfrak{F})$)  is the \emph{silhouette} of the Ferrers subdiagram $\mathfrak{F}(\Sigma)$ (resp. $\mathfrak{F}$).
	\end{proposition}
	\begin{proof}
		Consider a staircase path $\Sigma$ in $\overline{\mathrm{FF}}(n)$. Giving this path is equivalent to provide a sequence of integers $\lambda=(\lambda_1, \dots, \lambda_{n-1})$ such that, for every $1\leq i\leq n-1$, the value $0\leq \lambda_i\leq n-i$ counts the number of circle dots in the $i$-th row of $\overline{\mathrm{FF}}(n)$ that remain  to the right of $\Sigma$. As said in Remark \ref{rem: staircase silhouettes and Ferrers diagram}, these values satisfy $\lambda_i\geq \lambda_{i+1}\geq 0$, for every $1\leq i\leq n-2$. If the sequence $\lambda$ does not contain zeros, or equivalently $\lambda_{n-1}\neq 0$, then it is a partition and we have that $\mathfrak{F}(\Sigma)=\mathfrak{F}_\lambda$. On the other hand, if $\lambda_1=0$, then $\mathfrak{F}(\Sigma)=\mathfrak{F}_0$. This case corresponds to the staircase that flows horizontally up to the corner of $\overline{\mathrm{FF}}(n)$ and then comes down vertically as in the second diagram in Figure \ref{fig: staircase and diagrams}. Last, if $\lambda = (\lambda_1, \dots, \lambda_i, 0, \dots, 0)$, with $\lambda_i\neq 0$,  then $\mathfrak{F}(\Sigma)=\mathfrak{F}_{\lambda'}$ with $\lambda' = (\lambda_1, \dots, \lambda_i)$.
		
		Conversely, given a Ferrers subdiagram $\mathfrak{F}$ with associated embedded partition $\lambda'=(\lambda'_1, \dots, \lambda'_m)$, then $m\leq n-1$. If $m=n-1$, the partition $\lambda'$ is the one corresponding to $\Sigma(\mathfrak{F})$. Otherwise, it is enough to complete $\lambda'$ with $n-1-m$ extra zero components to obtain $(\lambda'_1, \dots, \lambda'_m, 0, \dots, 0)$ as the sequence of length $n-1$ that characterizes the silhouette $\Sigma(\mathfrak{F})$.
	\end{proof}
	
	The previous result can be read in terms of embedded partitions as follows.
	\begin{corollary}\label{cor: partition associated to staircase path}
		The set of staircase paths in $\overline{\mathrm{FF}}(n)$ is in one-to-one correspondence with the set of embedded partitions in $\mathrm{FF}(n)$, that is, for any staircase path $\Sigma$ there is a unique embedded partition $\lambda(\Sigma)$ such that $\mathfrak{F}(\Sigma)=\mathfrak{F}_{\lambda(\Sigma)}$.
	\end{corollary}

	\subsection{Coming back to the flag distance}
	In the previous subsection, we associated a couple of Ferrers diagrams to the distance support by adding a suitable collection of auxiliary red points. In that scenario, the set of staircase paths are characterized by using embedded partitions of integers. Once established these combinatorial tools, to re-connect them with our distance paths, we need to remove the auxiliary red structure somehow. This removing process will allows us describe distance paths associated to a flag distance value $d$ in terms of appropriate summand distributions of the corresponding codistance value $\bar{d}$. Let us explain this in more detail.
	
	Our first objective is to consistently retrieve distance paths from staircase paths. Recall that, as stated in Remark \ref{rem: staircase silhouettes and Ferrers diagram}, black and red dots alternate in a staircase path. Hence, in order to study distance paths obtained by removing red dots, it is important to observe how two consecutive black dots can be connected when the intermediate red point that appears between them in a staircase path is eliminated. There are four admissible local movements for this operation:

	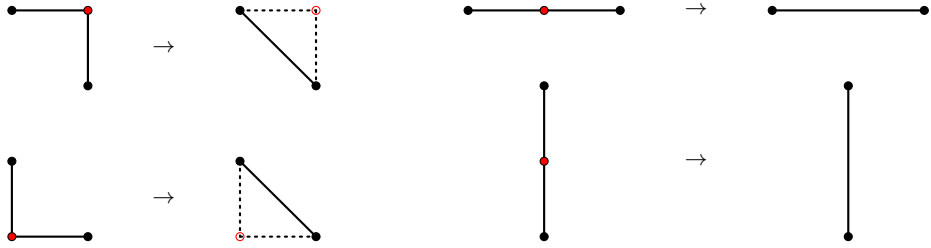
\begin{figure}[H]
		\centering
		\begin{tikzpicture}[line cap=round,line join=round,>=triangle 45,x=1cm,y=1cm]
			\draw [line width=0.8pt] (2,3)-- (3,3);
			\draw [line width=0.8pt] (3,3)-- (3,2);
			\draw [line width=0.8pt] (5,3)-- (6,2);
			\draw [line width=0.8pt,dotted] (5,3)-- (6,3);
			\draw [line width=0.8pt,dotted] (6,3)-- (6,2);
			\draw [line width=0.8pt] (2,1)-- (2,0);
			\draw [line width=0.8pt] (2,0)-- (3,0);
			\draw [line width=0.8pt] (5,1)-- (6,0);
			\draw [line width=0.8pt,dotted] (5,1)-- (5,0);
			\draw [line width=0.8pt,dotted] (5,0)-- (6,0);
			\draw [line width=0.8pt] (8,3)-- (10,3);
			\draw [line width=0.8pt] (12,3)-- (14,3);
			\draw [line width=0.8pt] (9,2)-- (9,0);
			\draw [line width=0.8pt] (13,2)-- (13,0);
			\begin{scriptsize}
				\draw [fill=black] (2,3) circle (1.5pt);
				\draw [fill=black] (3,2) circle (1.5pt);
				\draw[color=black] (4,2.5) node {$\rightarrow$};
				\draw [fill=black] (5,3) circle (1.5pt);
				\draw [fill=black] (6,2) circle (1.5pt);
				\draw [fill=red] (3,3) circle (1.5pt);
				\draw [color=red] (6,3) circle (1.5pt);
				\draw [fill=black] (2,1) circle (1.5pt);
				\draw [fill=black] (3,0) circle (1.5pt);
				\draw[color=black] (4,0.5) node {$\rightarrow$};
				\draw [fill=black] (5,1) circle (1.5pt);
				\draw [fill=black] (6,0) circle (1.5pt);
				\draw [fill=red] (2,0) circle (1.5pt);
				\draw [color=red] (5,0) circle (1.5pt);
				\draw [fill=black] (8,3) circle (1.5pt);
				\draw [fill=black] (10,3) circle (1.5pt);
				\draw[color=black] (11,3) node {$\rightarrow$};
				\draw [fill=black] (12,3) circle (1.5pt);
				\draw [fill=black] (14,3) circle (1.5pt);
				\draw [fill=black] (9,2) circle (1.5pt);
				\draw [fill=black] (9,0) circle (1.5pt);
				\draw [fill=black] (13,2) circle (1.5pt);
				\draw [fill=black] (13,0) circle (1.5pt);
				\draw [fill=red] (9,3) circle (1.5pt);
				\draw [fill=red] (9,1) circle (1.5pt);
				\draw[color=black] (11,1) node {$\rightarrow$};
			\end{scriptsize}
		\end{tikzpicture}
		\caption{Movements to locally recover a distance path by removing a red dot.}\label{fig: removing red dots}
	\end{figure}

	\begin{remark}
		The four motion patterns described in Figure \ref{fig: removing red dots} are also valid when they involve crossed points. 
	\end{remark}
	
\begin{figure}[H]
		\begin{center}
			\begin{tikzpicture}[line cap=round,line join=round,>=triangle 45,x=0.5cm,y=0.5cm]
				\draw [line width=0.8pt] (-8,4)-- (-4,4);
				\draw [line width=0.8pt] (-6,4)-- (-6,2);
				\draw [line width=0.8pt] (-6,2)-- (-8,2);
				\draw [line width=0.8pt] (-8,4)-- (-8,0);
				\begin{scriptsize}
					\draw [fill=black] (-8,4) circle (1.5pt);
					\draw [fill=black] (-4,4) circle (1.5pt);
					\draw[color=black] (-3.097321515756332,4.0725606143496735) node {$1$};
					\draw [fill=black] (-8,0) circle (1.5pt);
					\draw[color=black] (-7.276455203206154,-0.06348922065221829) node {$3$};
					\draw [fill=black] (-6,2) circle (1.5pt);
					\draw[color=black] (-5.251514138153147,2.0476195492966642) node {$2$};
					\draw [fill=red] (-8,2) circle (1.5pt);
					\draw [fill=red] (-6,4) circle (1.5pt);
				\end{scriptsize}
			\end{tikzpicture}
			\hspace{0.25cm}
			\begin{tikzpicture}[line cap=round,line join=round,>=triangle 45,x=0.5cm,y=0.5cm]
				\draw [line width=0.8pt] (-8,4)-- (-4,4);
				\draw [line width=0.8pt] (-8,4)-- (-6,2);
				\draw [line width=0.8pt] (-8,4)-- (-8,0);
				\draw [line width=0.8pt, dotted] (-6,4)-- (-6,2); 
				\draw [line width=0.8pt, dotted] (-8,2)-- (-6,2); 
				\begin{scriptsize}
					\draw [fill=black] (-8,4) circle (1.5pt);
					\draw [fill=black] (-4,4) circle (1.5pt);
					\draw[color=black] (-3.097321515756332,4.0725606143496735) node {$1$};
					\draw [fill=black] (-8,0) circle (1.5pt);
					\draw[color=black] (-7.276455203206154,-0.06348922065221829) node {$3$};
					\draw [fill=black] (-6,2) circle (1.5pt);
					\draw[color=black] (-5.251514138153147,2.0476195492966642) node {$2$};
					\draw [color=red] (-8,2) circle (1.5pt);
					\draw [color=red] (-6,4) circle (1.5pt);
				\end{scriptsize}
			\end{tikzpicture}
			\hspace{0.25cm}
			\begin{tikzpicture}[rotate=45,line cap=round,line join=round,>=triangle 45,x=0.5cm,y=0.5cm]
				\draw [line width=0.8pt] (-8,4)-- (-4,4);
				\draw [line width=0.8pt] (-8,4)-- (-6,2);
				\draw [line width=0.8pt] (-8,4)-- (-8,0);
				\begin{scriptsize}
					\draw [fill=black] (-8,4) circle (1.5pt);
					\draw [fill=black] (-4,4) circle (1.5pt);
					\draw[color=black] (-3.097321515756332,4.0725606143496735) node {$1$};
					\draw [fill=black] (-8,0) circle (1.5pt);
					\draw[color=black] (-7.276455203206154,-0.06348922065221829) node {$3$};
					\draw [fill=black] (-6,2) circle (1.5pt);
					\draw[color=black] (-5.251514138153147,2.0476195492966642) node {$2$};
				\end{scriptsize}
			\end{tikzpicture}
			\caption{Recovering the trident pattern after deletion of red points.}\label{fig: recovering the trident pattern}
		\end{center}
	\end{figure}
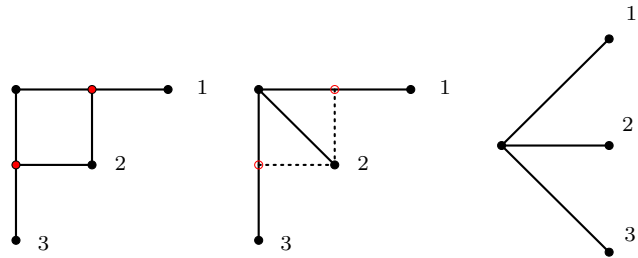

	With these four movements in mind, we do the following. We consider all the possible sequences of black-red-black points that are parts of staircase paths passing through a given black dot. We remove the red intermediate point and apply the corresponding movement in Figure \ref{fig: removing red dots} and get just three possibilities, labelled as $1, 2$ and $3$ in Figure \ref{fig: recovering the trident pattern}, starting from the given black dot. Observe that this figure corresponds to the rotation of the trident pattern exposed in Figure \ref{fig: trident pattern} and obtained in Proposition \ref{prop: allowed pattern}.
	
	As a consequence, we have straightforwardly the next result.
	
	\begin{proposition}
		The polygonal path $\Gamma(\Sigma)$ obtained after removing the red dots of any staircase path $\Sigma$ in $\overline{\mathrm{FF}}(n)$ by applying the movements in Figure \ref{fig: removing red dots} is a distance path in the distance support $\mathrm{S}(n)$ (after rotation). We call it the \emph{skeleton distance path} associated to the staircase path $\Sigma$. 
	\end{proposition}
	
	It is clear that different staircase paths in $\overline{\mathrm{FF}}(n)$ can share the same associated skeleton distance path in $\mathrm{S}(n)$ (after rotation). Hence, we can define an equivalence relation on the set of staircase paths.
	
	\begin{definition}
		We say that two staircase paths $\Sigma$ and $\Sigma'$ in $\overline{\mathrm{FF}}(n)$  are \emph{distance-equivalent} if they have the same associated skeleton distance path, i.e., $\Gamma(\Sigma)=\Gamma(\Sigma')$. Given a distance path $\Gamma$ in $\mathrm{S}(n)$, we denote the set of distance-equivalent staircase paths with $\Gamma$ as their skeleton distance path as $\Sigma(\Gamma)$.
	\end{definition}
	
	\begin{example}
		In Figure \ref{fig: distance pth in support and enriched support}, we can see a distance path $\Gamma$ in $\mathrm{S}(8)$ and the same path represented in the enriched support $\hat{\mathrm{S}}(8)$. Below, in Figure \ref{fig: 4 silhouettes with common skeleton}, we represent in red the four possible staircase paths in $\Sigma(\Gamma)$, i.e., those staircase paths having $\Gamma$ as their associated skeleton distance path.
		\begin{figure}[H]
			\begin{center}
				\begin{tikzpicture}[line cap=round,line join=round,>=triangle 45,x=0.6cm,y=0.3cm]
					\begin{scriptsize}
						\draw [line width=0.8pt,color=black] (0,0)-- (1,0);
						\draw [line width=0.8pt,color=black] (1,0)-- (2,0);
						\draw [line width=0.8pt,color=black] (2,0)-- (3,2);
						\draw [line width=0.8pt,color=black] (3,2)-- (4, 4);
						\draw [line width=0.8pt,color=black] (4,4)-- (5,4);
						\draw [line width=0.8pt,color=black] (5,4)-- (6,2);
						\draw [line width=0.8pt,color=black] (6,2)-- (7,2);
						\draw [line width=0.8pt,color=black] (7,2)--(8,0);
						\draw [color=black] (0,0) node[cross] {};
						\draw [color=black] (1,0) node[cross] {};
						\draw [color=black] (2,0) node[cross] {};
						\draw [color=black] (3,0) node[cross] {};
						\draw [color=black] (4,0) node[cross] {};
						\draw [color=black] (5,0) node[cross] {};
						\draw [color=black] (6,0) node[cross] {};
						\draw [color=black] (7,0) node[cross] {};
						\draw [color=black] (8,0) node[cross] {};
						\draw [fill=black] (1,2) circle (1.5pt);
						\draw [fill=black] (2,2) circle (1.5pt);
						\draw [fill=black] (2,4) circle (1.5pt);
						\draw [fill=black] (3,2) circle (1.5pt);
						\draw [fill=black] (3,4) circle (1.5pt);
						\draw [fill=black] (3,6) circle (1.5pt);
						\draw [fill=black] (4,2) circle (1.5pt);
						\draw [fill=black] (4,4) circle (1.5pt);
						\draw [fill=black] (4,6) circle (1.5pt);
						\draw [fill=black] (4,8) circle (1.5pt);
						\draw [fill=black] (5,2) circle (1.5pt);
						\draw [fill=black] (5,4) circle (1.5pt);
						\draw [fill=black] (5,6) circle (1.5pt);
						\draw [fill=black] (6,2) circle (1.5pt);
						\draw [fill=black] (6,4) circle (1.5pt);
						\draw [fill=black] (7,2) circle (1.5pt);
					\end{scriptsize}
				\end{tikzpicture}
				\begin{tikzpicture}[rotate=-45, line cap=round,line join=round,>=triangle 45,x=0.6cm,y=0.3cm]
					\begin{scriptsize}
						\draw [line width=0.8pt,color=black] (0,0)-- (1,0);
						\draw [line width=0.8pt,color=black] (1,0)-- (2,0);
						\draw [line width=0.8pt,color=black] (2,0)-- (3,2);
						\draw [line width=0.8pt,color=black] (3,2)-- (4, 4);
						\draw [line width=0.8pt,color=black] (4,4)-- (5,4);
						\draw [line width=0.8pt,color=black] (5,4)-- (6,2);
						\draw [line width=0.8pt,color=black] (6,2)-- (7,2);
						\draw [line width=0.8pt,color=black] (7,2)--(8,0);
						\draw [color=black] (0,0) node[cross] {};
						\draw [color=black] (1,0) node[cross] {};
						\draw [color=black] (2,0) node[cross] {};
						\draw [color=black] (3,0) node[cross] {};
						\draw [color=black] (4,0) node[cross] {};
						\draw [color=black] (5,0) node[cross] {};
						\draw [color=black] (6,0) node[cross] {};
						\draw [color=black] (7,0) node[cross] {};
						\draw [color=black] (8,0) node[cross] {};
						\draw [fill=black] (1,2) circle (1.5pt);
						\draw [fill=black] (2,2) circle (1.5pt);
						\draw [fill=black] (2,4) circle (1.5pt);
						\draw [fill=black] (3,2) circle (1.5pt);
						\draw [fill=black] (3,4) circle (1.5pt);
						\draw [fill=black] (3,6) circle (1.5pt);
						\draw [fill=black] (4,2) circle (1.5pt);
						\draw [fill=black] (4,4) circle (1.5pt);
						\draw [fill=black] (4,6) circle (1.5pt);
						\draw [fill=black] (4,8) circle (1.5pt);
						\draw [fill=black] (5,2) circle (1.5pt);
						\draw [fill=black] (5,4) circle (1.5pt);
						\draw [fill=black] (5,6) circle (1.5pt);
						\draw [fill=black] (6,2) circle (1.5pt);
						\draw [fill=black] (6,4) circle (1.5pt);
						\draw [fill=black] (7,2) circle (1.5pt);
						\draw [color=red] (0.5,1) node[cross, red] {};
						\draw [color=red] (1.5,1) node[cross, red] {};
						\draw [color=red] (2.5,1) node[cross, red] {};
						\draw [color=red] (3.5,1) node[cross, red] {};
						\draw [color=red] (4.5,1) node[cross, red] {};
						\draw [color=red] (5.5,1) node[cross, red] {};
						\draw [color=red] (6.5,1) node[cross, red] {};
						\draw [color=red] (7.5,1) node[cross, red] {};
						\draw [fill=red] (1.5,3) circle (1.5pt);
						\draw [fill=red] (2.5,3) circle (1.5pt);
						\draw [fill=red] (2.5,5) circle (1.5pt);
						\draw [fill=red] (3.5,3) circle (1.5pt);
						\draw [fill=red] (3.5,5) circle (1.5pt);
						\draw [fill=red] (3.5,7) circle (1.5pt);
						\draw [fill=red] (4.5,3) circle (1.5pt);
						\draw [fill=red] (4.5,5) circle (1.5pt);
						\draw [fill=red] (4.5,7) circle (1.5pt);
						\draw [fill=red] (5.5,3) circle (1.5pt);
						\draw [fill=red] (5.5,5) circle (1.5pt);
						\draw [fill=red] (6.5,3) circle (1.5pt);
					\end{scriptsize}
				\end{tikzpicture}
				\caption{A distance path $\Gamma$ in $\hat{\mathrm{S}}(8)$ (left) seen also  in $\overline{\mathrm{FF}}(8)$ (right).}\label{fig: distance pth in support and enriched support}
			\end{center}
		\end{figure}
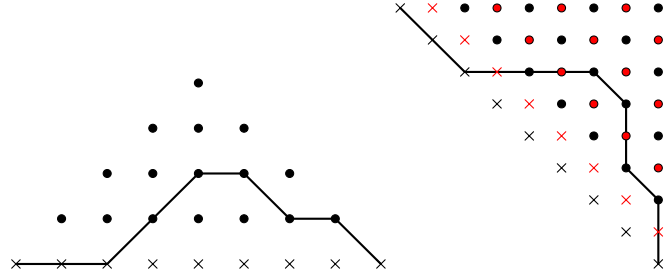
		
		\begin{figure}[H]
			\centering
			\begin{tikzpicture}[rotate=-45, line cap=round,line join=round,>=triangle 45,x=0.5cm,y=0.25cm]
				\fill[line width=2pt,color=red,fill=red,fill opacity=0.15] (0,0) -- (0.5,1) --(1,0) -- cycle;
				\fill[line width=2pt,color=red,fill=red,fill opacity=0.15] (1,0) -- (1.5,1) --(2,0) -- cycle;
				\fill[line width=2pt,color=red,fill=red,fill opacity=0.15] (4,4) -- (4.5,5) --(5,4) -- cycle;
				\fill[line width=2pt,color=red,fill=red,fill opacity=0.15] (6,2) -- (6.5,3) -- (7,2) -- cycle;			
				\begin{scriptsize}
					\draw [line width=0.8pt,color=red] (2,0)-- (3,2);
					\draw [line width=0.8pt,color=red] (3,2)-- (4, 4);
					\draw [line width=0.8pt,color=red] (5,4)-- (6,2);
					\draw [line width=0.8pt,color=red] (7,2)--(8,0);
					\draw [line width=0.8pt,color=red] (0,0)-- (0.5,1);
					\draw [line width=0.8pt,color=red] (1,0)-- (0.5,1);
					\draw [line width=0.8pt,color=red] (1,0)-- (1.5, 1);
					\draw [line width=0.8pt,color=red] (2,0)-- (1.5, 1);
					\draw [line width=0.8pt,color=red] (4,4)-- (4.5, 5);
					\draw [line width=0.8pt,color=red] (6,2)-- (6.5, 3);
					\draw [line width=0.8pt,color=red] (5,4)-- (4.5, 5);
					\draw [line width=0.8pt,color=red] (7,2)-- (6.5, 3);
					\draw [color=black] (0,0) node[cross] {};
					\draw [color=black] (1,0) node[cross] {};
					\draw [color=black] (2,0) node[cross] {};
					\draw [color=black] (3,0) node[cross] {};
					\draw [color=black] (4,0) node[cross] {};
					\draw [color=black] (5,0) node[cross] {};
					\draw [color=black] (6,0) node[cross] {};
					\draw [color=black] (7,0) node[cross] {};
					\draw [color=black] (8,0) node[cross] {};
					\draw [fill=black] (1,2) circle (1.5pt);
					\draw [fill=black] (2,2) circle (1.5pt);
					\draw [fill=black] (2,4) circle (1.5pt);
					\draw [fill=black] (3,2) circle (1.5pt);
					\draw [fill=black] (3,4) circle (1.5pt);
					\draw [fill=black] (3,6) circle (1.5pt);
					\draw [fill=black] (4,2) circle (1.5pt);
					\draw [fill=black] (4,4) circle (1.5pt);
					\draw [fill=black] (4,6) circle (1.5pt);
					\draw [fill=black] (4,8) circle (1.5pt);
					\draw [fill=black] (5,2) circle (1.5pt);
					\draw [fill=black] (5,4) circle (1.5pt);
					\draw [fill=black] (5,6) circle (1.5pt);
					\draw [fill=black] (6,2) circle (1.5pt);
					\draw [fill=black] (6,4) circle (1.5pt);
					\draw [fill=black] (7,2) circle (1.5pt);
					\draw [color=red] (0.5,1) node[cross, red] {};
					\draw [color=red] (1.5,1) node[cross, red] {};
					\draw [color=red] (2.5,1) node[cross, red] {};
					\draw [color=red] (3.5,1) node[cross, red] {};
					\draw [color=red] (4.5,1) node[cross, red] {};
					\draw [color=red] (5.5,1) node[cross, red] {};
					\draw [color=red] (6.5,1) node[cross, red] {};
					\draw [color=red] (7.5,1) node[cross, red] {};
					\draw [fill=red] (1.5,3) circle (1.5pt);
					\draw [fill=red] (2.5,3) circle (1.5pt);
					\draw [fill=red] (2.5,5) circle (1.5pt);
					\draw [fill=red] (3.5,3) circle (1.5pt);
					\draw [fill=red] (3.5,5) circle (1.5pt);
					\draw [fill=red] (3.5,7) circle (1.5pt);
					\draw [fill=red] (4.5,3) circle (1.5pt);
					\draw [fill=red] (4.5,5) circle (1.5pt);
					\draw [fill=red] (4.5,7) circle (1.5pt);
					\draw [fill=red] (5.5,3) circle (1.5pt);
					\draw [fill=red] (5.5,5) circle (1.5pt);
					\draw [fill=red] (6.5,3) circle (1.5pt);
				\end{scriptsize}
			\end{tikzpicture}
			\hspace{0.125cm}
			\begin{tikzpicture}[rotate=-45, line cap=round,line join=round,>=triangle 45,x=0.5cm,y=0.25cm]
				\fill[line width=2pt,color=red,fill=red,fill opacity=0.15] (0,0) -- (0.5,1) --(1,0) -- cycle;
				\fill[line width=2pt,color=red,fill=red,fill opacity=0.15] (1,0) -- (1.5,1) --(2,0) -- cycle;
				\fill[line width=2pt,color=red,fill=red,fill opacity=0.15] (4,4) -- (4.5,3) --(5,4) -- cycle;
				\fill[line width=2pt,color=red,fill=red,fill opacity=0.15] (6,2) -- (6.5,3) -- (7,2) -- cycle;		
				\begin{scriptsize}
					\draw [line width=0.8pt,color=red] (2,0)-- (3,2);
					\draw [line width=0.8pt,color=red] (3,2)-- (4, 4);
					\draw [line width=0.8pt,color=red] (5,4)-- (6,2);
					\draw [line width=0.8pt,color=red] (7,2)--(8,0);
					\draw [line width=0.8pt,color=red] (0,0)-- (0.5,1);
					\draw [line width=0.8pt,color=red] (1,0)-- (0.5,1);
					\draw [line width=0.8pt,color=red] (1,0)-- (1.5, 1);
					\draw [line width=0.8pt,color=red] (2,0)-- (1.5, 1);
					\draw [line width=0.8pt,color=red] (4,4)-- (4.5,3);
					\draw [line width=0.8pt,color=red] (6,2)-- (6.5, 3);
					\draw [line width=0.8pt,color=red] (5,4)-- (4.5, 3);
					\draw [line width=0.8pt,color=red] (7,2)-- (6.5, 3);
					\draw [color=black] (0,0) node[cross] {};
					\draw [color=black] (1,0) node[cross] {};
					\draw [color=black] (2,0) node[cross] {};
					\draw [color=black] (3,0) node[cross] {};
					\draw [color=black] (4,0) node[cross] {};
					\draw [color=black] (5,0) node[cross] {};
					\draw [color=black] (6,0) node[cross] {};
					\draw [color=black] (7,0) node[cross] {};
					\draw [color=black] (8,0) node[cross] {};
					\draw [fill=black] (1,2) circle (1.5pt);
					\draw [fill=black] (2,2) circle (1.5pt);
					\draw [fill=black] (2,4) circle (1.5pt);
					\draw [fill=black] (3,2) circle (1.5pt);
					\draw [fill=black] (3,4) circle (1.5pt);
					\draw [fill=black] (3,6) circle (1.5pt);
					\draw [fill=black] (4,2) circle (1.5pt);
					\draw [fill=black] (4,4) circle (1.5pt);
					\draw [fill=black] (4,6) circle (1.5pt);
					\draw [fill=black] (4,8) circle (1.5pt);
					\draw [fill=black] (5,2) circle (1.5pt);
					\draw [fill=black] (5,4) circle (1.5pt);
					\draw [fill=black] (5,6) circle (1.5pt);
					\draw [fill=black] (6,2) circle (1.5pt);
					\draw [fill=black] (6,4) circle (1.5pt);
					\draw [fill=black] (7,2) circle (1.5pt);
					\draw [color=red] (0.5,1) node[cross, red] {};
					\draw [color=red] (1.5,1) node[cross, red] {};
					\draw [color=red] (2.5,1) node[cross, red] {};
					\draw [color=red] (3.5,1) node[cross, red] {};
					\draw [color=red] (4.5,1) node[cross, red] {};
					\draw [color=red] (5.5,1) node[cross, red] {};
					\draw [color=red] (6.5,1) node[cross, red] {};
					\draw [color=red] (7.5,1) node[cross, red] {};
					\draw [fill=red] (1.5,3) circle (1.5pt);
					\draw [fill=red] (2.5,3) circle (1.5pt);
					\draw [fill=red] (2.5,5) circle (1.5pt);
					\draw [fill=red] (3.5,3) circle (1.5pt);
					\draw [fill=red] (3.5,5) circle (1.5pt);
					\draw [fill=red] (3.5,7) circle (1.5pt);
					\draw [fill=red] (4.5,3) circle (1.5pt);
					\draw [fill=red] (4.5,5) circle (1.5pt);
					\draw [fill=red] (4.5,7) circle (1.5pt);
					\draw [fill=red] (5.5,3) circle (1.5pt);
					\draw [fill=red] (5.5,5) circle (1.5pt);
					\draw [fill=red] (6.5,3) circle (1.5pt);
				\end{scriptsize}
			\end{tikzpicture}
			\hspace{0.125cm}
			\begin{tikzpicture}[rotate=-45, line cap=round,line join=round,>=triangle 45,x=0.5cm,y=0.25cm]
				\fill[line width=2pt,color=red,fill=red,fill opacity=0.15] (0,0) -- (0.5,1) --(1,0) -- cycle;
				\fill[line width=2pt,color=red,fill=red,fill opacity=0.15] (1,0) -- (1.5,1) --(2,0) -- cycle;				
				\fill[line width=2pt,color=red,fill=red,fill opacity=0.15] (4,4) -- (4.5,5) --(5,4) -- cycle;
				\fill[line width=2pt,color=red,fill=red,fill opacity=0.15] (6,2) -- (6.5,1) -- (7,2) -- cycle;				
				\begin{scriptsize}
					\draw [line width=0.8pt,color=red] (2,0)-- (3,2);
					\draw [line width=0.8pt,color=red] (3,2)-- (4, 4);
					\draw [line width=0.8pt,color=red] (5,4)-- (6,2);
					\draw [line width=0.8pt,color=red] (7,2)--(8,0);
					\draw [line width=0.8pt,color=red] (0,0)-- (0.5,1);
					\draw [line width=0.8pt,color=red] (1,0)-- (0.5,1);
					\draw [line width=0.8pt,color=red] (1,0)-- (1.5, 1);
					\draw [line width=0.8pt,color=red] (2,0)-- (1.5, 1);
					\draw [line width=0.8pt,color=red] (4,4)-- (4.5, 5);
					\draw [line width=0.8pt,color=red] (6,2)-- (6.5, 1);
					\draw [line width=0.8pt,color=red] (5,4)-- (4.5, 5);
					\draw [line width=0.8pt,color=red] (7,2)-- (6.5, 1);
					\draw [color=black] (0,0) node[cross] {};
					\draw [color=black] (1,0) node[cross] {};
					\draw [color=black] (2,0) node[cross] {};
					\draw [color=black] (3,0) node[cross] {};
					\draw [color=black] (4,0) node[cross] {};
					\draw [color=black] (5,0) node[cross] {};
					\draw [color=black] (6,0) node[cross] {};
					\draw [color=black] (7,0) node[cross] {};
					\draw [color=black] (8,0) node[cross] {};
					\draw [fill=black] (1,2) circle (1.5pt);
					\draw [fill=black] (2,2) circle (1.5pt);
					\draw [fill=black] (2,4) circle (1.5pt);
					\draw [fill=black] (3,2) circle (1.5pt);
					\draw [fill=black] (3,4) circle (1.5pt);
					\draw [fill=black] (3,6) circle (1.5pt);
					\draw [fill=black] (4,2) circle (1.5pt);
					\draw [fill=black] (4,4) circle (1.5pt);
					\draw [fill=black] (4,6) circle (1.5pt);
					\draw [fill=black] (4,8) circle (1.5pt);
					\draw [fill=black] (5,2) circle (1.5pt);
					\draw [fill=black] (5,4) circle (1.5pt);
					\draw [fill=black] (5,6) circle (1.5pt);
					\draw [fill=black] (6,2) circle (1.5pt);
					\draw [fill=black] (6,4) circle (1.5pt);
					\draw [fill=black] (7,2) circle (1.5pt);
					\draw [color=red] (0.5,1) node[cross, red] {};
					\draw [color=red] (1.5,1) node[cross, red] {};
					\draw [color=red] (2.5,1) node[cross, red] {};
					\draw [color=red] (3.5,1) node[cross, red] {};
					\draw [color=red] (4.5,1) node[cross, red] {};
					\draw [color=red] (5.5,1) node[cross, red] {};
					\draw [color=red] (6.5,1) node[cross, red] {};
					\draw [color=red] (7.5,1) node[cross, red] {};
					\draw [fill=red] (1.5,3) circle (1.5pt);
					\draw [fill=red] (2.5,3) circle (1.5pt);
					\draw [fill=red] (2.5,5) circle (1.5pt);
					\draw [fill=red] (3.5,3) circle (1.5pt);
					\draw [fill=red] (3.5,5) circle (1.5pt);
					\draw [fill=red] (3.5,7) circle (1.5pt);
					\draw [fill=red] (4.5,3) circle (1.5pt);
					\draw [fill=red] (4.5,5) circle (1.5pt);
					\draw [fill=red] (4.5,7) circle (1.5pt);
					\draw [fill=red] (5.5,3) circle (1.5pt);
					\draw [fill=red] (5.5,5) circle (1.5pt);
					\draw [fill=red] (6.5,3) circle (1.5pt);
				\end{scriptsize}
			\end{tikzpicture}
			\hspace{0.125cm}
			\begin{tikzpicture}[rotate=-45, line cap=round,line join=round,>=triangle 45,x=0.5cm,y=0.25cm]
				\fill[line width=2pt,color=red,fill=red,fill opacity=0.15] (0,0) -- (0.5,1) --(1,0) -- cycle;
				\fill[line width=2pt,color=red,fill=red,fill opacity=0.15] (1,0) -- (1.5,1) --(2,0) -- cycle;
				\fill[line width=2pt,color=red,fill=red,fill opacity=0.15] (4,4) -- (4.5,3) --(5,4) -- cycle;
				\fill[line width=2pt,color=red,fill=red,fill opacity=0.15] (6,2) -- (6.5,1) -- (7,2) -- cycle;				
				\begin{scriptsize}
					\draw [line width=0.8pt,color=red] (2,0)-- (3,2);
					\draw [line width=0.8pt,color=red] (3,2)-- (4, 4);
					\draw [line width=0.8pt,color=red] (5,4)-- (6,2);
					\draw [line width=0.8pt,color=red] (7,2)--(8,0);
					\draw [line width=0.8pt,color=red] (0,0)-- (0.5,1);
					\draw [line width=0.8pt,color=red] (1,0)-- (0.5,1);
					\draw [line width=0.8pt,color=red] (1,0)-- (1.5, 1);
					\draw [line width=0.8pt,color=red] (2,0)-- (1.5, 1);
					\draw [line width=0.8pt,color=red] (4,4)-- (4.5, 3);
					\draw [line width=0.8pt,color=red] (6,2)-- (6.5, 1);
					\draw [line width=0.8pt,color=red] (5,4)-- (4.5, 3);
					\draw [line width=0.8pt,color=red] (7,2)-- (6.5, 1);
					\draw [color=black] (0,0) node[cross] {};
					\draw [color=black] (1,0) node[cross] {};
					\draw [color=black] (2,0) node[cross] {};
					\draw [color=black] (3,0) node[cross] {};
					\draw [color=black] (4,0) node[cross] {};
					\draw [color=black] (5,0) node[cross] {};
					\draw [color=black] (6,0) node[cross] {};
					\draw [color=black] (7,0) node[cross] {};
					\draw [color=black] (8,0) node[cross] {};
					\draw [fill=black] (1,2) circle (1.5pt);
					\draw [fill=black] (2,2) circle (1.5pt);
					\draw [fill=black] (2,4) circle (1.5pt);
					\draw [fill=black] (3,2) circle (1.5pt);
					\draw [fill=black] (3,4) circle (1.5pt);
					\draw [fill=black] (3,6) circle (1.5pt);
					\draw [fill=black] (4,2) circle (1.5pt);
					\draw [fill=black] (4,4) circle (1.5pt);
					\draw [fill=black] (4,6) circle (1.5pt);
					\draw [fill=black] (4,8) circle (1.5pt);
					\draw [fill=black] (5,2) circle (1.5pt);
					\draw [fill=black] (5,4) circle (1.5pt);
					\draw [fill=black] (5,6) circle (1.5pt);
					\draw [fill=black] (6,2) circle (1.5pt);
					\draw [fill=black] (6,4) circle (1.5pt);
					\draw [fill=black] (7,2) circle (1.5pt);
					\draw [color=red] (0.5,1) node[cross, red] {};
					\draw [color=red] (1.5,1) node[cross, red] {};
					\draw [color=red] (2.5,1) node[cross, red] {};
					\draw [color=red] (3.5,1) node[cross, red] {};
					\draw [color=red] (4.5,1) node[cross, red] {};
					\draw [color=red] (5.5,1) node[cross, red] {};
					\draw [color=red] (6.5,1) node[cross, red] {};
					\draw [color=red] (7.5,1) node[cross, red] {};
					\draw [fill=red] (1.5,3) circle (1.5pt);
					\draw [fill=red] (2.5,3) circle (1.5pt);
					\draw [fill=red] (2.5,5) circle (1.5pt);
					\draw [fill=red] (3.5,3) circle (1.5pt);
					\draw [fill=red] (3.5,5) circle (1.5pt);
					\draw [fill=red] (3.5,7) circle (1.5pt);
					\draw [fill=red] (4.5,3) circle (1.5pt);
					\draw [fill=red] (4.5,5) circle (1.5pt);
					\draw [fill=red] (4.5,7) circle (1.5pt);
					\draw [fill=red] (5.5,3) circle (1.5pt);
					\draw [fill=red] (5.5,5) circle (1.5pt);
					\draw [fill=red] (6.5,3) circle (1.5pt);
				\end{scriptsize}
			\end{tikzpicture}
			\caption{The four elements in $\Sigma(\Gamma)$.}\label{fig: 4 silhouettes with common skeleton}
		\end{figure}
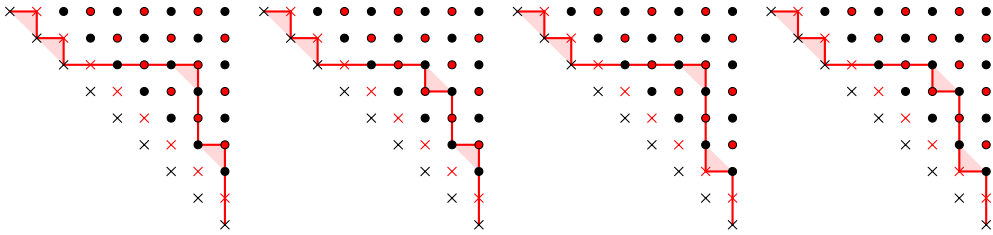		
	\end{example}
	
	The next result gives the exact number of different staircase paths that lie in a given coset of the distance-equivalence relationship, in terms of the number of \textit{plateaus} of their associated skeleton distance path (recall Definition \ref{def: plateu}).
	\begin{proposition}
		Consider a distance path $\Gamma$ in the distance support $\mathrm{S}(n)$ with $p$ \textit{plateaus} of positive height. Then the number of staircase paths in $\Sigma(\Gamma)$ is exactly $2^p$.
	\end{proposition}
	\begin{proof}
		Let $\Gamma$ a the distance path in the distance support $\mathrm{S}(n)$. Consider an arbitrary edge $e$ of $\Gamma$, connecting two consecutive vertices $(i, \delta_i)$ and $(i+1, \delta_{i+1})$, for some $0\leq i\leq n-1$, and assume that $\Sigma$ is a staircase path passing through these two black points too. Let us study the possibilities for the red point in $\Sigma$ connecting these two black ones. By virtue of Proposition \ref{prop: allowed pattern}, we know that $e$ has slope either $-1, 0$ or $1$.  If it has slope equal to $-1$ (resp. $1$), then after rotation we obtain a vertical (resp. horizontal) segment that already determines the unique red point connecting the starting vertices, as we can see in the next figure.
		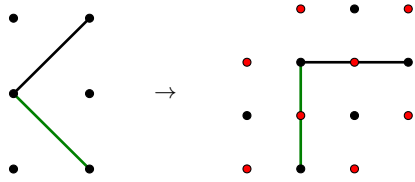
\begin{figure}[H]
			\centering
			\begin{tikzpicture}[line cap=round,line join=round,>=triangle 45,x=1cm,y=1cm]
				\draw [line width=1pt] (4,3)-- (5,4);
				\draw [line width=1pt,color=ao(english)] (4,3)-- (5,2);
				\begin{scriptsize}
					\draw [fill=black] (4,3) circle (1.5pt);
					\draw [fill=black] (5,4) circle (1.5pt);
					\draw [fill=black] (5,3) circle (1.5pt);
					\draw [fill=black] (5,2) circle (1.5pt);
					\draw [fill=black] (4,2) circle (1.5pt);
					\draw [fill=black] (4,4) circle (1.5pt);
					\draw[color=black] (6,3) node {$\rightarrow$};
				\end{scriptsize}
			\end{tikzpicture}
			\hspace{0.5cm}
			\begin{tikzpicture}[rotate=-45,line cap=round,line join=round,>=triangle 45,x=1cm,y=1cm]
				\draw [line width=1pt] (4,3)-- (5,4);
				\draw [line width=1pt,color=ao(english)] (4,3)-- (5,2);
				\begin{scriptsize}
					\draw [fill=black] (4,3) circle (1.5pt);
					\draw [fill=black] (5,4) circle (1.5pt);
					\draw [fill=black] (5,3) circle (1.5pt);
					\draw [fill=black] (5,2) circle (1.5pt);
					\draw [fill=black] (4,2) circle (1.5pt);
					\draw [fill=black] (4,4) circle (1.5pt);
					\draw [fill=red] (3.5,2.5) circle (1.5pt);
					\draw [fill=red] (4.5,2.5) circle (1.5pt);
					\draw [fill=red] (5.5,2.5) circle (1.5pt);
					\draw [fill=red] (5.5,3.5) circle (1.5pt);
					\draw [fill=red] (4.5,3.5) circle (1.5pt);
					\draw [fill=red] (3.5,3.5) circle (1.5pt);
					\draw [fill=red] (4.5,4.5) circle (1.5pt);
					\draw [fill=red] (4.5,1.5) circle (1.5pt);
				\end{scriptsize}
			\end{tikzpicture}
			\caption{Segments of slope $1$ (in black) and $-1$ (in green) in $\mathrm{S}(n)$ (left). The same segments seen in $\overline{\mathrm{FF}}(n)$ (right).}\label{fig: prueba plateaus 1}
		\end{figure}
		
		On the other hand, if $e$ has slope equal to $0$, i.e., it is a \textit{plateau}, then it is transformed (after rotation) into a segment with slope $-1$ in  $\overline{\mathrm{FF}}(n)$. It can be replaced by two sequences of movements in $\Sigma$: either right-down or down-right, marked in red and green, respectively in the picture below. These sequences correspond to use the middle red point with coordinates $(i+ \frac{1}{2}, \delta_i +\frac{1}{2})$ or $(i+ \frac{1}{2}, \delta_i - \frac{1}{2})$, respectively. Hence, there are two possibilities for replacing $e$, unless $\delta_i=\delta_{i+1}=0$. In this case, only the crossed red point $(i+ \frac{1}{2}, \frac{1}{2})$ can be used. As a result, if $p$ counts the number of \textit{plateaus} of $\Gamma$ with positive height, then there are exactly $2^p$ different staircase paths with $\Gamma$ as their skeleton.
		\begin{figure}[H]
			\centering
			\begin{tikzpicture}[line cap=round,line join=round,>=triangle 45,x=1cm,y=1cm]
				\draw [line width=1pt] (4,4)-- (5,4);
				\begin{scriptsize}
					\draw [fill=black] (4,3) circle (1.5pt);
					\draw [fill=black] (5,3) circle (1.5pt);
					\draw [fill=black] (5,4) circle (1.5pt);
					\draw [fill=black] (4,4) circle (1.5pt);
					\draw [fill=black] (5,5) circle (1.5pt);
					\draw [fill=black] (4,5) circle (1.5pt);
					\draw[color=black] (6,4) node {$\rightarrow$};
				\end{scriptsize}
			\end{tikzpicture}
			\hspace{0.5cm}
			\begin{tikzpicture}[rotate=-45,line cap=round,line join=round,>=triangle 45,x=1cm,y=1cm]
				\draw [line width=1pt] (4,4)-- (5,4);
				\draw [line width=1pt,color=red] (4,4)-- (4.5,4.5);
				\draw [line width=1pt,color=red] (4.3029754755557255,4.302975475555725) -- (4.3135705706668706,4.1864294293331294);
				\draw [line width=1pt,color=red] (4.3029754755557255,4.302975475555725) -- (4.18642942933313,4.31357057066687);
				\draw [line width=1pt,color=red] (4.5,4.5)-- (5,4);
				\draw [line width=1pt,color=red] (4.8029754755557255,4.1970245244442745) -- (4.68642942933313,4.1864294293331294);
				\draw [line width=1pt,color=red] (4.8029754755557255,4.1970245244442745) -- (4.813570570666871,4.31357057066687);
				\draw [line width=1pt,color= ao(english)] (4,4)-- (4.5,3.5);
				\draw [line width=1pt,color= ao(english)] (4.3029754755557255,3.6970245244442745) -- (4.18642942933313,3.68642942933313);
				\draw [line width=1pt,color= ao(english)] (4.3029754755557255,3.6970245244442745) -- (4.3135705706668706,3.81357057066687);
				\draw [line width=1pt,color= ao(english)] (4.5,3.5)-- (5,4);
				\draw [line width=1pt,color= ao(english)] (4.8029754755557255,3.802975475555725) -- (4.813570570666871,3.68642942933313);
				\draw [line width=1pt,color= ao(english)] (4.8029754755557255,3.802975475555725) -- (4.68642942933313,3.81357057066687);
				\begin{scriptsize}
					\draw [fill=black] (4,3) circle (1.5pt);
					\draw [fill=black] (5,3) circle (1.5pt);
					\draw [fill=black] (5,4) circle (1.5pt);
					\draw [fill=black] (4,4) circle (1.5pt);
					\draw [fill=red] (5.5,3.5) circle (1.5pt);
					\draw [fill=red] (3.5,3.5) circle (1.5pt);
					\draw [fill=red] (4.5,4.5) circle (1.5pt);
					\draw [fill=red] (4.5,3.5) circle (1.5pt);
					\draw [fill=red] (3.5,4.5) circle (1.5pt);
					\draw [fill=red] (5.5,4.5) circle (1.5pt);
					\draw [fill=black] (5,5) circle (1.5pt);
					\draw [fill=black] (4,5) circle (1.5pt);
				\end{scriptsize}
			\end{tikzpicture}
			\caption{A \textit{plateau} in $\mathrm{S}(n)$ (left) and its associated staircase paths in $\overline{\mathrm{FF}}(n)$ (right).}\label{fig: prueba plateaus 2}
		\end{figure}
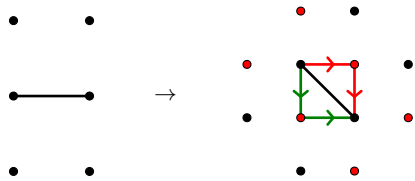
	\end{proof}
	
	Notice that, since distance-equivalent staircase paths have the same associated skeleton distance path, in particular, they are associated to the same flag distance value, which becomes a numerical invariant that can be assigned to staircase paths. On the other hand, a staircase path always contains $n+1$ black dots and $n$ red ones. Using these facts along with Remark \ref{rem: distance 2 number of dots}, we obtain the next result.
	\begin{corollary}\label{cor: d is invariant for staircase paths}
		Given a staircase path $\Sigma$, the number of circle black dots in $\Sigma$ or to its left is constant for all staircase paths distance-equivalent to $\Sigma$. This value is exactly $d_{\Gamma(\Sigma)}$.
	\end{corollary}
	
	The same idea can be applied to the set of points to the right of $\Sigma$, i.e., the  points in  $\mathfrak{F}(\Sigma)$, the Ferrers subdiagram of $\mathrm{FF}(n)$ having $\Sigma$ as their silhouette.
	
	\begin{definition}
		Let $\mathfrak{F}$ be a Ferrers subdiagram in $\mathrm{FF}(n)$. The \emph{underlying black diagram} of $\mathfrak{F}$ is the set of black points $U_\mathfrak{F}$ obtained after removing the set of red points on it. 
	\end{definition}
	
	\begin{remark}
		Observe that the the underlying black diagram $U_\mathfrak{F}$ of the Ferrers subdiagram $\mathfrak{F}$ might be empty. This happens if, and only if $\mathfrak{F}$ does not contain any circle black point. This happens if either $\mathfrak{F}=\mathfrak{F}_0$ (for every value of $n$) or $\mathfrak{F}=\mathfrak{F}_{(1)}$ and $n$ is odd.
	\end{remark}

	\begin{definition}
		Two Ferrers subdiagrams of $\mathrm{FF}(n)$ are said to be \emph{distance-equivalent} if they have the same underlying black diagram. Analogously, two embedded partitions $\lambda$ and $\lambda'$ are said to be \emph{distance-equivalent} if their associated Ferrers subdiagrams $\mathfrak{F}_\lambda$ and $\mathfrak{F}_{\lambda'}$ are.
	\end{definition}
	
	\begin{figure}[H]
		\centering
		\begin{tikzpicture}[rotate=-45, line cap=round,line join=round,>=triangle 45,x=0.6cm,y=0.3cm]
			\fill[line width=2pt,color=red,fill=red,fill opacity=0.1] (1.5,4) -- (2.5,2) -- (4.5,6) -- (6.5,2) --(7,3) --(4,9) -- cycle;
			\begin{scriptsize}
				\draw [color=black] (0,0) node[cross] {};
				\draw [color=black] (1,0) node[cross] {};
				\draw [color=black] (2,0) node[cross] {};
				\draw [color=black] (3,0) node[cross] {};
				\draw [color=black] (4,0) node[cross] {};
				\draw [color=black] (5,0) node[cross] {};
				\draw [color=black] (6,0) node[cross] {};
				\draw [color=black] (7,0) node[cross] {};
				\draw [color=black] (8,0) node[cross] {};
				\draw [fill=black] (1,2) circle (1.5pt);
				\draw [fill=black] (2,2) circle (1.5pt);
				\draw [fill=black] (2,4) circle (1.5pt);
				\draw [fill=black] (3,2) circle (1.5pt);
				\draw [fill=black] (3,4) circle (1.5pt);
				\draw [fill=black] (3,6) circle (1.5pt);
				\draw [fill=black] (4,2) circle (1.5pt);
				\draw [fill=black] (4,4) circle (1.5pt);
				\draw [fill=black] (4,6) circle (1.5pt);
				\draw [fill=black] (4,8) circle (1.5pt);
				\draw [fill=black] (5,2) circle (1.5pt);
				\draw [fill=black] (5,4) circle (1.5pt);
				\draw [fill=black] (5,6) circle (1.5pt);
				\draw [fill=black] (6,2) circle (1.5pt);
				\draw [fill=black] (6,4) circle (1.5pt);
				\draw [fill=black] (7,2) circle (1.5pt);
				\draw [color=red] (0.5,1) node[cross, red] {};
				\draw [color=red] (1.5,1) node[cross, red] {};
				\draw [color=red] (2.5,1) node[cross, red] {};
				\draw [color=red] (3.5,1) node[cross, red] {};
				\draw [color=red] (4.5,1) node[cross, red] {};
				\draw [color=red] (5.5,1) node[cross, red] {};
				\draw [color=red] (6.5,1) node[cross, red] {};
				\draw [color=red] (7.5,1) node[cross, red] {};
				\draw [fill=red] (1.5,3) circle (1.5pt);
				\draw [fill=red] (2.5,3) circle (1.5pt);
				\draw [fill=red] (2.5,5) circle (1.5pt);
				\draw [fill=red] (3.5,3) circle (1.5pt);
				\draw [fill=red] (3.5,5) circle (1.5pt);
				\draw [fill=red] (3.5,7) circle (1.5pt);
				\draw [fill=red] (4.5,3) circle (1.5pt);
				\draw [fill=red] (4.5,5) circle (1.5pt);
				\draw [fill=red] (4.5,7) circle (1.5pt);
				\draw [fill=red] (5.5,3) circle (1.5pt);
				\draw [fill=red] (5.5,5) circle (1.5pt);
				\draw [fill=red] (6.5,3) circle (1.5pt);
			\end{scriptsize}
		\end{tikzpicture}
		\hspace{0.1cm}
		\begin{tikzpicture}[rotate=-45, line cap=round,line join=round,>=triangle 45,x=0.6cm,y=0.3cm]
			\fill[line width=2pt,color=red,fill=red,fill opacity=0.1] (1,3) -- (1.5,2) -- (2,3) --(2.5,2)--(4,5) -- (4.5,4) --(5,5) --(6,3) --(6.5,4) --(4,9) -- cycle;
			\begin{scriptsize}
				\draw [color=black] (0,0) node[cross] {};
				\draw [color=black] (1,0) node[cross] {};
				\draw [color=black] (2,0) node[cross] {};
				\draw [color=black] (3,0) node[cross] {};
				\draw [color=black] (4,0) node[cross] {};
				\draw [color=black] (5,0) node[cross] {};
				\draw [color=black] (6,0) node[cross] {};
				\draw [color=black] (7,0) node[cross] {};
				\draw [color=black] (8,0) node[cross] {};
				\draw [fill=black] (1,2) circle (1.5pt);
				\draw [fill=black] (2,2) circle (1.5pt);
				\draw [fill=black] (2,4) circle (1.5pt);
				\draw [fill=black] (3,2) circle (1.5pt);
				\draw [fill=black] (3,4) circle (1.5pt);
				\draw [fill=black] (3,6) circle (1.5pt);
				\draw [fill=black] (4,2) circle (1.5pt);
				\draw [fill=black] (4,4) circle (1.5pt);
				\draw [fill=black] (4,6) circle (1.5pt);
				\draw [fill=black] (4,8) circle (1.5pt);
				\draw [fill=black] (5,2) circle (1.5pt);
				\draw [fill=black] (5,4) circle (1.5pt);
				\draw [fill=black] (5,6) circle (1.5pt);
				\draw [fill=black] (6,2) circle (1.5pt);
				\draw [fill=black] (6,4) circle (1.5pt);
				\draw [fill=black] (7,2) circle (1.5pt);
				\draw [color=red] (0.5,1) node[cross, red] {};
				\draw [color=red] (1.5,1) node[cross, red] {};
				\draw [color=red] (2.5,1) node[cross, red] {};
				\draw [color=red] (3.5,1) node[cross, red] {};
				\draw [color=red] (4.5,1) node[cross, red] {};
				\draw [color=red] (5.5,1) node[cross, red] {};
				\draw [color=red] (6.5,1) node[cross, red] {};
				\draw [color=red] (7.5,1) node[cross, red] {};
				\draw [fill=red] (1.5,3) circle (1.5pt);
				\draw [fill=red] (2.5,3) circle (1.5pt);
				\draw [fill=red] (2.5,5) circle (1.5pt);
				\draw [fill=red] (3.5,3) circle (1.5pt);
				\draw [fill=red] (3.5,5) circle (1.5pt);
				\draw [fill=red] (3.5,7) circle (1.5pt);
				\draw [fill=red] (4.5,3) circle (1.5pt);
				\draw [fill=red] (4.5,5) circle (1.5pt);
				\draw [fill=red] (4.5,7) circle (1.5pt);
				\draw [fill=red] (5.5,3) circle (1.5pt);
				\draw [fill=red] (5.5,5) circle (1.5pt);
				\draw [fill=red] (6.5,3) circle (1.5pt);
			\end{scriptsize}
		\end{tikzpicture}
		\hspace{0.1cm}
		\begin{tikzpicture}[rotate=-45, line cap=round,line join=round,>=triangle 45,x=0.6cm,y=0.3cm]
			\fill[line width=2pt,color=red,fill=red,fill opacity=0.1] (1.5,4) -- (2.5,2) -- (4.5,6) -- (6.5,2) --(7,3) --(4,9) -- cycle;
			\begin{scriptsize}
				\draw [color=black] (0,0) node[cross] {};
				\draw [color=black] (1,0) node[cross] {};
				\draw [color=black] (2,0) node[cross] {};
				\draw [color=black] (3,0) node[cross] {};
				\draw [color=black] (4,0) node[cross] {};
				\draw [color=black] (5,0) node[cross] {};
				\draw [color=black] (6,0) node[cross] {};
				\draw [color=black] (7,0) node[cross] {};
				\draw [color=black] (8,0) node[cross] {};
				\draw [fill=black] (1,2) circle (1.5pt);
				\draw [fill=black] (2,2) circle (1.5pt);
				\draw [fill=black] (2,4) circle (1.5pt);
				\draw [fill=black] (3,2) circle (1.5pt);
				\draw [fill=black] (3,4) circle (1.5pt);
				\draw [fill=black] (3,6) circle (1.5pt);
				\draw [fill=black] (4,2) circle (1.5pt);
				\draw [fill=black] (4,4) circle (1.5pt);
				\draw [fill=black] (4,6) circle (1.5pt);
				\draw [fill=black] (4,8) circle (1.5pt);
				\draw [fill=black] (5,2) circle (1.5pt);
				\draw [fill=black] (5,4) circle (1.5pt);
				\draw [fill=black] (5,6) circle (1.5pt);
				\draw [fill=black] (6,2) circle (1.5pt);
				\draw [fill=black] (6,4) circle (1.5pt);
				\draw [fill=black] (7,2) circle (1.5pt);
			\end{scriptsize}
		\end{tikzpicture}
		\caption{Two distance-equivalent Ferrers subdiagrams in $\mathrm{FF}(8)$ and their common underlying black diagram.}\label{fig: dist-equiv subdiagrams}
	\end{figure}
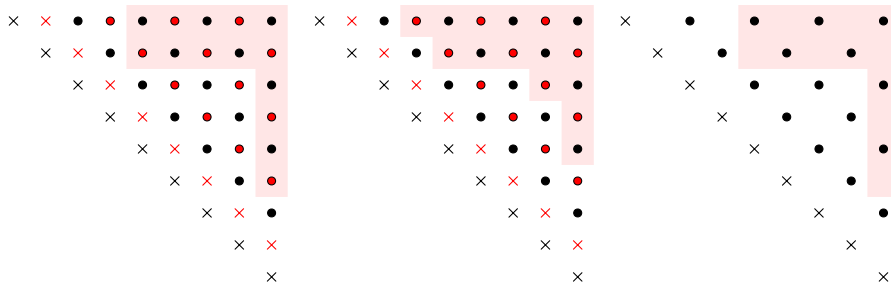 
	
	It is also possible to determine algebraically whether two different subdiagrams in $\mathrm{FF}(n)$ are distance-equivalent. To do so, we make use of the corresponding embedded partitions.
	
	\begin{remark}
		Observe that, given $\lambda=(\lambda_1, \dots, \lambda_m)$ an embedded partition in $\mathrm{FF}(n)$, in order to compute the number of black (or red) dots in the $i$-th row of the corresponding Ferrers subdiagram $\mathfrak{F}_\lambda$, there are two possibilities depending on the parity of both $i$ and $n$:
		
		\begin{enumerate}
			\item In case $i$ is even, the number of black dots in the $i$-th row is
			$$ \left\lfloor\frac{\lambda_i}{2}\right\rfloor \ \text{for} \ n \ \text{even, and} \ \left\lceil\frac{\lambda_i}{2}\right\rceil \ \text{for }  n \text{ odd.} $$
			\item In case $i$ is odd, the number of black dots in the $i$-th row is
			$$ \left\lceil\frac{\lambda_i}{2}\right\rceil \ \text{for } \ n \ \text{even,} \text{ and } \left\lfloor\frac{\lambda_i}{2}\right\rfloor \text{for } \ n \ \text{ odd.} $$
		\end{enumerate}
		Clearly, the number of red points at each row is just $\lambda_i$ minus the corresponding number of black points given above.
	\end{remark}

	\begin{definition}\label{def: underlying distribution}
		Given $\lambda=(\lambda_1, \dots, \lambda_m)$ an embedded partition  in $\mathrm{FF}(n)$, we define its \emph{underlying distribution} as the vector
		$$
		U_\lambda=\left\lbrace
		\begin{array}{cccl}
			\left( \left\lceil \frac{\lambda_1}{2}\right\rceil, \left\lfloor \frac{\lambda_2}{2}\right\rfloor, \left\lceil \frac{\lambda_3}{2}\right\rceil,  \dots \right) & \text{if} & n & \text{is even,}\\
			& & & \\[-1em]
			\left( \left\lfloor \frac{\lambda_1}{2}\right\rfloor, \left\lceil \frac{\lambda_2}{2}\right\rceil, \left\lfloor \frac{\lambda_3}{2}\right\rfloor, \dots \right) & \text{if} & n & \text{is odd.}
		\end{array}
		\right.
		$$
	\end{definition}
	
	Notice that the $i$-th component of the underlying distribution $U_\lambda$ represents the number of black points in the $i$-th row of the underlying black diagram $U_{\mathfrak{F}_\lambda}$. Nevertheless, the underlying distribution $U_{\lambda}$ of a partition $\lambda$ is not necessarily a partition itself as we can see in the following example.
	
	\begin{example}
		In $\mathrm{FF}(7)$, for the embedded partition $\lambda=(6, 3, 2)$, we have  $U_{\lambda}=(3, 2, 1)$, which is, in turn, a partition of $6$. Nevertheless, $\lambda'=(6,5,2,1,1)$ gives us $U_{\lambda'}=(3,2,1,0,1,0)$, which is not a partition. On the other side, we can have partitions with distance-equivalent Ferrers subdiagrams but with different underlying distributions. Just look Figure \ref{fig: dist-equiv subdiagrams}, where we have  $\lambda=(5,5,1,1,1,1)$ and $\lambda'=(6,5,2,1,1)$, both partitions of $14$, such that $U_{\mathfrak{F}_\lambda}=U_{\mathfrak{F}_{\lambda'}}$ whereas $U_{\lambda}=(3,2,1,0,1,0) \neq U_{\lambda'}=(3,2,1,0,1).$
	\end{example}
	
	In the next result we present a criterion to characterize those distance-equivalent embedded partitions also in terms of their (maybe different) associated underlying distributions.
	
	\begin{theorem}
		Let $\lambda=(\lambda_1, \dots, \lambda_m)$ and $\lambda'=(\lambda'_1, \dots, \lambda'_{m'})$ be  two embedded partitions  in $\mathrm{FF}(n)$ and assume $m\leq m'$. Then $\lambda$ and $\lambda'$ are distance-equivalent if, and only if, we have the following:
		\begin{enumerate}
			\item One of these two conditions hold:
			\begin{enumerate}
				\item $m=m'$.
				\item $m+n$ is odd, $m'=m+1$ and $\lambda'_{m'}=1$.
			\end{enumerate}
			\item In any of the cases above,
			$$\left\lbrace
			\begin{array}{cccl}
				\left\lceil \frac{\lambda_i}{2}\right\rceil = \left\lceil \frac{\lambda'_i}{2}\right\rceil      & \text{if} & n+i &   \text{is odd;}\\
				& & & \\[-1em]
				\left\lfloor \frac{\lambda_i}{2}\right\rfloor = \left\lfloor \frac{\lambda'_i}{2}\right\rfloor  & \text{if} & n+i &   \text{is even.}
			\end{array}
			\right.
			$$
			for every $1\leq i\leq m$.
		\end{enumerate} 
	\end{theorem} 
	\begin{proof}
		Assume that $\lambda$ and $\lambda'$ are distance-equivalent, i.e., their associated Ferrers subdiagrams $\mathfrak{F}_\lambda$  and  $\mathfrak{F}_{\lambda'}$ are distance-equivalent. Hence, their underlying black diagrams must coincide, that is, $U_{\mathfrak{F}_{\lambda}}=U_{\mathfrak{F}_{\lambda'}}$. Since one out of two rows in $\mathrm{FF}(n)$ starts with a black point, the number of rows of distance-equivalent Ferrers subdiagrams can differ by, at most, one unit. Moreover, the extra row can only contain a red point. In terms of the partitions $\lambda$ and $\lambda'$, we conclude that $m'\in\{m, m+1\}$. Moreover, in case $m'=m+1$, the first  dot (from the right) in the $(m+1)$-th row of $\mathfrak{F}_{\lambda'}$ must be red. That happens if, and only if, $m+n$ is odd and the part $\lambda'_{m'}=1.$ In addition, for the first $m$ rows, the number of black dots in both diagrams have to coincide. Equivalently, for every $1\leq i\leq m$, it must satisfy
		$$ 	
		\left\lbrace
		\begin{array}{cccl}
			\left\lceil \frac{\lambda_i}{2}\right\rceil = \left\lceil \frac{\lambda'_i}{2}\right\rceil      & \text{if} & n+i &   \text{is odd;}\\
			& & & \\[-1em]
			\left\lfloor \frac{\lambda_i}{2}\right\rfloor = \left\lfloor \frac{\lambda'_i}{2}\right\rfloor  & \text{if} & n+i &   \text{is even.}
		\end{array}
		\right.
		$$
		The converse holds immediately by using the same arguments.
	\end{proof}
	At this point we are able to establish in a consistent way the necessary link between embedded partitions in $\mathrm{FF}(n)$ and flag codistance values which will permit us to study properties of flag codes in terms of suitable partitions. The next result follows straightforwardly:
	
	\begin{proposition}\label{prop: splitting codistance}
		Let $\Sigma$ be a staircase path with associated flag distance $d=d_{\Gamma(\Sigma)}$ and Ferrers subdiagram $\mathfrak{F}(\Sigma)$. Take $\lambda=(\lambda_1, \dots, \lambda_m)$ an embedded partition such that  $\mathfrak{F}(\Sigma)=\mathfrak{F}_\lambda$. Then the codistance $\bar{d}=D^n-d$ coincides with the value $u_{\lambda}$ where 
		\begin{equation}\label{eq: splitting}
			u_\lambda = 
			\left\lbrace
			\begin{array}{cccl}
				\left\lceil \frac{\lambda_1}{2}\right\rceil + \left\lfloor \frac{\lambda_2}{2}\right\rfloor + \left\lceil \frac{\lambda_3}{2}\right\rceil + \dots & \text{if} & n & \text{is even;}\\
				& & & \\[-1em]
				\left\lfloor \frac{\lambda_1}{2}\right\rfloor + \left\lceil \frac{\lambda_2}{2}\right\rceil + \left\lfloor \frac{\lambda_3}{2}\right\rfloor + \dots & \text{if} & n & \text{is odd.}
			\end{array}
			\right.
		\end{equation}
	\end{proposition}
	
	Due to the previous proposition, we can introduce a new concept that relates embedded partitions and codistance.
	
	\begin{definition} Given $\lambda=(\lambda_1, \dots, \lambda_m)$ an embedded partition in $\mathrm{FF}(n)$, we say that its underlying distribution $U_{\lambda}$ \emph{splits} the value $u_{\lambda}$ defined in (\ref{eq: splitting}), or that it is an \emph{splitting} of $u_{\lambda}$. This value $u_\lambda$ is common for $\mathfrak{F}_\lambda$ and all its distance-equivalent Ferrers subdiagrams.
	\end{definition}
	\begin{remark}
		By extension, if we are in the conditions of Proposition \ref{prop: splitting codistance}, we have that $u_{\lambda}=\bar{d}$, and we say that $U_\lambda$ is an \emph{splitting} of the codistance $\bar{d}$. Notice that these splittings are not \emph{codistance vectors}. Given full flags $\cF, \cF'$ on $\bbF_q^n,$ their codistance vector is the sequence of (subspace) codistances between their subspaces. More precisely, the $i$-th component of this vector is
		$$
		\min\{ i, n-i\} - d_I(\cF_i, \cF'_i),
		$$
		for every $0\leq i\leq n.$ Moreover, if $d=d_f(\cF, \cF')$, then $\bar{d}$ is also obtained as the sum of the previous values. However, the notions of splitting and codistance vector (associated to $\bar{d}$) represent different ideas.
	\end{remark}
	
	Finally, the next result provides the bridge to translate the information given by distance paths to the embedded partitions level and conversely.
	\begin{theorem}
		Let $n \geq 2$ be an integer and $0\leq d \leq D^n$ a flag distance value. Then there is a bijection between the set of distance paths of distance $d$ in $\mathrm{S}(n)$ and the set of splittings of the codistance $\bar{d}=D^n-d$.
	\end{theorem}
	\begin{proof}	
		It suffices to summarize all the results provided along this section. We start from a distance path $\Gamma$ and consider the flag distance value $d=d_\Gamma$. This value is an invariant of all the distance-equivalent staircase paths in $\Sigma(\Gamma)$ by Corollary \ref{cor: d is invariant for staircase paths}. Take now all the Ferrers subdiagrams $\mathrm{FF}(n)$ whose silhouettes are in $\Sigma(\Gamma)$, i.e., those ones of the form $\mathfrak{F}_{\Sigma}$ for any $\Sigma \in \Sigma(\Gamma)$. Each $\mathfrak{F}_{\Sigma}$ is also associated to the corresponding partition $\lambda(\Sigma)$. Notice that all these Ferrers diagrams are distance-equivalent since they have the set of black points that remains at right of $\Gamma$ as their common underlying black diagram $U$, which has exactly $\bar{d}$ black points. Hence, all the partitions  $\{\lambda(\Sigma) \ | \ \Sigma\in\Sigma(\Gamma)\}$ are distance-equivalent and their common underlying distribution is, by means of Proposition \ref{prop: splitting codistance}, a splitting of the codistance $\bar{d}$.

		On the other hand,  consider a splitting $U_\lambda$ of the codistance $\bar{d}$, induced by an embedded partition $\lambda$ (or any other distance-equivalent partition). The distribution $U_\lambda$ determines the underlying black diagram  $U_{\mathfrak{F}_\lambda}$ of $\mathfrak{F}_\lambda$ (and of all its distance-equivalent Ferrers subdiagrams). The silhouette $\Sigma=\Sigma(\mathfrak{F}_\lambda)$ has a skeleton $\Gamma=\Gamma(\Sigma)$ that is a distance path associated to the distance $d$. This skeleton $\Gamma$ is common for all the staircase paths that are silhouettes of subdiagrams distance-equivalent to $\mathfrak{F}_\lambda$. Hence, every partition providing the distribution $U_\lambda$ leads to the same distance path $\Gamma$.
	\end{proof}

	\section{Applications and examples}\label{sec: applications}
	
	In this section we show how this new dictionary between flag distance values and underlying distributions of certain partitions can be applied to establish connections between the parameters of a given full flag code and the ones of its projected codes. To this end, we start with a lemma that counts the number of circle black dots of a rectangular Ferrers subdiagram.
	
	\begin{lemma}\label{lem: points in rectangle}
		Let $\mathfrak{R}$ be a Ferrers subdiagram in $\mathrm{FF}(n)$ with rectangular shape. If $\mathfrak{R}$ has $a$ rows and $b$ columns, then the number of circle black dots in $\mathfrak{R}$ is  
		$$
		\left\lbrace
		\begin{array}{cccl}
			\left\lceil\frac{ab}{2}\right\rceil   & \text{if}  & n & \text{is even,}\\
			& & & \\[-1em]
			\left\lfloor\frac{ab}{2}\right\rfloor & \text{if}  & n & \text{is odd.}\\
		\end{array}
		\right.
		$$
	\end{lemma}
	\begin{proof}
		Note that,  as $\mathfrak{R}$ has $a$ rows and $b$ columns, it is the Ferrers subdiagram associated to the partition $\lambda=(b, \overset{(a)}{\ldots}, b)$ where $\lambda=ab$. Then, the number of black points in $\mathfrak{R}$ is the value $u_\lambda$ (see (\ref{eq: splitting})), given as the sum of the components of the underlying distribution $U_\lambda$. According to Definition \ref{def: underlying distribution}, the expression of $U_\lambda$ is
		
		$$
		U_\lambda = \left\lbrace
		\begin{array}{cccl}
			\big( \left\lceil\frac{b}{2} \right\rceil, \left\lfloor\frac{b}{2} \right\rfloor  , \left\lceil\frac{b}{2} \right\rceil, \left\lfloor\frac{b}{2} \right\rfloor , \dots \big) & \text{if}  & n & \text{is even and}\\
			& & & \\[-1em]
			\big(\left\lfloor\frac{b}{2} \right\rfloor  , \left\lceil\frac{b}{2} \right\rceil, \left\lfloor\frac{b}{2} \right\rfloor , \left\lceil\frac{b}{2} \right\rceil,  \dots \big) & \text{if}  & n & \text{is odd.}
		\end{array}
		\right.
		$$
		Hence, for even values of $n$, we have:
		\begin{equation}\label{eq: number points rectangle 1}
			u_\lambda = \left\lceil\frac{a}{2} \right\rceil\cdot\left\lceil\frac{b}{2} \right\rceil + \left\lfloor\frac{a}{2} \right\rfloor\cdot\left\lfloor\frac{b}{2} \right\rfloor.
		\end{equation}
		In general, note that for every positive integer $c$, we have $c=\left\lfloor\frac{c}{2} \right\rfloor+\left\lceil\frac{c}{2} \right\rceil$. Thus, in case that either $a$ or $b$ is even, it holds $u_\lambda = \frac{ab}{2}= \left\lceil\frac{ab}{2}\right\rceil $.
		On the other hand, if both $a$ and $b$ are odd, then expression (\ref{eq: number points rectangle 1}) becomes
		$$
		u_\lambda = \frac{a+1}{2}\cdot\frac{b+1}{2} + \frac{a-1}{2}\cdot\frac{b-1}{2} = \frac{ab+1}{2}=  \left\lceil\frac{ab}{2}\right\rceil
		$$
		and the result is true for $n$ even. If $n$ is odd, the result follows by using the same arguments and taking into account that
		\begin{equation}\label{eq: number points rectangle}
			u_\lambda = \left\lfloor\frac{a}{2} \right\rfloor\cdot\left\lceil\frac{b}{2} \right\rceil + \left\lceil\frac{a}{2} \right\rceil\cdot\left\lfloor\frac{b}{2} \right\rfloor.
		\end{equation}
	\end{proof}
	
	Next we apply this lemma in order to relate the cardinality of a given flag code to the ones of its projected codes, by counting circle black dots in specific rectangles in the Ferrers diagram frame. 	
	
	\begin{theorem}\label{theo: separabilidad en funcion codistancia}
		Consider a full flag code $\cC$ on $\bbF_q^{n}$ with codistance $\bar{d}_f(\cC)$ and take a dimension $1\leq i\leq \left\lfloor \frac{n}{2}\right\rfloor$. If the codistance satisfies 
		\begin{equation}\label{eq: condicion codistancia}
			\bar{d}_f(\cC)< \left\lceil\frac{i(n-i)}{2}\right\rceil,
		\end{equation}
		then $|\cC|=|\cC_i|= \cdots=|\cC_{n-i}|$.
	\end{theorem}	
	\begin{proof}
		Assume that $|\cC|\neq|\cC_i|$ and that $\bar{d}_f(\cC)$ satisfies (\ref{eq: condicion codistancia}). In this case, there exist different flags $\cF, \cF'$ in $\cC$ with $\cF_i=\cF'_i$. Equivalently, the distance path $\Gamma(\cF, \cF')\in\Gamma(\cC)$ passes through the crossed point $(i, 0)$ of S(n), which determines, in turn, a rectangle with $i$ rows and $n-i$ columns over it in the enriched distance support $\hat{S}(n)$. By means of Lemma \ref{lem: points in rectangle}, this rectangle contains exactly
		$$
		p=\left\lbrace 
		\begin{array}{cccc}
			\left\lceil\frac{i(n-i)}{2}\right\rceil   & \text{if} & n & \text{is even,}\\
			&           &   &  \\[-1em]
			\left\lfloor\frac{i(n-i)}{2}\right\rfloor & \text{if} & n & \text{is odd,}
		\end{array}	
		\right.
		$$
		circle black dots. Moreorver, notice that, if $n$ is odd, then $i(n-i)$ is even and we can simply write
		$$
		p=\left\lceil\frac{i(n-i)}{2}\right\rceil,
		$$
		for every value of $n$. Notice that, at least all these $p$ circle black points remain over the distance path $\Gamma(\cF, \cF')$, and then they do not contribute to the computation of $d_f(\cF, \cF')$. Hence, we have
		$$
		d_f(\cC)\leq d_f(\cF, \cF) \leq D^n - p.
		$$
		Consequently, we obtain $\bar{d}_f(\cC)=D^n - d_f(\cC) \geq p$, which is a contradiction, and it must hold $|\cC|=|\cC_i|$. The same arguments, but considering a rectangle with $n-i$ rows and $i$ columns, lead to $|\cC|=|\cC_{n-i}|.$
		On the other hand, if we take a dimension $i\leq j\leq\left\lfloor\frac{n}{2}\right\rfloor$, and we write $j=i+k$ for some integer $k\geq 0$, then it holds
		$$
		j(n-j)= (i+k)(n-i-k) = i(n-i) + k( n-2i-k) \geq i(n-i) + k (n-2j) \geq i(n-i).
		$$
		Hence, if $\bar{d}_f(\cC)$ satisfies the stated condition, in particular, we also have
		$$ 
		\bar{d}_f(\cC)< \left\lceil\frac{j(n-j)}{2}\right\rceil 
		$$
		and, arguing as above, we get $|\cC|= |\cC_j|=|\cC_{n-j}|,$ for every $i\leq j\leq\left\lfloor\frac{n}{2}\right\rfloor.$
	\end{proof}
	
	On the other hand, the connection between distance paths and Ferrers subdiagrams established in the previous section, enables us to associate the following combinatorial objects to a flag code $\cC$.
	
	\begin{definition}\label{def: ferrers diagrams of a code}
		Let $\cC$ be a full flag code on $\bbF_q^n$. The set of \emph{Ferrers subdiagrams of $\cC$} is 
		$$
		\mathfrak{F}(\cC)= \{ \mathfrak{F} \ \text{subdiagram of} \ \mathrm{FF}(n) \ | \  \Gamma(\Sigma(\mathfrak{F})) \in \Gamma(\cC) \}.
		$$
		In other words, subdiagrams in $\mathfrak{F}(\cC)$ are those ones whose silhouettes have a distance path in $\Gamma(\cC)$ as their distance skeleton.
	\end{definition}
	
	Associated to this set of Ferrers subdiagrams, we can consider, in turn, their sets of Durfee rectangles as follows.

	\begin{definition}
		Let $\cC$ be a full flag code on $\bbF_q^n$. For every $0\leq k \leq n-2$, the set of \emph{Durfee $k$-rectangles of $\cC$} is given by
		$$
		D_k(\cC)= \{ D_k(\mathfrak{F}) \ | \  \mathfrak{F}\in \mathfrak{F}(\cC) \}.
		$$
		In particular, for $k=0$, we simply write $D(\cC)=D_0(\cC)$ and speak about the set of \emph{Durfee squares of $\cC$.}
	\end{definition}

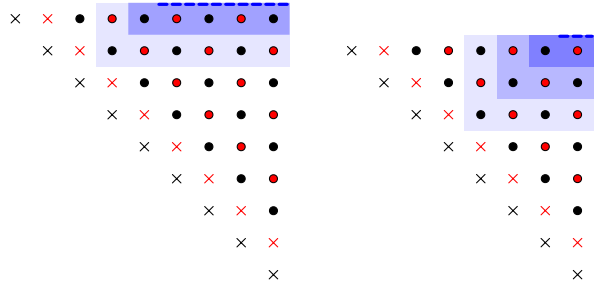
\begin{figure}[H]
		\centering
		\begin{tikzpicture}[rotate=-45, line cap=round,line join=round,>=triangle 45,x=0.6cm,y=0.3cm]
			\draw[densely dashed,line width=1.2pt,color=blue] (2,4.9)-- (4,8.9);
			\fill[line width=2pt,color=blue,fill=blue,fill opacity=0.1]  (2,1) -- (1,3) -- (4,9) -- (5,7) -- cycle;
			\fill[line width=2pt,color=blue,fill=blue,fill opacity=0.3]  (2,3) -- (1.5,4) -- (4,9) -- (4.5,8) -- cycle;
			\begin{scriptsize}
				\draw [color=black] (0,0) node[cross] {};
				\draw [color=black] (1,0) node[cross] {};
				\draw [color=black] (2,0) node[cross] {};
				\draw [color=black] (3,0) node[cross] {};
				\draw [color=black] (4,0) node[cross] {};
				\draw [color=black] (5,0) node[cross] {};
				\draw [color=black] (6,0) node[cross] {};
				\draw [color=black] (7,0) node[cross] {};
				\draw [color=black] (8,0) node[cross] {}; 
				\draw [fill=black] (1,2) circle (1.5pt);
				\draw [fill=black] (2,2) circle (1.5pt);
				\draw [fill=black] (2,4) circle (1.5pt);
				\draw [fill=black] (3,2) circle (1.5pt);
				\draw [fill=black] (3,4) circle (1.5pt);
				\draw [fill=black] (3,6) circle (1.5pt);
				\draw [fill=black] (4,2) circle (1.5pt);
				\draw [fill=black] (4,4) circle (1.5pt);
				\draw [fill=black] (4,6) circle (1.5pt);
				\draw [fill=black] (4,8) circle (1.5pt);
				\draw [fill=black] (5,2) circle (1.5pt);
				\draw [fill=black] (5,4) circle (1.5pt);
				\draw [fill=black] (5,6) circle (1.5pt);
				\draw [fill=black] (6,2) circle (1.5pt);
				\draw [fill=black] (6,4) circle (1.5pt);
				\draw [fill=black] (7,2) circle (1.5pt);
				\draw [color=red] (0.5,1) node[cross, red] {};
				\draw [color=red] (1.5,1) node[cross, red] {};
				\draw [color=red] (2.5,1) node[cross, red] {};
				\draw [color=red] (3.5,1) node[cross, red] {};
				\draw [color=red] (4.5,1) node[cross, red] {};
				\draw [color=red] (5.5,1) node[cross, red] {};
				\draw [color=red] (6.5,1) node[cross, red] {};
				\draw [color=red] (7.5,1) node[cross, red] {};
				\draw [fill=red] (1.5,3) circle (1.5pt);
				\draw [fill=red] (2.5,3) circle (1.5pt);
				\draw [fill=red] (2.5,5) circle (1.5pt);
				\draw [fill=red] (3.5,3) circle (1.5pt);
				\draw [fill=red] (3.5,5) circle (1.5pt);
				\draw [fill=red] (3.5,7) circle (1.5pt);
				\draw [fill=red] (4.5,3) circle (1.5pt);
				\draw [fill=red] (4.5,5) circle (1.5pt);
				\draw [fill=red] (4.5,7) circle (1.5pt);
				\draw [fill=red] (5.5,3) circle (1.5pt);
				\draw [fill=red] (5.5,5) circle (1.5pt);
				\draw [fill=red] (6.5,3) circle (1.5pt);
			\end{scriptsize}
		\end{tikzpicture}
		\hspace{0.5cm}
		\begin{tikzpicture}[rotate=-45, line cap=round,line join=round,>=triangle 45,x=0.6cm,y=0.3cm]
			\draw[densely  dashed, line width=1.2pt,color=blue] (3,6.9)-- (3.5,7.9);
			\fill[line width=2pt,color=blue,fill=blue,fill opacity=0.3]  (3.5,8) -- (2.5,6) -- (3,5) -- (4,7) -- cycle;
			\fill[line width=2pt,color=blue,fill=blue,fill opacity=0.2]  (3.5,8) -- (2,5) -- (3,3) -- (4.5,6) -- cycle;
			\fill[line width=2pt,color=blue,fill=blue,fill opacity=0.1]  (3.5,8) -- (1.5,4) -- (3,1) -- (5,5) -- cycle;
			\begin{scriptsize}
				\draw [color=black] (0,0) node[cross] {};
				\draw [color=black] (1,0) node[cross] {};
				\draw [color=black] (2,0) node[cross] {};
				\draw [color=black] (3,0) node[cross] {};
				\draw [color=black] (4,0) node[cross] {};
				\draw [color=black] (5,0) node[cross] {};
				\draw [color=black] (6,0) node[cross] {};
				\draw [color=black] (7,0) node[cross] {};
				\draw [fill=black] (1,2) circle (1.5pt);
				\draw [fill=black] (2,2) circle (1.5pt);
				\draw [fill=black] (2,4) circle (1.5pt);
				\draw [fill=black] (3,2) circle (1.5pt);
				\draw [fill=black] (3,4) circle (1.5pt);
				\draw [fill=black] (3,6) circle (1.5pt);
				\draw [fill=black] (4,2) circle (1.5pt);
				\draw [fill=black] (4,4) circle (1.5pt);
				\draw [fill=black] (4,6) circle (1.5pt);
				\draw [fill=black] (5,2) circle (1.5pt);
				\draw [fill=black] (5,4) circle (1.5pt);
				\draw [fill=black] (6,2) circle (1.5pt);
				\draw [color=red] (0.5,1) node[cross, red] {};
				\draw [color=red] (1.5,1) node[cross, red] {};
				\draw [color=red] (2.5,1) node[cross, red] {};
				\draw [color=red] (3.5,1) node[cross, red] {};
				\draw [color=red] (4.5,1) node[cross, red] {};
				\draw [color=red] (5.5,1) node[cross, red] {};
				\draw [color=red] (6.5,1) node[cross, red] {};
				\draw [fill=red] (1.5,3) circle (1.5pt);
				\draw [fill=red] (2.5,3) circle (1.5pt);
				\draw [fill=red] (2.5,5) circle (1.5pt);
				\draw [fill=red] (3.5,3) circle (1.5pt);
				\draw [fill=red] (3.5,5) circle (1.5pt);
				\draw [fill=red] (3.5,7) circle (1.5pt);
				\draw [fill=red] (4.5,3) circle (1.5pt);
				\draw [fill=red] (4.5,5) circle (1.5pt);
				\draw [fill=red] (5.5,3) circle (1.5pt);
			\end{scriptsize}
		\end{tikzpicture}
		\caption{Possible Durfee $4$-rectangles in $\mathrm{FF}(8)$ and $1$-rectangles in $\mathrm{FF}(7),$ included those ones with zero rows (in dashed lines).}\label{fig: durfee rect of a code}
	\end{figure}

	Notice that, given an embedded partition $\lambda=(\lambda_1, \dots, \lambda_m)$ and an integer $0\leq k\leq n-2$, the Durfee $k$-rectangle $D_k(\mathfrak{F}_\lambda)$  has, at least one row if, and only if, it holds $\lambda_1\geq k+1$. On the other hand, for those Ferrers diagrams $\mathfrak{F}$ having no more than $k$ points in their first row, we put $D_k(\mathfrak{F})$ as the ``empty'' \emph{Durfee $k$-rectangle}, which has zero rows (see Figure \ref{fig: durfee rect of a code}). Considering this special case makes the set $D_k(\cC)$ contain at least one element, for every $0\leq k\leq n-2$.

	\begin{remark}\label{rem: sides Durfee rectangles}
		Observe that the set $D_k(\cC)$ can be encoded as a set of integers as follows. Assume that 
		$$
		D_k(\cC)=\{ \mathfrak{R}_1^k, \dots, \mathfrak{R}_{m_k}^k\},
		$$
		where each $\mathfrak{R}_j^k$ is a Durfee $k$-rectangle having $0\leq r^k_j\leq \left\lfloor \frac{n-k}{2}\right\rfloor$ rows, thus $r^k_j\times (r^k_j+k)$ points. Then, to know $D_k(\cC)$, we just need to store the list of integers $\{r^k_1, \dots, r^k_{m_k}\}$. Moreover, without loss of generality, we can assume that $r^k_1 > r^k_2 > \dots > r^k_{m_k}\geq 0$ so that $ \mathfrak{R}_1^k$ is the biggest Durfee $k$-rectangle in $D_k(\cC)$. As said before, we are also contemplating the possibility of having Durfee $k$-rectangles with $0$ rows. Hence, in any case, the value $r^k_1$ associated to $\cC$ always exists.
	\end{remark}
	
	Let us finally see how can we use Durfee rectangles to relate the parameters of $\cC$ to the ones of its projected codes.

	\subsection{From the flag distance to subspace distances}
	
	By construction, a Durfee $k$-rectangle in $\mathrm{FF}(n)$ having $r$ rows (and then $r+k$ columns) has its left down vertex at the point with coordinates
	$$
	\left( \frac{n-k}{2}, \frac{n-k}{2}-(r-1)  \right).
	$$
	in the enriched distance support $\hat{\mathrm{S}}(n)$. This vertex will help us to obtain information about the projected codes of a given flag code whenever it is a point in the distance support $\mathrm{S}(n)$, i.e., if it is a circle black point. Observe also that this happens if, and only if, $\frac{n-k}{2}$ is an integer or, equivalently, when $n$ and $k$ have the same parity. In this case, $k$-rectangles give information about the projected code of dimension $\frac{n-k}{2}$. The following picture illustrates this fact. 
	
	At left, for $n=8$, we have three Durfee $4$-rectangles with $0, 1$ and $2$ rows, respectively. Their left down vertices, after rotating back the diagram, are in the column corresponding to the dimension $\frac{8-4}{2}=2$. At right, different Durfee $1$-rectangles on $\mathrm{FF}(7)$ are shown. All of them have their left down vertices in the column related to the dimension $\frac{7-1}{2}=3$.	Having this in mind, we start from  our full flag code $\cC$ on $\bbF_q^n$ and take $0\leq k\leq n-2$ with the same parity than $n$. We use the set of $k$-rectangles $D_k(\cC)$ in order to derive information about the cardinality and minimum distance of the projected code of dimension $i=\frac{n-k}{2}$ with $i \in \{1, \dots, \left\lfloor \frac{n}{2}\right\rfloor\}$. Conversely, if we are specifically interested in the projected code  $\cC_i$, where $1\leq i\leq  \left\lfloor \frac{n}{2}\right\rfloor$, then we just need to consider the set of $(n-2i)$-rectangles of $\cC$.

	\begin{theorem}\label{lem: distance and rectangle}
		Let $\cC$ be a full flag code on $\bbF_q^{n}$ and consider a dimension $1\leq i\leq  \left\lfloor \frac{n}{2}\right\rfloor$. They are equivalent:
		\begin{enumerate}[(i)]
			\item There are Durfee $(n-2i)$-rectangles in $D_{n-2i}(\cC)$ with $r$ rows, where $0\leq r\leq i$.
			\item There exist flags $\cF, \cF'\in\cC$ such that $d_I(\cF_{i}, \cF'_{i})=i-r$.
		\end{enumerate}
	\end{theorem}
	\begin{proof}
		Notice that any $(n-2i)$-Durfee rectangle in the Ferrers diagram frame has its lower left vertex in a circle black dot, corresponding to the dimension $i$ in the distance support. Moreover, the number of rows of the rectangle coincides with the number of dots being simultaneously in the rectangle and in $\mathrm{S}(i, n)$, i.e., in the $i$-th column of the distance support  $\mathrm{S}(n)$ (recall (\ref{def: injection distance support}) and (\ref{def: injection distance support2})). 
		
		Hence, a Durfee $(n-2i)$-rectangle in $D_{n-2i}(\cC)$ with $r$ rows is determined by the existence of Ferrers subdiagrams in $\mathfrak{F}(\cC)$ whose silhouettes pass through the black circle point $(i, i-r)$. This happens if, and only if, the skeleton of each of these staircase paths is a distance path in $\Gamma(\cC)$ passing through the vertex $(i, i-r)$ as well. Equivalently, there must exist flags $\cF, \cF'\in\cC$ with $d_I(\cF_{i}, \cF'_{i})= i-r$.
	\end{proof}

	Now, we focus on the biggest Durfee $(n-2i)$-rectangle of $\cC$, denoted in Remark \ref{rem: sides Durfee rectangles} as $\mathfrak{R}^{n-2i}_1$, and that contains $0\leq r^{n-2i}_1 \leq i$ rows. In addition, since we will always work with rectangles in $D_{n-2i}(\cC)$, we drop the superscripts and simply write $\mathfrak{R}_1$ and $r_1$, respectively. In light of the previous theorem, we analyze two possible situations concluding the following results. 
	\begin{theorem}\label{theo: dist proj durfee even 1}
		Assume that $\cC$ is a full flag code on $\bbF_q^n$. Consider a dimension $1\leq i\leq \left\lfloor\frac{n}{2}\right\rfloor$ and the biggest rectangle $\mathfrak{R}_1$ in $D_{n-2i}(\cC)$. The following statements are equivalent:
		\begin{enumerate}
			\item The number of rows of $\mathfrak{R}_1$, that is $r_1$, satisfies $0\leq r_1 <i$.
			\item It holds $|\cC|=|\cC_{i}|$ and $d_I(\cC_{i}) = i-r_1$.
		\end{enumerate}
	\end{theorem}
	\begin{proof}
		By application of Theorem \ref{lem: distance and rectangle} and the maximality of $r_1$, we know that the existence of Durfee $(n-2i)$-rectangles with $0\leq r_1<i$ rows in $D_{n-2i}(\cC)$ is equivalent to say that, for any choice of different flags $\cF,\cF'\in\cC$, we have
		$$
		d_I(\cF_{i}, \cF'_{i})  \geq i-r_1 > 0
		$$
		and the equality holds for some pair of flags in the code. In other words, $|\cC_{i}|=|\cC|$ because a couple of flags in $\cC$ never share their $i$-th subspaces and, clearly, $d_I(\cC_{i})=i-r_1$.
	\end{proof}	
	
	Now we study the remaining case, in which $r_1$ attains its maximum possible value, that is, $r_1=i$. In that situation, we obtain valuable information about the projected code $\cC_i$ in terms of the second largest rectangle $\mathfrak{R}_2$ in $D_{n-2i}(\cC)$, whenever it exists. More precisely:
	\begin{theorem}\label{theo: dist proj durfee even 2}
		Let $\cC$ be a full flag code on $\bbF_q^n$. Fix a dimension $1\leq i \leq \left\lfloor\frac{n}{2}\right\rfloor$ and assume that the largest rectangle $\mathfrak{R}_1\in D_{n-2i}(\cC)$ has $r_1=i$ rows. Hence, the following statements hold:
		\begin{enumerate}
			\item $D_{n-2i}(\cC)=\{\mathfrak{R}_1\}$ if, and only if, $|\cC_{i}|=1$ or, equivalently, $d_I(\cC_{i})=0$.
			\item $D_{n-2i}(\cC)\neq\{\mathfrak{R}_1\}$ if, and only if, $|\cC|>|\cC_{k}| > 1$ and $d_I(\cC_{k})= k - r_2 > 0,$ where $0\leq r_2 < r_1=i$ is the number of rows of $\mathfrak{R}_2$.
		\end{enumerate}
	\end{theorem}
	\begin{proof}
		By means of Theorem \ref{lem: distance and rectangle}, the condition $r_1= i$ is equivalent to say that $d_I(\cF_{i}, \cF'_{i})=0$  for a pair of different flags $\cF, \cF'\in\cC$. This happens if, and only if, $\cF_{i}=\cF'_{i}$, i.e, if $|\cC_{i}|<|\cC|$. Let us distinguish two cases:
		\begin{enumerate}
			\item If $\mathfrak{R}_1$ is the only $(n-2i)$-rectangle of the code, by Theorem \ref{lem: distance and rectangle}, we have that  $d_I(\cF_{i}, \cF'_{i})=0$ for every pair of different flags $\cF, \cF'\in\cC$. In other words, all the flags in $\cC$ share their common $i$-th subspace. Equivalently, the projected code $\cC_{i}$ consists of a single subspace and $d_I(\cC_{i})=0$.
			\item On the other hand, if $D_{n-2i}(\cC)\neq\{\mathfrak{R}_1\}$, we can consider the second largest $(n-2i)$-rectangle $\mathfrak{R}_2$ of $\cC$, which has $0\leq r_2 <  r_1= i$ rows. Hence, for every pair of different flags $\cF, \cF'\in\cC$ it holds either $d_I(\cF_{i}, \cF'_{i})=0$  or $d_I(\cF_{i}, \cF'_{i})\geq i-r_2,$ and the last inequality becomes an equality for some choice of flags in $\cC$ by means of Theorem \ref{lem: distance and rectangle}. Equivalently,  $d_I(\cC_{i})=i-r_2>i-r_1 =0$. 
		\end{enumerate}
	\end{proof}

	\begin{remark}\label{rem: higher dimensions}
		Observe that Theorems \ref{theo: dist proj durfee even 1} and \ref{theo: dist proj durfee even 2} give us information about the parameters of all the projected codes of dimensions up to $\left\lfloor\frac{n}{2}\right\rfloor$. For the remaining dimensions, it suffices to reason in the same way by simply considering maximal $k$-rectangles with $c$ columns and $c+k$ rows. Moreover, it is important to point out that we do not even need to know all the rectangles in $D_{n-2i}(\cC)$; it suffices to know the largest one and, in the worst case, also the second one.
	\end{remark}
	
	From the results stated along this subsection, and applying Lemma \ref{lem: points in rectangle}, we derive bounds for the minimum distance of the projected codes  of $\cC$ in terms of its codistance $\bar{d}_f(\cC)$.
	
	\begin{corollary}\label{cor: from df to dist projected}
		Let $\cC$ be a full flag code on $\bbF_q^{n}$ with associated codistance $\bar{d}_f(\cC)$. Take a dimension  $1\leq i\leq \lfloor\frac{n}{2}\rfloor$ and consider an integer $0\leq r \leq i$. Hence, whenever 
		\begin{equation}\label{eq: codistance condition corolario}
			\bar{d}_f(\cC) < \left\lceil\frac{r(r+n-2i)}{2}\right\rceil,
		\end{equation}
		then $|\cC_i|=|\cC|$ and $d_I(\cC_{i}) > i-r.$
	\end{corollary}
	\begin{proof}
		It suffices to see that, by means of Lemma \ref{lem: points in rectangle}, the number of circle black points in a rectangle with $r$ rows and $r+n-2i$ columns is exactly 
		$$
		\left\lbrace
		\begin{array}{cccl}
			\left\lceil\frac{r(r+n-2i)}{2}\right\rceil   & \text{if} & n & \text{is even,}\\
			& & & \\[-1em]
			\left\lfloor\frac{r(r+n-2i)}{2}\right\rfloor & \text{if} & n & \text{is odd.}
		\end{array}
		\right.
		$$
		Morover, notice that, for odd values of $n$, the product $r(r+n-2i)$ is always even and then
		$$
		\left\lfloor\frac{r(r+n-2i)}{2}\right\rfloor = \frac{r(r+n-2i)}{2} = \left\lceil\frac{r(r+n-2i)}{2}\right\rceil.
		$$
		Hence, under condition (\ref{eq: codistance condition corolario}), the number of rows of any Durfee $(n-2i)$-rectangle in $D_{n-2i}(\cC)$ is upper bounded by $r_1 < r\leq i$ and we conclude, by means of Theorem \ref{theo: dist proj durfee even 1}, that $|\cC_i|=|\cC|$ and $d_I(\cC_i)=i-r_1> i-r$.
	\end{proof}
	
	\subsection{From subspace distances to the flag distance}
	In this subsection we deal with the converse problem: given a full flag code on $\bbF_q^n$, we obtain information of its parameters from the ones of its projected codes. As before we restrict our study to dimensions $1\leq i\leq \left\lfloor\frac{n}{2}\right\rfloor$ since, for higher dimensions, it suffices to consider rectangles with more rows than columns.
	
	\begin{theorem}\label{theo: from subspace distance to flag dist n even}
		Let $\cC$ be a full flag code on $\bbF_q^{n}$ and take $1\leq i \leq \left\lfloor\frac{n}{2}\right\rfloor$.
		\begin{enumerate}
			\item If $|\cC_{i}|=|\cC|,$ then 
			$$
			d_I(\cC_{i})^2 \leq d_f(\cC) \leq D^n - \left\lceil \frac{(i-d_I(\cC_{i}))(n-i-d_I(\cC_i))}{2}\right\rceil.
			$$
			\item If $|\cC_{i}|<|\cC|,$ then 
			$$
			0 \leq d_f(\cC) \leq  D^n - \left\lceil \frac{i(n-i)}{2}\right\rceil.
			$$
		\end{enumerate}
	\end{theorem}
	\begin{proof}
		Let us start assuming that $|\cC_{i}|=|\cC|$. If we write $d_I(\cC_i)=i-r$ for some $0\leq r\leq i$, then Corollary \ref{cor: from df to dist projected} leads to
		$$
		\bar{d}_f(\cC) \geq \left\lceil\frac{r(r+n-2i)}{2}\right\rceil = \left\lceil\frac{(i-d_I(\cC_i))(n-i-d_I(\cC_i))}{2}\right\rceil.
		$$
		As a consequence, it holds $d_f(\cC)=D^n-\bar{d}_f(\cC) \leq D^n  - \left\lceil\frac{(i-d_I(\cC_i))(n-i-d_I(\cC_i))}{2}\right\rceil$. On the other hand, given arbitrary different flags $\cF, \cF'\in\cC$, we have that $\cF_i\neq\cF'_i$ and then $d_I(\cF_{i}, \cF'_{i})\geq d_I(\cC_{i})$. Hence, every distance path in $\Gamma(\cC)$ passes either through the vertex $(i, d_I(\cC_{i}))$  (green vertex in the next figure) or above it. As a result, such a vertex determines a set of circle black dots that always remain under distance paths in $\Gamma(\cC)$. This set is exactly the triangle in red in the picture below, which top vertex is the point $(i, d_I(\cC_{i}))$ and edges with slope $1$ (left) and $-1$ (right).
		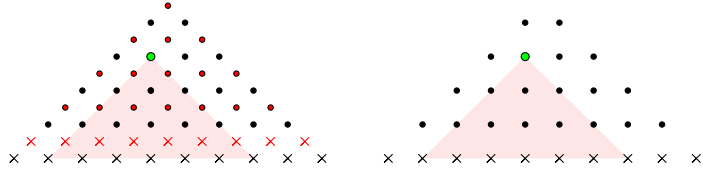
\begin{figure}[H]
			\begin{center}
				\begin{tikzpicture}[line cap=round,line join=round,>=triangle 45,x=0.45cm,y=0.45cm]
					\fill[line width=2pt,fill=red,fill opacity=0.1] (1,0)--(4,3) -- (7,0) -- cycle;
					\begin{scriptsize}
						\draw [fill=black] (1,1) circle (1pt);
						\draw [fill=black] (2,1) circle (1pt);
						\draw [fill=black] (6,1) circle (1pt);
						\draw [fill=black] (6,2) circle (1pt);
						\draw [fill=black] (6,3) circle (1pt);
						\draw [fill=black] (2,2) circle (1pt);
						\draw [fill=black] (4,2) circle (1pt);
						\draw [fill=black] (4,1) circle (1pt);
						\draw [fill=green] (4,3) circle (1.5pt);
						\draw [fill=black] (4,4) circle (1pt);
						\draw [fill=black] (8,1) circle (1pt);
						\draw [color=black] (0,0)-- ++(-1.5pt,-1.5pt) -- ++(3pt,3pt) ++(-3pt,0) -- ++(3pt,-3pt);
						\draw [color=black] (1,0)-- ++(-1.5pt,-1.5pt) -- ++(3pt,3pt) ++(-3pt,0) -- ++(3pt,-3pt);
						\draw [color=black] (2,0)-- ++(-1.5pt,-1.5pt) -- ++(3pt,3pt) ++(-3pt,0) -- ++(3pt,-3pt);
						\draw [color=black] (3,0)-- ++(-1.5pt,-1.5pt) -- ++(3pt,3pt) ++(-3pt,0) -- ++(3pt,-3pt);
						\draw [color=black] (4,0)-- ++(-1.5pt,-1.5pt) -- ++(3pt,3pt) ++(-3pt,0) -- ++(3pt,-3pt);
						\draw [color=black] (6,0)-- ++(-1.5pt,-1.5pt) -- ++(3pt,3pt) ++(-3pt,0) -- ++(3pt,-3pt);
						\draw [color=black] (8,0)-- ++(-1.5pt,-1.5pt) -- ++(3pt,3pt) ++(-3pt,0) -- ++(3pt,-3pt);
						\draw [color=black] (9,0)-- ++(-1.5pt,-1.5pt) -- ++(3pt,3pt) ++(-3pt,0) -- ++(3pt,-3pt);
						\draw [fill=black] (4,2) circle (1pt);
						\draw [fill=black] (3,1) circle (1pt);
						\draw [fill=black] (3,2) circle (1pt);
						\draw [fill=black] (3,3) circle (1pt);
						\draw [fill=black] (5,4) circle (1pt);
						\draw [fill=black] (5,3) circle (1pt);
						\draw [fill=black] (5,2) circle (1pt);
						\draw [fill=black] (5,1) circle (1pt);
						\draw [fill=black] (7,2) circle (1pt);
						\draw [fill=black] (7,1) circle (1pt);
						\draw [fill=red] (1.5,1.5) circle (1pt);
						\draw [fill=red] (2.5,1.5) circle (1pt);
						\draw [fill=red] (3.5,1.5) circle (1pt);
						\draw [fill=red] (4.5,1.5) circle (1pt);
						\draw [fill=red] (5.5,1.5) circle (1pt);
						\draw [fill=red] (6.5,1.5) circle (1pt);
						\draw [fill=red] (2.5,2.5) circle (1pt);
						\draw [fill=red] (3.5,2.5) circle (1pt);
						\draw [fill=red] (4.5,2.5) circle (1pt);
						\draw [fill=red] (5.5,2.5) circle (1pt);
						\draw [fill=red] (7.5,1.5) circle (1pt);
						\draw [fill=red] (6.5,2.5) circle (1pt);
						\draw [fill=red] (4.5,3.5) circle (1pt);
						\draw [fill=red] (5.5,3.5) circle (1pt);
						\draw [fill=red] (3.5,3.5) circle (1pt);
						\draw [fill=red] (4.5,4.5) circle (1pt);
						\draw [color=red] (0.5,0.5)-- ++(-1.5pt,-1.5pt) -- ++(3pt,3pt) ++(-3pt,0) -- ++(3pt,-3pt);
						\draw [color=red] (1.5,0.5)-- ++(-1.5pt,-1.5pt) -- ++(3pt,3pt) ++(-3pt,0) -- ++(3pt,-3pt);
						\draw [color=red] (2.5,0.5)-- ++(-1.5pt,-1.5pt) -- ++(3pt,3pt) ++(-3pt,0) -- ++(3pt,-3pt);
						\draw [color=red] (3.5,0.5)-- ++(-1.5pt,-1.5pt) -- ++(3pt,3pt) ++(-3pt,0) -- ++(3pt,-3pt);
						\draw [color=red] (4.5,0.5)-- ++(-1.5pt,-1.5pt) -- ++(3pt,3pt) ++(-3pt,0) -- ++(3pt,-3pt);
						\draw [color=red] (5.5,0.5)-- ++(-1.5pt,-1.5pt) -- ++(3pt,3pt) ++(-3pt,0) -- ++(3pt,-3pt);
						\draw [color=red] (6.5,0.5)-- ++(-1.5pt,-1.5pt) -- ++(3pt,3pt) ++(-3pt,0) -- ++(3pt,-3pt);
						\draw [color=red] (7.5,0.5)-- ++(-1.5pt,-1.5pt) -- ++(3pt,3pt) ++(-3pt,0) -- ++(3pt,-3pt);
						\draw [color=red] (8.5,0.5)-- ++(-1.5pt,-1.5pt) -- ++(3pt,3pt) ++(-3pt,0) -- ++(3pt,-3pt);
						\draw [color=black] (7,0)-- ++(-1.5pt,-1.5pt) -- ++(3pt,3pt) ++(-3pt,0) -- ++(3pt,-3pt);
						\draw [color=black] (5,0)-- ++(-1.5pt,-1.5pt) -- ++(3pt,3pt) ++(-3pt,0) -- ++(3pt,-3pt);
					\end{scriptsize}
				\end{tikzpicture}
				\hspace{0.5cm}
				\begin{tikzpicture}[line cap=round,line join=round,>=triangle 45,x=0.45cm,y=0.45cm]
					\fill[line width=2pt,fill=red,fill opacity=0.1] (1,0)--(4,3) -- (7,0) -- cycle;
					\begin{scriptsize}
						\draw [fill=black] (1,1) circle (1pt);
						\draw [fill=black] (2,1) circle (1pt);
						\draw [fill=black] (6,1) circle (1pt);
						\draw [fill=black] (6,2) circle (1pt);
						\draw [fill=black] (6,3) circle (1pt);
						\draw [fill=black] (2,2) circle (1pt);
						\draw [fill=black] (4,2) circle (1pt);
						\draw [fill=black] (4,1) circle (1pt);
						\draw [fill=black] (4,4) circle (1pt);
						\draw [fill=black] (8,1) circle (1pt);
						\draw [color=black] (0,0)-- ++(-1.5pt,-1.5pt) -- ++(3pt,3pt) ++(-3pt,0) -- ++(3pt,-3pt);
						\draw [color=black] (1,0)-- ++(-1.5pt,-1.5pt) -- ++(3pt,3pt) ++(-3pt,0) -- ++(3pt,-3pt);
						\draw [color=black] (2,0)-- ++(-1.5pt,-1.5pt) -- ++(3pt,3pt) ++(-3pt,0) -- ++(3pt,-3pt);
						\draw [color=black] (3,0)-- ++(-1.5pt,-1.5pt) -- ++(3pt,3pt) ++(-3pt,0) -- ++(3pt,-3pt);
						\draw [color=black] (4,0)-- ++(-1.5pt,-1.5pt) -- ++(3pt,3pt) ++(-3pt,0) -- ++(3pt,-3pt);
						\draw [color=black] (6,0)-- ++(-1.5pt,-1.5pt) -- ++(3pt,3pt) ++(-3pt,0) -- ++(3pt,-3pt);
						\draw [color=black] (8,0)-- ++(-1.5pt,-1.5pt) -- ++(3pt,3pt) ++(-3pt,0) -- ++(3pt,-3pt);
						\draw [color=black] (9,0)-- ++(-1.5pt,-1.5pt) -- ++(3pt,3pt) ++(-3pt,0) -- ++(3pt,-3pt);
						\draw [fill=green] (4,3) circle (1.5pt);
						\draw [fill=black] (3,1) circle (1pt);
						\draw [fill=black] (3,2) circle (1pt);
						\draw [fill=black] (3,3) circle (1pt);
						\draw [fill=black] (5,4) circle (1pt);
						\draw [fill=black] (5,3) circle (1pt);
						\draw [fill=black] (5,2) circle (1pt);
						\draw [fill=black] (5,1) circle (1pt);
						\draw [fill=black] (7,2) circle (1pt);
						\draw [fill=black] (7,1) circle (1pt);
						\draw [color=black] (7,0)-- ++(-1.5pt,-1.5pt) -- ++(3pt,3pt) ++(-3pt,0) -- ++(3pt,-3pt);
						\draw [color=black] (5,0)-- ++(-1.5pt,-1.5pt) -- ++(3pt,3pt) ++(-3pt,0) -- ++(3pt,-3pt);
					\end{scriptsize}
				\end{tikzpicture}
				\caption{Vertex $(i, d_I(\cC_{i}))$ (in green) and the triangle ``under'' it (in red).}
			\end{center}
		\end{figure}
		\noindent The number of circle black points in such a triangle coincides with the number of circle dots in the distance support $\mathrm{S}(2d_I(\cC_{i}))$, which is $d_I(\cC_{i})^2$. Hence, Remark \ref{rem: distance 2 number of dots} leads to $d_f(\cC)\geq  d_I(\cC_{i})^2$. Now, if we suppose that $|\cC_{i}| < |\cC|$, the result follows straightforwardly by Theorem \ref{theo: separabilidad en funcion codistancia}.
	\end{proof}

	\subsection{Some examples}
	We finish this section by applying the previous results to three representative situations.
	\begin{example}
		For $n=8$, the  full flag distance takes values in the interval $[0, 16]$. Consider an arbitrary full flag code $\cC$ with minimum distance $d_f(\cC)=12$ or, equivalently, codistance $\bar{d}_f(\cC)= 16-12=4$. For this value of the codistance, and by application of Theorem \ref{theo: separabilidad en funcion codistancia}, we can derive that $|\cC|= |\cC_2|=\dots=|\cC_6|,$ since $\bar{d}_f(\cC)=4 < \frac{2(8-2)}{2} = 6$. Concerning the projected codes distances, for dimension $4$, we look at the number of points in Durfee squares ($0$-rectangles). Since a Ferrers subdiagram with a Durfee square with $3$ rows contains, at least $\lceil 9/2\rceil=5 > 4$ circle black points, by means of Corollary \ref{cor: from df to dist projected}, we can ensure that $d_I(\cC_4) \geq 2$. 
		\begin{figure}[H]
			\centering
			\begin{tikzpicture}[rotate=-45, line cap=round,line join=round,>=triangle 45,x=0.6cm,y=0.3cm]
				\fill[line width=2pt,color=blue,fill=blue,fill opacity=0.3] (4,5) --  (3,7)  -- (4,9) -- (5,7) -- cycle;
				\begin{scriptsize}
					\draw [color=black] (0,0) node[cross] {};
					\draw [color=black] (1,0) node[cross] {};
					\draw [color=black] (2,0) node[cross] {};
					\draw [color=black] (3,0) node[cross] {};
					\draw [color=black] (4,0) node[cross] {};
					\draw [color=black] (5,0) node[cross] {};
					\draw [color=black] (6,0) node[cross] {};
					\draw [color=black] (7,0) node[cross] {};
					\draw [color=black] (8,0) node[cross] {}; 
					\draw [fill=black] (1,2) circle (1.5pt);
					\draw [fill=black] (2,2) circle (1.5pt);
					\draw [fill=black] (2,4) circle (1.5pt);
					\draw [fill=black] (3,2) circle (1.5pt);
					\draw [fill=black] (3,4) circle (1.5pt);
					\draw [fill=black] (3,6) circle (1.5pt);
					\draw [fill=black] (4,2) circle (1.5pt);
					\draw [fill=black] (4,4) circle (1.5pt);
					\draw [fill=black] (4,6) circle (1.5pt);
					\draw [fill=black] (4,8) circle (1.5pt);
					\draw [fill=black] (5,2) circle (1.5pt);
					\draw [fill=black] (5,4) circle (1.5pt);
					\draw [fill=black] (5,6) circle (1.5pt);
					\draw [fill=black] (6,2) circle (1.5pt);
					\draw [fill=black] (6,4) circle (1.5pt);
					\draw [fill=black] (7,2) circle (1.5pt);
					\draw [color=red] (0.5,1) node[cross, red] {};
					\draw [color=red] (1.5,1) node[cross, red] {};
					\draw [color=red] (2.5,1) node[cross, red] {};
					\draw [color=red] (3.5,1) node[cross, red] {};
					\draw [color=red] (4.5,1) node[cross, red] {};
					\draw [color=red] (5.5,1) node[cross, red] {};
					\draw [color=red] (6.5,1) node[cross, red] {};
					\draw [color=red] (7.5,1) node[cross, red] {};
					\draw [fill=red] (1.5,3) circle (1.5pt);
					\draw [fill=red] (2.5,3) circle (1.5pt);
					\draw [fill=red] (2.5,5) circle (1.5pt);
					\draw [fill=red] (3.5,3) circle (1.5pt);
					\draw [fill=red] (3.5,5) circle (1.5pt);
					\draw [fill=red] (3.5,7) circle (1.5pt);
					\draw [fill=red] (4.5,3) circle (1.5pt);
					\draw [fill=red] (4.5,5) circle (1.5pt);
					\draw [fill=red] (4.5,7) circle (1.5pt);
					\draw [fill=red] (5.5,3) circle (1.5pt);
					\draw [fill=red] (5.5,5) circle (1.5pt);
					\draw [fill=red] (6.5,3) circle (1.5pt);
				\end{scriptsize}
			\end{tikzpicture}
			\begin{tikzpicture}[rotate=-45, line cap=round,line join=round,>=triangle 45,x=0.6cm,y=0.3cm]
				\fill[line width=2pt,color=blue,fill=blue,fill opacity=0.3]  (3,3) -- (2,5) -- (4,9) -- (5,7) -- cycle;
				\begin{scriptsize}
					\draw [color=black] (0,0) node[cross] {};
					\draw [color=black] (1,0) node[cross] {};
					\draw [color=black] (2,0) node[cross] {};
					\draw [color=black] (3,0) node[cross] {};
					\draw [color=black] (4,0) node[cross] {};
					\draw [color=black] (5,0) node[cross] {};
					\draw [color=black] (6,0) node[cross] {};
					\draw [color=black] (7,0) node[cross] {};
					\draw [color=black] (8,0) node[cross] {}; 
					\draw [fill=black] (1,2) circle (1.5pt);
					\draw [fill=black] (2,2) circle (1.5pt);
					\draw [fill=black] (2,4) circle (1.5pt);
					\draw [fill=black] (3,2) circle (1.5pt);
					\draw [fill=black] (3,4) circle (1.5pt);
					\draw [fill=black] (3,6) circle (1.5pt);
					\draw [fill=black] (4,2) circle (1.5pt);
					\draw [fill=black] (4,4) circle (1.5pt);
					\draw [fill=black] (4,6) circle (1.5pt);
					\draw [fill=black] (4,8) circle (1.5pt);
					\draw [fill=black] (5,2) circle (1.5pt);
					\draw [fill=black] (5,4) circle (1.5pt);
					\draw [fill=black] (5,6) circle (1.5pt);
					\draw [fill=black] (6,2) circle (1.5pt);
					\draw [fill=black] (6,4) circle (1.5pt);
					\draw [fill=black] (7,2) circle (1.5pt);
					\draw [color=red] (0.5,1) node[cross, red] {};
					\draw [color=red] (1.5,1) node[cross, red] {};
					\draw [color=red] (2.5,1) node[cross, red] {};
					\draw [color=red] (3.5,1) node[cross, red] {};
					\draw [color=red] (4.5,1) node[cross, red] {};
					\draw [color=red] (5.5,1) node[cross, red] {};
					\draw [color=red] (6.5,1) node[cross, red] {};
					\draw [color=red] (7.5,1) node[cross, red] {};
					\draw [fill=red] (1.5,3) circle (1.5pt);
					\draw [fill=red] (2.5,3) circle (1.5pt);
					\draw [fill=red] (2.5,5) circle (1.5pt);
					\draw [fill=red] (3.5,3) circle (1.5pt);
					\draw [fill=red] (3.5,5) circle (1.5pt);
					\draw [fill=red] (3.5,7) circle (1.5pt);
					\draw [fill=red] (4.5,3) circle (1.5pt);
					\draw [fill=red] (4.5,5) circle (1.5pt);
					\draw [fill=red] (4.5,7) circle (1.5pt);
					\draw [fill=red] (5.5,3) circle (1.5pt);
					\draw [fill=red] (5.5,5) circle (1.5pt);
					\draw [fill=red] (6.5,3) circle (1.5pt);
				\end{scriptsize}
			\end{tikzpicture}
			\begin{tikzpicture}[rotate=-45, line cap=round,line join=round,>=triangle 45,x=0.6cm,y=0.3cm]
				\fill[line width=2pt,color=blue,fill=blue,fill opacity=0.3]  (2,3) -- (1.5,4) -- (4,9) -- (4.5,8) -- cycle;
				\begin{scriptsize}
					\draw [color=black] (0,0) node[cross] {};
					\draw [color=black] (1,0) node[cross] {};
					\draw [color=black] (2,0) node[cross] {};
					\draw [color=black] (3,0) node[cross] {};
					\draw [color=black] (4,0) node[cross] {};
					\draw [color=black] (5,0) node[cross] {};
					\draw [color=black] (6,0) node[cross] {};
					\draw [color=black] (7,0) node[cross] {};
					\draw [color=black] (8,0) node[cross] {}; 
					\draw [fill=black] (1,2) circle (1.5pt);
					\draw [fill=black] (2,2) circle (1.5pt);
					\draw [fill=black] (2,4) circle (1.5pt);
					\draw [fill=black] (3,2) circle (1.5pt);
					\draw [fill=black] (3,4) circle (1.5pt);
					\draw [fill=black] (3,6) circle (1.5pt);
					\draw [fill=black] (4,2) circle (1.5pt);
					\draw [fill=black] (4,4) circle (1.5pt);
					\draw [fill=black] (4,6) circle (1.5pt);
					\draw [fill=black] (4,8) circle (1.5pt);
					\draw [fill=black] (5,2) circle (1.5pt);
					\draw [fill=black] (5,4) circle (1.5pt);
					\draw [fill=black] (5,6) circle (1.5pt);
					\draw [fill=black] (6,2) circle (1.5pt);
					\draw [fill=black] (6,4) circle (1.5pt);
					\draw [fill=black] (7,2) circle (1.5pt);
					\draw [color=red] (0.5,1) node[cross, red] {};
					\draw [color=red] (1.5,1) node[cross, red] {};
					\draw [color=red] (2.5,1) node[cross, red] {};
					\draw [color=red] (3.5,1) node[cross, red] {};
					\draw [color=red] (4.5,1) node[cross, red] {};
					\draw [color=red] (5.5,1) node[cross, red] {};
					\draw [color=red] (6.5,1) node[cross, red] {};
					\draw [color=red] (7.5,1) node[cross, red] {};
					\draw [fill=red] (1.5,3) circle (1.5pt);
					\draw [fill=red] (2.5,3) circle (1.5pt);
					\draw [fill=red] (2.5,5) circle (1.5pt);
					\draw [fill=red] (3.5,3) circle (1.5pt);
					\draw [fill=red] (3.5,5) circle (1.5pt);
					\draw [fill=red] (3.5,7) circle (1.5pt);
					\draw [fill=red] (4.5,3) circle (1.5pt);
					\draw [fill=red] (4.5,5) circle (1.5pt);
					\draw [fill=red] (4.5,7) circle (1.5pt);
					\draw [fill=red] (5.5,3) circle (1.5pt);
					\draw [fill=red] (5.5,5) circle (1.5pt);
					\draw [fill=red] (6.5,3) circle (1.5pt);
				\end{scriptsize}
			\end{tikzpicture}
			\caption{Largest $k$-rectangles with at most $4$ black points for $k=0,2,4$.}
		\end{figure}
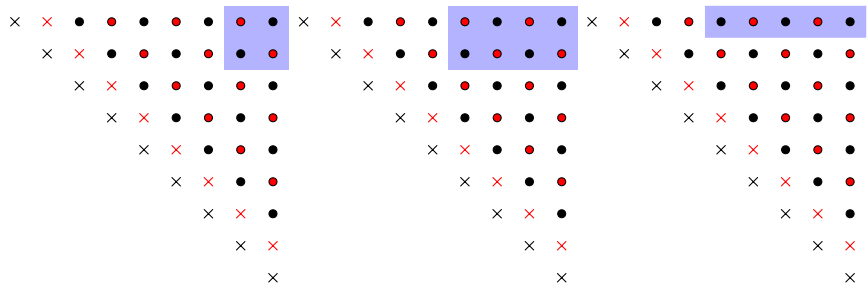
		
		Similarly, for dimensions $2$ and $3$, we consider $4$-rectangles and $2$-rectangles, respectively and obtain $d_I(\cC_2)\geq 1$ and $d_I(\cC_3)\geq 1$. These properties are transferred to the projected codes of dimensions $5$ and $6$ by symmetry of the Ferrers diagram frame.	On the other hand, the  knowledge of just the code codistance does not give any information about the projected codes $\cC_1$ and $\cC_7$. To extract more information in these last cases it is also necessary to know if $D_6(\cC)$ contains either one or two elements. There are three possible situations:
		$$
		D_6(\cC) = 
		\left\lbrace
		\begin{array}{lll}
			\{ \mathfrak{R}_1\}      & \text{and} & r_1=0, \\
			\{ \mathfrak{R}_1\}      & \text{and} & r_1=1, \\
			\{ \mathfrak{R}_1, \mathfrak{R}_2\} & \text{and} & r_1=1 > r_2=0.
		\end{array}
		\right.
		$$
		In the first case, by means of Theorem \ref{theo: dist proj durfee even 1}, we know that $|\cC_1|=|\cC|$ and $d_I(\cC_1)=1$. In the remaining cases, we have $|\cC_1|<|\cC|$. Moreover, Theorem  \ref{theo: dist proj durfee even 2} leads to $d_I(\cC_1)=0$ in the second situation and to $d_I(\cC_1)=1$ in the third one.
	\end{example}
	
	\begin{example}
		Also for $n=8$, now we consider a full flag code $\cC'$ and assume that the projected code $\cC'_4$ satisfies $|\cC'_4|=|\cC'|$ and $d_I( \cC'_4)=2$. In this case, by application of Theorem \ref{theo: from subspace distance to flag dist n even}, we know that $4\leq d_f(\cC')\leq 14.$ Moreover, if,  in addition,  $|\cC'_3|=|\cC|$ and $d_I(\cC'_3)=1$, then Theorem \ref{theo: from subspace distance to flag dist n even} leads also to $1\leq d_f(\cC')\leq 12$ and we conclude that $4\leq d_f(\cC')\leq 12.$
		
		In general, the more conditions on the projected codes, the more information on the flag code. However, at times we obtain redundant information. For instance, if we require the projected code $\cC'_2$ to fulfill
		$ |\cC'_2|=|\cC'| \ \text{and} \ \ d_I(\cC'_2)=1,$
		we obtain $1\leq d_f(\cC')\leq 13$, which does not provide any new data. Nevertheless, if $|\cC'_2|<|\cC'|$, we improve our knowledge about $d_f(\cC')$ since, by virtue of Theorem \ref{theo: from subspace distance to flag dist n even}, it must hold $4\leq d_f(\cC')\leq 10.$
	\end{example} 
	\begin{remark}
		As pointed out in Remark \ref{rem: relation distance projected}, determining the exact value of $d_f(\cC')$ from just the list of distances $\{d_I(\cC_i)\}_{i=1}^{7}$, and conversely, is not always possible. Nevertheless, we have seen that just with the data of at most two Durfee $(n-2i)$-rectangles for any dimension $i$, which is considerably less than knowing the whole set of distance paths of $\cC$, we are able to establish interesting connections between flag codes and their projected codes.
	\end{remark}
	
	\begin{example}
		We finish this subsection by looking at the special case of full flag codes with the maximum possible distance. These codes were characterized in \cite[Th. 3.11]{CasoPlanarCHAP9} in terms of their projected codes, which must attain the maximum possible distance for their dimensions and have the same cardinality than the flag code. Here below, we translate that result to the setting introduced in this work to state an alternative combinatorial characterization of optimum distance full flag codes.
		
		\begin{theorem}
			Let $\cC$ be a full flag code on $\bbF_q^n$. They are equivalent:
			\begin{enumerate}
				\item $d_f(\cC)=D^n$ (or $\bar{d}_f(\cC)=0$).
				\item The set  $\Gamma(\cC)$ consists of the only distance path passing either through the point $(\frac{n}{2},\frac{n}{2})$, if $n$ is even, or through the points $(\left\lfloor \frac{n}{2}\right\rfloor, \left\lfloor \frac{n}{2}\right\rfloor)$ and $(\left\lceil \frac{n}{2}\right\rceil, \left\lfloor \frac{n}{2}\right\rfloor)$, if $n$ is odd.
				\item The set of Ferrers subdiagrams associated to $\cC$ is
				$$
				\mathfrak{F}(\cC)
				=
				\left\lbrace
				\begin{array}{llll}
					\{ \mathfrak{F}_0\}             & \text{if} & n & \text{is even or}\\
					\{ \mathfrak{F}_0, \mathfrak{F}_{(1)} \} & \text{if} & n & \text{is odd.}
				\end{array}
				\right.
				$$
			\end{enumerate}
		\end{theorem}
		\begin{proof}
			Observe that, by means of Corollary \ref{cor: codistance}, the condition $\bar{d}_f(\cC)=0$ holds if, and only if, distance paths in $\Gamma(\cC)$ leave no point above them. In other words, $\cC$ has a unique distance path: the one that passes trough the points of $S(n)$ having coordinates $(i, \min\{i ,n-i\})$, for all $0\leq i\leq n$ (see the picture below). By means of the trident rules (\ref{eq: rule 1}) and (\ref{eq: rule 2}), this is exactly the only distance path passing through  $(\frac{n}{2},\frac{n}{2})$, if $n$ is even, or through $(\left\lfloor \frac{n}{2}\right\rfloor, \left\lfloor \frac{n}{2}\right\rfloor)$ and $(\left\lceil \frac{n}{2}\right\rceil, \left\lfloor \frac{n}{2}\right\rfloor)$, if $n$ is odd.
			\begin{figure}[H]
				\centering
				\begin{tikzpicture}[line cap=round,line join=round,>=triangle 45,x=0.5cm,y=0.25cm]
					\draw [line width=1pt,color=red] (0,0)-- (1,2);
					\draw [line width=1pt,color=red] (1,2)-- (2,4);
					\draw [line width=1pt,color=red] (2,4)-- (3,6);
					\draw [line width=1pt,color=red] (3,6)-- (4,8);
					\draw [line width=1pt,color=red] (4,8)-- (5,6);
					\draw [line width=1pt,color=red] (5,6)-- (6,4);
					\draw [line width=1pt,color=red] (6,4)-- (7,2);
					\draw [line width=1pt,color=red] (7,2)-- (8,0);		
					\begin{scriptsize}
						\draw [color=black] (0,0) node[cross] {};
						\draw [color=black] (1,0) node[cross] {};
						\draw [color=black] (2,0) node[cross] {};
						\draw [color=black] (3,0) node[cross] {};
						\draw [color=black] (4,0) node[cross] {};
						\draw [color=black] (5,0) node[cross] {};
						\draw [color=black] (6,0) node[cross] {};
						\draw [color=black] (7,0) node[cross] {};
						\draw [color=black] (8,0) node[cross] {}; 
						\draw [fill=black] (1,2) circle (1.5pt);
						\draw [fill=black] (2,2) circle (1.5pt);
						\draw [fill=black] (2,4) circle (1.5pt);
						\draw [fill=black] (3,2) circle (1.5pt);
						\draw [fill=black] (3,4) circle (1.5pt);
						\draw [fill=black] (3,6) circle (1.5pt);
						\draw [fill=black] (4,2) circle (1.5pt);
						\draw [fill=black] (4,4) circle (1.5pt);
						\draw [fill=black] (4,6) circle (1.5pt);
						\draw [fill=black] (4,8) circle (1.5pt);
						\draw [fill=black] (5,2) circle (1.5pt);
						\draw [fill=black] (5,4) circle (1.5pt);
						\draw [fill=black] (5,6) circle (1.5pt);
						\draw [fill=black] (6,2) circle (1.5pt);
						\draw [fill=black] (6,4) circle (1.5pt);
						\draw [fill=black] (7,2) circle (1.5pt);
					\end{scriptsize}
				\end{tikzpicture}
				\hspace{0.5cm}
				\begin{tikzpicture}[line cap=round,line join=round,>=triangle 45,x=0.5cm,y=0.25cm]
					\draw [line width=1pt,color=red] (0,0)-- (1,2);
					\draw [line width=1pt,color=red] (1,2)-- (2,4);
					\draw [line width=1pt,color=red] (2,4)-- (3,6);
					\draw [line width=1pt,color=red] (3,6)-- (4,6);
					\draw [line width=1pt,color=red] (4,6)-- (5,4);
					\draw [line width=1pt,color=red] (5,4)-- (6,2);
					\draw [line width=1pt,color=red] (6,2)-- (7,0);
					\begin{scriptsize}
						\draw [color=black] (0,0) node[cross] {};
						\draw [color=black] (1,0) node[cross] {};
						\draw [color=black] (2,0) node[cross] {};
						\draw [color=black] (3,0) node[cross] {};
						\draw [color=black] (4,0) node[cross] {};
						\draw [color=black] (5,0) node[cross] {};
						\draw [color=black] (6,0) node[cross] {};
						\draw [color=black] (7,0) node[cross] {};
						\draw [fill=black] (1,2) circle (1.5pt);
						\draw [fill=black] (2,2) circle (1.5pt);
						\draw [fill=black] (2,4) circle (1.5pt);
						\draw [fill=black] (3,2) circle (1.5pt);
						\draw [fill=black] (3,4) circle (1.5pt);
						\draw [fill=black] (3,6) circle (1.5pt);
						\draw [fill=black] (4,2) circle (1.5pt);
						\draw [fill=black] (4,4) circle (1.5pt);
						\draw [fill=black] (4,6) circle (1.5pt);
						\draw [fill=black] (5,2) circle (1.5pt);
						\draw [fill=black] (5,4) circle (1.5pt);
						\draw [fill=black] (6,2) circle (1.5pt);
					\end{scriptsize}
				\end{tikzpicture}
				\caption{Maximum distance paths with $n$ even (left) and $n$ odd (right).}
			\end{figure}
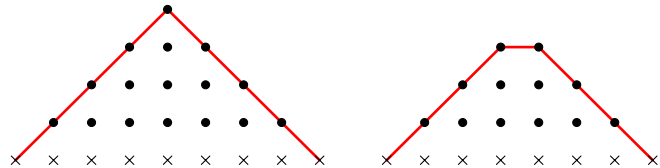
As a result, statements $(1)$ and $(2)$ are equivalent. Moreover, assuming $(2)$, and according to Definition \ref{def: ferrers diagrams of a code}, we conclude that only Ferrers diagrams $\mathfrak{F}_0$ and $\mathfrak{F}_{(1)}$ (if $n$ is odd) appear in $\mathfrak{F}(\cC)$. On the other hand, if condition $(3)$ holds, the underlying black diagrams associated to the flag code $\cC$ are empty and then, the associated value of the codistance $\bar{d}_f(\cC)$ is zero, which finishes the proof.
	\end{proof}
	\end{example}

	\begin{remark}
	The result given in \cite[Th. 3.11]{CasoPlanarCHAP9} can also be proved in our new combinatorial terms. Observe that condition $(3)$ in the previous theorem is equivalent to say that, for every $1\leq i\leq \lfloor \frac{n}{2}\rfloor$, the set $D_{n-2i}(\cC)$ contains just one element: the Durfee $(n-2i)$-rectangle with zero rows. Hence, by means of Theorem \ref{theo: dist proj durfee even 1}, we conclude that every projected code of dimension up to $\lfloor \frac{n}{2}\rfloor$ attains the maximum possible distance, i.e., $d_I(\cC_i)=i$, and has size $|\cC_i|=|\cC|$. For those projected codes of higher dimensions, as stated in Remark \ref{rem: higher dimensions}, it suffices to consider rectangles with more columns than rows. 
	\end{remark}
	
	\section{Conclusions and future work}
	
	In this paper, we have undertaken a detailed study of the flag distance in terms of different combinatorial objects. To this end, we have first devised a nice way to graphically represent this numerical parameter through distance paths drawn in a distance support whose particular shape has led us to design an associated Ferrers diagram frame. Hence, we have established a one-to-one correspondence between the set of distance paths (associated to a precise distance value $d$) and the set of underlying distributions of Ferrers subdiagrams having exactly $\bar{d}$ (the corresponding codistance value) black points. This fact allows us to perfectly translate properties related to the flag distance into the language of integer partitions. Moreover, we take advantage of this dictionary to suitably associate a family of Ferrers subdiagrams (and their corresponding Durfee rectangles) to a given full flag code. Finally, we show how these objects result very useful to make connections between the parameters of the flag code (minimum distance and cardinality) and the ones of the corresponding projected codes.
	
	In future research, we want apply the dictionary established in this paper to derive new results concerning specific families of full flag codes. Moreover, we would also like to adapt the ideas in the current work to the flag variety of general type vector, where the distance support loses its triangular shape.

\end{document}